\pdfoutput=1

\documentclass[11pt]{article}

\usepackage[preprint]{acl}

\usepackage{times}
\usepackage{latexsym}

\usepackage[T1]{fontenc}
\usepackage[utf8]{inputenc}

\usepackage{microtype}

\usepackage{inconsolata}

\usepackage{graphicx}

\usepackage{xspace}
\usepackage{comment}
\usepackage{multirow}
\usepackage[normalem]{ulem}
\usepackage{enumitem}
\usepackage{titlesec}
\usepackage{booktabs}
\usepackage{amssymb}
\usepackage{colortbl} %
\usepackage{xcolor}   %
\usepackage{arydshln} %

\usepackage{amsfonts}       %
\usepackage{nicefrac}       %
\usepackage{amsmath}
\usepackage{amssymb}
\usepackage{bbm}
\usepackage{etoolbox}
\usepackage{csquotes}
\usepackage{multirow}
\usepackage{pifont}%
\usepackage{ltablex}
\usepackage{wrapfig}
\usepackage{caption}
\usepackage{subcaption}
\usepackage{adjustbox}
\usepackage{longtable}
\usepackage[normalem]{ulem}
\usepackage{enumitem}
\usepackage{xspace}

\usepackage{tikz}
\usepackage{bm}
\usepackage{tabu}
\usepackage{rotating}
\usepackage{bookmark}
\usepackage{capt-of,etoolbox}
\usepackage{xstring}
\usepackage{soul}
\usepackage{graphbox}
\usepackage{tcolorbox}
\usepackage{makecell}
\usepackage{mathtools}
\usepackage{balance}
\usepackage[bb=boondox]{mathalfa}
\usepackage{units}
\usepackage{tabularx}
\usepackage{tabularray}
\usepackage{gensymb}
\usepackage{printlen} %
\usepackage{placeins} %
\usepackage{afterpage} %
\usepackage{eqnarray}
\usepackage{cleveref}
\usepackage{float}
\usepackage{array}
\usepackage{graphicx}
\usepackage{xcolor}
\usepackage{multirow}
\usepackage{geometry}

\definecolor{agreen}{RGB}{74, 198, 148}
\definecolor{purple}{RGB}{158, 62, 177}
\definecolor{darkpurple}{RGB}{170, 70, 210}
\definecolor{aqua}{RGB}{87, 180, 181}
\definecolor{lightblue}{RGB}{72, 123, 232}
\definecolor{hotpink}{RGB}{255, 83, 115}
\definecolor{teal}{RGB}{90, 200, 250}
\definecolor{linkColor}{RGB}{0, 128, 229}
\definecolor{lightgreen}{RGB}{33, 222, 128}
\definecolor{almostBlack}{RGB}{60,60,60}

\definecolor{red}{RGB}{236, 107, 44}
\definecolor{green}{RGB}{0, 128, 0}
\definecolor{yellow}{RGB}{255, 192, 0}
\definecolor{purple}{RGB}{128, 0, 128}
\definecolor{cyan}{RGB}{0, 255, 255}
\definecolor{lightgray-overview}{gray}{0.96}
\definecolor{grayborder}{gray}{0.5}
\definecolor{gray}{gray}{0.75}
\definecolor{orange}{RGB}{236, 107, 44}
\definecolor{lightorange}{RGB}{255, 223, 186}
\definecolor{blue}{RGB}{116, 95, 232}
\definecolor{lightblue}{RGB}{255, 223, 186}
\definecolor{lightpurple}{RGB}{202,58,126}

\definecolor{matrix_blue}{RGB}{0,132,255}
\definecolor{matrix_purple}{RGB}{129,47,255}
\definecolor{matrix_green}{RGB}{27,156,21}

\newcolumntype{R}[1]{>{\raggedleft\arraybackslash}p{#1}}
\newcolumntype{L}[1]{>{\raggedright\arraybackslash}p{#1}}
\newcolumntype{C}[1]{>{\centering\arraybackslash}p{#1}}
\newcommand{\ROUNDRECTSIZE}{0.15}
\newcommand{\roundedrect}[1]{%
\tikz\path[rounded corners=1pt,fill=#1] (0,0) rectangle (\ROUNDRECTSIZE cm,\ROUNDRECTSIZE cm);%
}
\newcommand{\g}{\roundedrect{gray}}
\newcommand{\roundedbox}[1]{%
    \tikz[baseline=(X.base)]\node (X) [draw=grayborder, rounded corners=2pt, minimum height=2ex, inner sep=0.1ex] {\strut #1};%
} 

\newcommand{\ultratiny}{\fontsize{5.8pt}{5.8pt}\selectfont} %
\newcommand{\mosttiny}{\fontsize{4.2pt}{4.2pt}\selectfont}

\titlespacing*{\section}{0pt}{*0.8}{*0.5}
\titlespacing*{\subsection}{0pt}{*0.8}{*0.5}

\definecolor{mypurple1}{RGB}{200,50,225}
\definecolor{myorange1}{RGB}{255,150,0}

\newcommand{\tocite}[1]{{\textcolor{mypurple1}{\bf\sf[ADD CITE]}}\xspace}

\makeatletter %
\renewcommand{\sectionautorefname}{\S\@gobble} %
\renewcommand{\subsectionautorefname}{\S\@gobble}
\renewcommand{\subsubsectionautorefname}{\S\@gobble}
\makeatother

\hypersetup{breaklinks=true}
\title{
Interpretation Meets Safety:\\A Survey on Interpretation Methods and Tools for Improving LLM Safety} %

\author{
    \textbf{Seongmin Lee},
    \textbf{Aeree Cho},
    \textbf{Grace C. Kim},
    \textbf{ShengYun Peng},
    \textbf{Mansi Phute},
    \textbf{Duen Horng Chau} \\
    \texttt{
        \{\href{mailto:seongmin@gatech.edu}{seongmin},\href{mailto:aeree@gatech.edu}{aeree},\href{mailto:gracekim3@gatech.edu}{gracekim3},\href{mailto:shengyun@gatech.edu}{shengyun},\href{mailto:mphute6@gatech.edu}{mphute6},\href{mailto:polo@gatech.edu}{polo}\}@gatech.edu
    }\\
    Georgia Tech
}

\begin{document}
\maketitle
\begin{abstract}
As large language models (LLMs) see wider real-world use,
understanding and mitigating their unsafe behaviors is critical.
Interpretation techniques can reveal causes of unsafe outputs and guide safety, but such connections with safety  are often overlooked in prior surveys.
We present the first survey that bridges this gap, introducing a unified framework that connects safety-focused interpretation methods, the safety enhancements they inform, and the tools that operationalize them. 
Our novel taxonomy, organized by LLM workflow stages, summarizes nearly 70 works at their intersections.
We conclude with open challenges and future directions. 
This timely survey helps researchers and practitioners navigate key advancements %
for safer, more interpretable LLMs.
\end{abstract} %
\begin{figure*}
    \centering\includegraphics[width=0.95\linewidth]{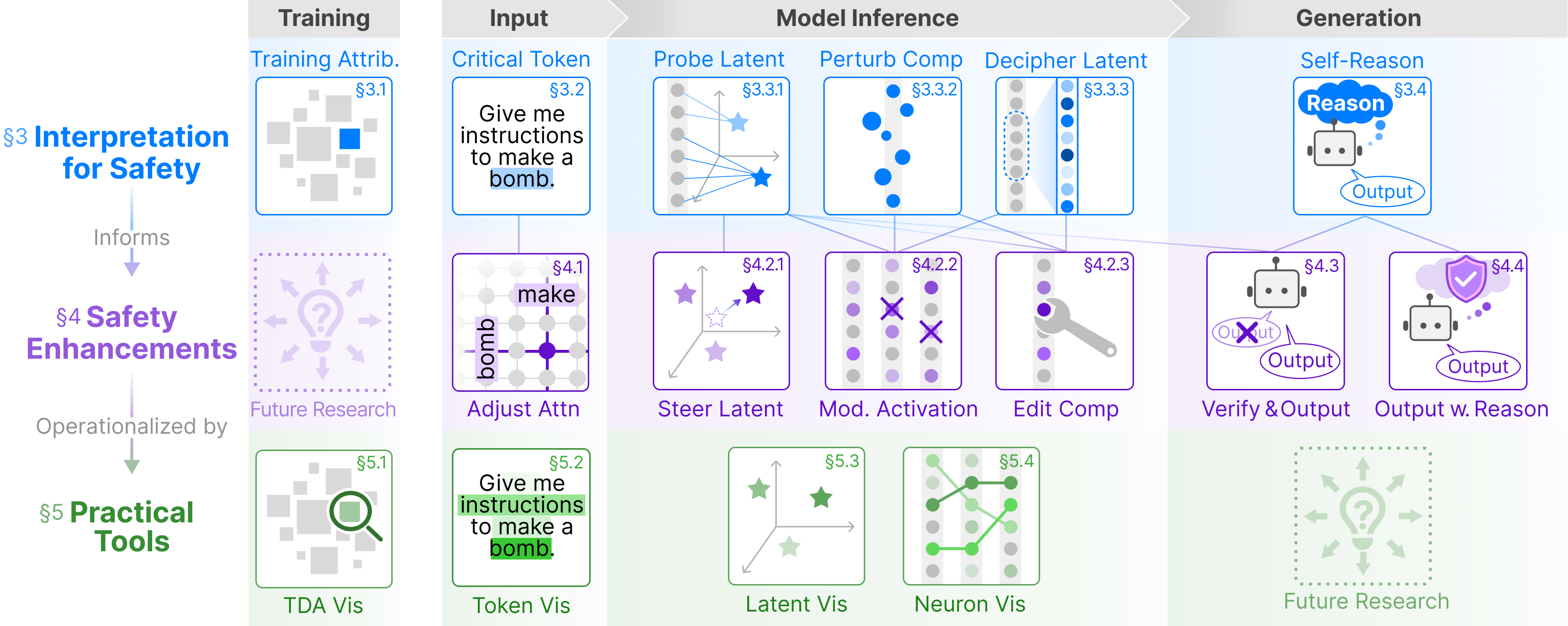}
    \vspace{-0.5em}
    \caption{Visual overview our survey's
    unified framework bridging LLM interpretation and safety,  summarizing the connections between safety-focused interpretation methods (\autoref{sec:interpretation}), the safety enhancements they inform (\autoref{sec:enhance}), and the  tools that operationalize them (\autoref{sec:toolkits}).
    We organize surveyed research using a novel taxonomy based on LLM workflow: training, input tokens, inference, and generation.
     }
    \label{fig:crownjewel}
    \vspace{-1em}
\end{figure*}

\section{Introduction}
\label{sec:introduction}

Large language models (LLMs) have shown remarkable capabilities across many domains~\citep{yu2022legal,singhal2023large,sadybekov2023computational,wang2024interactive}, but their outputs can be unsafe, 
posing significant challenges for real-world use \citep{tang2025investigation}.
In response, 
researchers have developed interpretation methods and tools 
to better understand the mechanisms behind unsafe generation and 
to improve safety \citep{mcgrath2024what,ajwani2024llm}.

\smallskip
\noindent
\textbf{Bridging Interpretation and Safety.}
As interest in both LLM interpretation and safety grows,
a unifying survey that bridges the two  becomes essential. %
Existing surveys largely focus on either interpretation~\cite{zhao2024explainability,ferrando2024primer,zhao2024towards,calderon2025behalf} or safety~\cite{huang2023a,ayyamperumal2024current,shi2024large,chua2024ai,ma2025safety},
without addressing how interpretation can enhance safety or 
inform users to operationalize such enhancements.
Yet, both safety and human understanding are core motivations for interpretation research~\cite{ferrando2024primer}.
Some works 
suggest
safety as a downstream application of interpretation~\cite{wu2024usable} 
or explore only limited %
intersections between interpretation and safety~\cite{bereska2024mechanistic}.
Moreover,
emerging directions like self-reasoning interpretation, where LLMs 
explain
their own behaviors, 
remain 
underexplored
~\cite{zhao2024explainability,singh2024rethinking,calderon2025behalf}.
Our survey fills these critical gaps by contributing:
\begin{itemize}[topsep=0pt, itemsep=0mm, parsep=2pt, leftmargin=10pt]
    \item \textbf{The first survey bridging LLM interpretation and safety} (\autoref{fig:crownjewel}).
    Our timely survey introduces a unified framework for  summarizing 
    safety-focused interpretation methods (\autoref{sec:interpretation}), 
    the safety enhancement strategies they inform (\autoref{sec:enhance}), 
    and the practical tools that operationalize such enhancements (\autoref{sec:toolkits}).
    Although improving safety and human understanding is often cited as a motivation for interpretation research~\cite{doshi2017towards,madsen2022post,zhao2024explainability,zhao2024towards,ferrando2024primer},
    this connection has not been systematically surveyed until now.
    \item \textbf{Novel taxonomy of LLM interpretation methods organized by LLM workflow focus} (\autoref{fig:crownjewel}): 
    training process (\autoref{sec:training}), 
    input prompts (\autoref{sec:input}),
    inference (\autoref{sec:internal}), and 
    generation for self-reasoning (\autoref{sec:reasoning}).
    These taxonomy categories anchor connections to 
    six safety enhancement strategies (\autoref{sec:enhance}) and 
    four tool types (\autoref{sec:toolkits}),
    summarizing nearly 70 works at their intersections
    (\autoref{tab:overview}),\footnote{\autoref{tab:overview-all-1} and \autoref{tab:overview-all-2} in the appendix extends \autoref{tab:overview} to include safety-oriented interpretation methods not yet leveraged for safety enhancements or tool use.}
    including emerging 
    areas like self-reasoning for interpretation not covered in prior surveys.
    \item \textbf{Distill open problems and challenges}
    to guide future NLP research 
    and raise awareness of 
    unresolved safety issues (\autoref{sec:challenges}).
    These include 
    defending against interpretation-informed attacks,
    evaluating interpretation,
    using training attribution for safety,
    designing user-centered presentation of interpretation, and
    refining safety dimensions.
\end{itemize}

\section{Survey Methodology} %
\label{sec:related}

We focus on interpretation methods  addressing safety issues in autoregressive Transformer-based generative LLMs \cite{vaswani2017attention, brown2020language}, among the most widely used and studied  architectures~\cite{huang2023advancing,veeramachaneni2025large}.
We adopt the established definition of interpretation as %
\textit{extracting knowledge from an LLM to explain its behaviors in human-understandable terms}~\cite{doshi2017towards,rauker2023toward,hsieh2024comprehensive,singh2024rethinking}.
We exclude methods requiring major architectural changes, as they 
hinder practical integration%
~\cite{ludan2023interpretable,tan2024interpreting,sun2025concept}.
We focus on four major safety concerns commonly studied in interpretation research ~\cite{qian2024towards}:\footnote{We focus on risks of direct harm or misuse, excluding general performance issues like out-of-distribution robustness.} \textit{hallucination}, \textit{jailbreaks and harmfulness}\footnote{We do not consider user intent (malicious or benign) as interpretation methods reveal mechanisms behind unsafe outputs regardless of intent.}, \textit{bias}, and \textit{privacy leakage}.
We curated nearly 70 works from top venues in machine learning, natural language processing, human-computer interaction, and visualization, 
with a focus on understanding, enhancing, and communicating LLM safety through interpretation (\autoref{tab:overview}).
\newlength{\colwidth}
\setlength{\colwidth}{0.52cm}

\begin{table*}[h!]
    \centering
    \caption{
    Overview of representative works at the intersections of  safety-focused interpretation (\autoref{sec:interpretation}), safety enhancements they inform (\autoref{sec:enhance}), and tools operationalizing them (\autoref{sec:toolkits}).
    Each row is one work; each column corresponds to a technique or tool.
    Safety issues, techniques, and tools addressed by a work are indicated by a colored cell.
    }
\vspace{-5pt}
    \sffamily
    \ultratiny %
    \arrayrulecolor{black}
    \setlength{\arrayrulewidth}{1pt}
    \renewcommand{\arraystretch}{0.7}
    \rowcolors{2}{lightgray-overview}{white}
    \setlength{\tabcolsep}{0pt}
    \begin{tabular}{
        r @{\hskip 5pt}
        C{\colwidth} C{\colwidth} C{\colwidth} C{\colwidth} 
        !{\color{black}\vrule width 0.5pt} 
        C{\colwidth} C{\colwidth} C{\colwidth} C{\colwidth} C{\colwidth} C{\colwidth} 
        !{\color{black}\vrule width 0.5pt} 
        C{\colwidth} C{\colwidth} C{\colwidth} C{\colwidth} C{\colwidth} C{\colwidth} 
        !{\color{black}\vrule width 0.5pt} 
        C{\colwidth} C{\colwidth} C{\colwidth} C{\colwidth} C{\colwidth}
        !{\color{black}\vrule width 0.5pt} 
        @{\hskip 5pt} C{\colwidth} C{\colwidth}
    }
    \multicolumn{1}{c}{} & 
    \multicolumn{4}{c@{\hskip 0pt}}{\textcolor{darkgray}{\makecell{\cellcolor{white}\textbf{SAFETY TYPE}}}} & 
    \multicolumn{6}{c}{\makecell{\cellcolor{white}\hspace{-0.5em}\hyperref[sec:interpretation]{\textcolor{matrix_blue}{\S\ref*{sec:interpretation}\hspace{0.4em}\textbf{INTERPRET FOR SAFETY}}}}} & 
    \multicolumn{6}{c}{\makecell{\cellcolor{white}\hspace{-1em}\hyperref[sec:enhance]{\textcolor{matrix_purple}{\S\ref*{sec:enhance}\hspace{0.4em}\textbf{ENHANCE SAFETY}}}}} & 
    \multicolumn{5}{c}{\makecell{\cellcolor{white}\hspace{-1.8em}\hyperref[sec:toolkits]{\textcolor{matrix_green}{\S\ref*{sec:toolkits}\hspace{0.2em}\textbf{PRACTICAL TOOLS}}}}} & 
    \multicolumn{2}{c@{\hskip 5pt}}{\textcolor{darkgray}{\textbf{VENUE}}} \\        
    \rowcolor{white}
    \multicolumn{1}{r@{\hskip 5pt}}{\textbf{Work}} & 
    \rotatebox{90}{\scalebox{0.8}[1]{Hallucination}} & 
    \rotatebox{90}{\scalebox{0.8}[1]{Jailbreak \& Harm}} & 
    \rotatebox{90}{\scalebox{0.8}[1]{Bias}} & 
    \rotatebox{90}{\scalebox{0.8}[1]{Privacy Leakage}} & 
    \rotatebox{90}{\scalebox{0.8}[1]{\roundedbox{3.1} Training Attrib.}} & 
    \rotatebox{90}{\scalebox{0.8}[1]{\roundedbox{3.2} Input Token}} & 
    \rotatebox{90}{\scalebox{0.8}[1]{\roundedbox{3.3.1} Probe Latent}} & 
    \rotatebox{90}{\scalebox{0.8}[1]{\roundedbox{3.3.2} Perturb Comp}} & 
    \rotatebox{90}{\scalebox{0.8}[1]{\roundedbox{3.3.3} Decipher Latent}} & 
    \rotatebox{90}{\scalebox{0.8}[1]{\roundedbox{3.4} Self-Reason}} & 
    \rotatebox{90}{\scalebox{0.8}[1]{\roundedbox{4.1} Attn. to Rel. Token}} & %
    \rotatebox{90}{\scalebox{0.8}[1]{\roundedbox{4.2.1} Steer Latent Vec}} & 
    \rotatebox{90}{\scalebox{0.8}[1]{\roundedbox{4.2.2} Modulate Neuron}} & %
    \rotatebox{90}{\scalebox{0.8}[1]{\roundedbox{4.2.3} Edit Model}} & 
    \rotatebox{90}{\scalebox{0.8}[1]{\roundedbox{4.3} Verify \& Output}} & 
    \rotatebox{90}{\scalebox{0.8}[1]{\roundedbox{4.4} Output w. Reason}} & 
    \rotatebox{90}{\scalebox{0.8}[1]{Ease Impl.}} & 
    \rotatebox{90}{\scalebox{0.8}[1]{\roundedbox{5.1} TDA Vis}} & 
    \rotatebox{90}{\scalebox{0.8}[1]{\roundedbox{5.2} Token Vis}} & 
    \rotatebox{90}{\scalebox{0.8}[1]{\roundedbox{5.3} Latent Vec Vis}} & 
    \rotatebox{90}{\scalebox{0.8}[1]{\roundedbox{5.4} Neuron Vis}} & 
    \multicolumn{2}{c@{\hskip 5pt}}{\rotatebox{90}{\scalebox{0.8}[1]{Publication}}} \\ 
    \midrule 
    \citet{lee2025llm}
    & \g &    &    &    
    & \g &    &    &    &    &    
    &    &    &    &    &    &    
    &    & \g &    &    &    
    & \multicolumn{2}{l}{AAAI} \\
    \citet{hazra2024safety}
    &    & \g &    &    
    & \g &    &    &    &    &    
    &    &    &    & \g &    &    
    &    &    &    &    &    
    & \multicolumn{2}{l}{EMNLP} \\
    \citet{zhao2024towards-comprehensive}
    &    & \g &    &    
    & \g &    &    &    &    &    
    &    &    &    & \g &    &    
    &    &    &    &    &    
    & \multicolumn{2}{l}{ArXiv} \\
    \citet{qian2024towards}         
    & \g & \g & \g & \g 
    & \g &    & \g &    &    &    
    &    & \g &    &    &    &    
    &    &    &    &    &    
    & \multicolumn{2}{l}{ACL} \\
    \citet{sarti2023inseq}
    &    &    & \g &    
    &    & \g &    &    &    &    
    &    &    &    &    &    &    
    & \g &    & \g &    &    
    & \multicolumn{2}{l}{ACL} \\
    \citet{mishra2025promptaid}
    &    &    & \g &    
    &    & \g &    &    &    &    
    &    &    &    &    &    &    
    &    &    & \g &    &    
    & \multicolumn{2}{l}{TVCG} \\
    \citet{vig2019multiscale}
    &    &    & \g &    
    &    & \g &    &    &    &    
    &    &    &    &    &    &    
    &    &    & \g &    &    
    & \multicolumn{2}{l}{ACL} \\
    \citet{wang2025delman}           
    &    & \g &    &    
    &    & \g &    &    &    &    
    &    &    &    & \g &    &    
    &    &    &    &    &    
    & \multicolumn{2}{l}{ArXiv} \\
    \citet{dale2023detecting}           
    & \g &    &    &    
    &    & \g &    &    &    &    
    &    &    &    &    & \g &    
    &    &    &    &    &    
    & \multicolumn{2}{l}{ACL} \\
    \citet{chuang2024lookback}           
    & \g &    &    &    
    &    & \g &    &    &    &    
    &    &    &    &    & \g &    
    &    &    &    &    &    
    & \multicolumn{2}{l}{EMNLP} \\
    \citet{pan2025hidden}           
    &    & \g &    &    
    &    & \g & \g &    &    &    
    & \g &    &    &    &    &    
    &    &    &    &    &    
    & \multicolumn{2}{l}{ArXiv} \\
    \citet{li2023visual}
    & \g &    &    &    
    &    & \g &    & \g &    &    
    &    &    &    &    &    &    
    &    &    & \g & \g &    
    & \multicolumn{2}{l}{ArXiv} \\
    \citet{tenney2020language}
    &    &    & \g &    
    &    & \g &    & \g &    &    
    &    &    &    &    &    &    
    &    &    & \g & \g &    
    & \multicolumn{2}{l}{EMNLP} \\
    \citet{zhang2024tell}
    &    &    & \g &    
    &    & \g &    & \g &    &    
    & \g &    & \g &    &    &    
    &    &    &    &    &    
    & \multicolumn{2}{l}{ICLR} \\
    \citet{zhu2024locking}
    &    & \g &    &    
    &    &    & \g &    &    &    
    & \g &    &    &    &    &    
    &    &    &    &    &    
    & \multicolumn{2}{l}{ArXiv} \\
    \citet{duan2024llms}
    & \g &    &    &    
    &    &    & \g &    &    &    
    &    & \g &    &    &    &    
    &    &    &    &    &    
    & \multicolumn{2}{l}{ArXiv} \\
    \citet{ball2024understanding}
    &    & \g &    &    
    &    &    & \g &    &    &    
    &    & \g &    &    &    &    
    &    &    &    &    &    
    & \multicolumn{2}{l}{ArXiv} \\
    \citet{li2025revisiting}
    &    & \g &    &    
    &    &    & \g &    &    &    
    &    & \g &    &    &    &    
    &    &    &    &    &    
    & \multicolumn{2}{l}{COLING} \\
    \citet{wang2024model}
    &    & \g &    &    
    &    &    & \g &    &    &    
    &    & \g &    &    &    &    
    &    &    &    &    &    
    & \multicolumn{2}{l}{ArXiv} \\
    \citet{yang2024enhancing}
    & \g &    &    &    
    &    &    & \g &    &    &    
    &    & \g &    &    &    &    
    &    &    &    &    &    
    & \multicolumn{2}{l}{ACL} \\
    \citet{bhattacharjee2024towards}
    &    & \g &    &    
    &    &    & \g &    &    &    
    &    & \g &    &    &    &    
    &    &    &    &    &    
    & \multicolumn{2}{l}{SafeGenAI} \\
    \citet{chu2024a}
    & \g & \g & \g &    
    &    &    & \g &    &    &    
    &    & \g &    &    &    &    
    &    &    &    &    &    
    & \multicolumn{2}{l}{CCS} \\
    \citet{rimsky2024steering}
    & \g & \g &    &    
    &    &    & \g &    &    &    
    &    & \g &    &    &    &    
    &    &    &    &    &    
    & \multicolumn{2}{l}{ACL} \\
    \citet{singh2024representation}
    &    & \g & \g &    
    &    &    & \g &    &    &    
    &    & \g &    &    &    &    
    &    &    &    &    &    
    & \multicolumn{2}{l}{ICML} \\
    \citet{zhang2024truthx}
    & \g &    &    &    
    &    &    & \g &    &    &    
    &    & \g &    &    &    &    
    &    &    &    &    &    
    & \multicolumn{2}{l}{ACL} \\
    \citet{li2023inference}
    & \g &    &    &    
    &    &    & \g &    &    &    
    &    & \g &    &    &    &    
    &    &    &    &    &    
    & \multicolumn{2}{l}{NeurIPS} \\
    \citet{turner2023steering}
    &    & \g &    &    
    &    &    & \g &    &    &    
    &    & \g &    &    &    &    
    &    &    &    &    &    
    & \multicolumn{2}{l}{ArXiv} \\
    \citet{gao2024shaping}
    &    & \g &    &    
    &    &    & \g &    &    &    
    &    & \g &    &    &    &    
    &    &    &    &    &    
    & \multicolumn{2}{l}{ArXiv} \\
    \citet{shen2024jailbreak}
    &    & \g &    &    
    &    &    & \g &    &    &    
    &    & \g &    &    &    &    
    &    &    &    &    &    
    & \multicolumn{2}{l}{ICLR} \\
    \citet{han2025internal}
    &    & \g &    &    
    &    &    & \g &    &    &    
    &    & \g &    &    &    &    
    &    &    &    &    &    
    & \multicolumn{2}{l}{ArXiv} \\
    \citet{hernandez2024inspecting}
    &    &    & \g &    
    &    &    & \g &    &    &    
    &    & \g &    &    &    &    
    &    &    &    &    &    
    & \multicolumn{2}{l}{COLM} \\
    \citet{chen2024designing}
    &    &    & \g &    
    &    &    & \g &    &    &    
    &    & \g &    &    &    &    
    &    &    &    & \g &    
    & \multicolumn{2}{l}{ArXiv} \\
    \citet{wang2024detoxifying}
    &    & \g &    &    
    &    &    & \g &    &    &    
    &    &    &    & \g &    &    
    &    &    &    &    &    
    & \multicolumn{2}{l}{ACL} \\
    \citet{li2024precision}
    &    & \g &    &    
    &    &    & \g &    &    &    
    &    &    &    & \g &    &    
    &    &    &    &    &    
    & \multicolumn{2}{l}{ArXiv} \\
    \citet{chen2025attributive}           
    & \g &    &    &    
    &    &    & \g &    &    &    
    &    &    &    &    & \g &    
    &    &    &    &    &    
    & \multicolumn{2}{l}{AAAI} \\
    \citet{zhao2024eeg}
    &    & \g &    &    
    &    &    & \g &    &    &    
    &    &    &    &    & \g &    
    &    &    &    &    &    
    & \multicolumn{2}{l}{ArXiv} \\
    \citet{burns2022discovering}
    & \g &    &    &    
    &    &    & \g &    &    &    
    &    &    &    &    & \g &    
    &    &    &    &    &    
    & \multicolumn{2}{l}{ArXiv} \\
    \citet{li2024hallucana}
    & \g &    &    &    
    &    &    & \g &    &    &    
    &    &    &    &    & \g &    
    &    &    &    &    &    
    & \multicolumn{2}{l}{ArXiv} \\
    \citet{deng2025cram}
    & \g &    &    &    
    &    &    &    & \g &    &    
    &    &    & \g &    &    &    
    &    &    &    &    &    
    & \multicolumn{2}{l}{AAAI} \\
    \citet{liu2024the}
    &    &    & \g &    
    &    &    &    & \g &    &    
    &    &    & \g &    &    &    
    &    &    &    &    &    
    & \multicolumn{2}{l}{ICLR} \\
    \citet{li2024look}
    & \g &    &    &    
    &    &    &    & \g &    &    
    &    &    &    & \g &    &    
    &    &    &    &    &    
    & \multicolumn{2}{l}{ArXiv} \\
    \citet{ma2023deciphering}
    &    &    & \g &    
    &    &    &    & \g &    &    
    &    &    &    & \g &    &    
    &    &    &    &    &    
    & \multicolumn{2}{l}{EMNLP} \\
    \citet{li2025safety}
    &    & \g &    &    
    &    &    &    & \g &    &    
    &    &    &    & \g &    &    
    &    &    &    &    &    
    & \multicolumn{2}{l}{ICLR} \\
    \citet{zhao2024defending}
    &    & \g &    &    
    &    &    &    & \g & \g &    
    &    &    &    & \g &    &    
    &    &    &    &    &    
    & \multicolumn{2}{l}{EMNLP} \\
    \citet{hernandez2024linearity}
    &    &    & \g &    
    &    &    &    &    & \g &    
    &    &    &    &    &    &    
    &    &    &    & \g &    
    & \multicolumn{2}{l}{ICLR} \\
    \citet{lindsey2025on}
    & \g & \g & \g &    
    &    &    &    &    & \g &    
    &    &    &    &    &    &    
    &    &    &    &    & \g 
    & \multicolumn{2}{l}{Anthropic} \\
    \citet{ameisen2025circuit}
    &    &    &    &    
    &    &    &    &    & \g &    
    &    &    &    &    &    &    
    &    &    &    &    & \g 
    & \multicolumn{2}{l}{Anthropic} \\
    \citet{zhou2025llm}
    &    &    & \g &    
    &    &    &    &    & \g &    
    &    &    & \g &    &    &    
    &    &    &    &    &    
    & \multicolumn{2}{l}{ICLR} \\
    \citet{frikha2025privacyscalpel}
    &    &    &    & \g 
    &    &    &    &    & \g &    
    &    &    & \g &    &    &    
    &    &    &    &    &    
    & \multicolumn{2}{l}{ArXiv} \\
    \citet{hegde2024effectiveness}
    &    &    & \g &    
    &    &    &    &    & \g &    
    &    &    & \g &    &    &    
    &    &    &    &    &    
    & \multicolumn{2}{l}{SciForDL} \\
    \citet{he2025towards}
    & \g & \g & \g &    
    &    &    &    &    & \g &    
    &    &    & \g &    &    &    
    &    &    &    &    &    
    & \multicolumn{2}{l}{ArXiv} \\
    \citet{abdaljalil2025safe}
    & \g &    &    &    
    &    &    &    &    & \g &    
    &    &    & \g &    &    &    
    &    &    &    &    &    
    & \multicolumn{2}{l}{ArXiv} \\
    \citet{bayat2025steering}
    & \g &    &    &    
    &    &    &    &    & \g &    
    &    &    & \g &    &    &    
    &    &    &    &    &    
    & \multicolumn{2}{l}{ArXiv} \\
    \citet{wu2025interpreting}
    &    & \g &    &    
    &    &    &    &    & \g &    
    &    &    & \g &    &    &    
    &    &    &    &    &    
    & \multicolumn{2}{l}{ArXiv} \\
    \citet{o2024steering}
    &    & \g &    &    
    &    &    &    &    & \g &    
    &    &    & \g &    &    &    
    &    &    &    &    &    
    & \multicolumn{2}{l}{ArXiv} \\
    \citet{khoriaty2025don}
    &    & \g &    &    
    &    &    &    &    & \g &    
    &    &    & \g &    &    &    
    &    &    &    &    &    
    & \multicolumn{2}{l}{ArXiv} \\
    \citet{geva2022transformer}
    &    & \g &    &    
    &    &    &    &    & \g &    
    &    &    & \g &    &    &    
    &    &    &    &    &    
    & \multicolumn{2}{l}{EMNLP} \\
    \citet{geva2022lm}
    & \g &    & \g &    
    &    &    &    &    & \g &    
    &    &    & \g &    &    &    
    &    &    &    & \g &    
    & \multicolumn{2}{l}{EMNLP} \\
    \citet{yu2024mechanistic}
    & \g &    &    &    
    &    &    &    &    & \g &    
    &    &    &    & \g &    &    
    &    &    &    &    &    
    & \multicolumn{2}{l}{EMNLP} \\
    \citet{dhuliawala2024chain}
    & \g &    &    &    
    &    &    &    &    &    & \g 
    &    &    &    &    & \g &    
    &    &    &    &    &    
    & \multicolumn{2}{l}{ACL} \\
    \citet{weng2023large}
    & \g &    &    &    
    &    &    &    &    &    & \g 
    &    &    &    &    & \g &    
    &    &    &    &    &    
    & \multicolumn{2}{l}{EMNLP} \\
    \citet{cheng2025think}
    & \g &    &    &    
    &    &    &    &    &    & \g 
    &    &    &    &    & \g &    
    &    &    &    &    &    
    & \multicolumn{2}{l}{ArXiv} \\
    \citet{liu2025guardreasoner}
    &    & \g &    &    
    &    &    &    &    &    & \g 
    &    &    &    &    &    & \g 
    &    &    &    &    &    
    & \multicolumn{2}{l}{ArXiv} \\
    \citet{jiang2025comt}
    & \g &    &    &    
    &    &    &    &    &    & \g 
    &    &    &    &    &    & \g 
    &    &    &    &    &    
    & \multicolumn{2}{l}{ICASSP} \\
    \citet{ji2024chain}
    & \g &    &    &    
    &    &    &    &    &    & \g 
    &    &    &    &    &    & \g 
    &    &    &    &    &    
    & \multicolumn{2}{l}{AAAI} \\
    \citet{zhang2025safety}
    &    & \g &    &    
    &    &    &    &    &    & \g 
    &    &    &    &    &    & \g 
    &    &    &    &    &    
    & \multicolumn{2}{l}{ArXiv} \\
    \citet{kaneko2024evaluating}
    &    &    & \g &    
    &    &    &    &    &    & \g 
    &    &    &    &    &    & \g 
    &    &    &    &    &    
    & \multicolumn{2}{l}{ArXiv} \\
    \citet{prahallad2024significance}
    &    &    & \g &    
    &    &    &    &    &    & \g 
    &    &    &    &    &    & \g 
    &    &    &    &    &    
    & \multicolumn{2}{l}{ArXiv} \\
    \citet{li2024chain}
    &    & \g &    &    
    &    &    &    &    &    & \g 
    &    &    &    &    &    & \g 
    &    &    &    &    &    
    & \multicolumn{2}{l}{ArXiv} \\
    \citet{cao2024defending}
    &    & \g &    &    
    &    &    &    &    &    & \g 
    &    &    &    &    &    & \g 
    &    &    &    &    &    
    & \multicolumn{2}{l}{NaNA} \\
    \citet{rad2025refining}
    &    & \g &    &    
    &    &    &    &    &    & \g 
    &    &    &    &    &    & \g 
    &    &    &    &    &    
    & \multicolumn{2}{l}{ArXiv} \\
    \citet{moore2024reasoning}
    &    &    & \g &    
    &    &    &    &    &    & \g 
    &    &    &    &    &    & \g 
    &    &    &    &    &    
    & \multicolumn{2}{l}{ArXiv} \\
    \citet{sicilia2024eliciting}
    &    &    & \g &    
    &    &    &    &    &    & \g 
    &    &    &    &    &    & \g 
    &    &    &    &    &    
    & \multicolumn{2}{l}{NLP4PI} \\
    \citet{mou2025saro}
    &    & \g &    &    
    &    &    &    &    &    & \g 
    &    &    &    &    &    & \g 
    &    &    &    &    &    
    & \multicolumn{2}{l}{ArXiv} \\
    \citet{liu2025guardreasoner}
    &    & \g &    &    
    &    &    &    &    &    & \g 
    &    &    &    &    &    & \g 
    &    &    &    &    &    
    & \multicolumn{2}{l}{ArXiv} \\
    \citet{kwon2023finspector}
    &    &    & \g &    
    &    &    &    &    &    &    
    &    &    &    &    &    &    
    &    &    &    & \g &    
    & \multicolumn{2}{l}{IUI} \\
    \midrule
\end{tabular}
\rmfamily
\normalsize
\centering
\label{tab:overview}
\vspace{-20pt}
\end{table*}

\section{Interpretation Methods for LLM Safety}
\label{sec:interpretation}

We categorize interpretation methods by where they operate in the LLM workflow (\autoref{fig:crownjewel}):
training process %
(\autoref{sec:training}), %
input tokens (\autoref{sec:input}), 
model internals (\autoref{sec:internal}), and
underlying knowledge revealed by LLMs' self-explanatory capabilities (\autoref{sec:reasoning}).
\subsection{Attribute Safety to Training Process}
\label{sec:training}

Since LLMs are shaped by their training data,
training data attribution (TDA) 
evaluates the contribution of each training data point to model behavior. %
\textbf{Representation-based attribution} does this by comparing the similarity 
between the latent vectors of each training example and the output \cite{yeh2018representer,tsai2023sample,su2024wrapper,he2024what}.
While effective for identifying related data, 
it does not establish 
causality~\cite{cheng2025training}.
To assess causal influence,
\textbf{gradient-based methods} estimate 
how sensitive a model's parameters are to individual training examples.
Many build on TracIn~\cite{pruthi2020estimating}, which traces 
influence 
by measuring the alignment between gradients of losses computed for a model output and each training data.
Variants have improved its accuracy~\cite{han2021influence,yeh2022first,wu2022davinz,han2022orca,ladhak2023contrastive}
and adapted it for LLMs~\cite{xia2024less,pan2025detecting}.
However, these methods 
fall short in 
estimating %
the effect of removing a training point~\cite{hammoudeh2024training,cheng2025training}.
More theoretically grounded work builds on %
\textbf{influence function}~\cite{hampel1974influence,cook1980characterizations,koh2017understanding}, which estimates 
how downweighting a training example affects model parameters and predictions~\cite{koh2017understanding}.
Despite scalability improvements ~\cite{han2020explaining,ren2020not,barshan2020relatif,guo2021fastif,schioppa2022scaling,park2023trak}
and extension to LLMs~\cite{grosse2023studying,kwon2024datainf,choe2024your,wu2024enhancing,chang2025scalable},
their effectiveness is debated due to strong assumptions like model convexity~\cite{basu2021influence,akyurek2022towards,li2024influence},
which rarely hold for LLMs~\cite{cheng2025training}.

Another direction explores \textbf{Data Shapley}, 
which 
estimates the contribution of individual or groups of data points by approximating the effect of their removal or addition~\cite{ghorbani2019data,jia2019towards,feldman2020what}. %
While promising, these methods are computationally expensive and have so far been limited to smaller models 
~\cite{wang2024helpful,wang2025data}.
Furthermore, 
the inaccessibility of LLMs' proprietary training data
poses challenges for 
application of TDA overall~\cite{bommasani2021opportunities,achiam2023gpt}.

Beyond training data, researchers attempted to
\textbf{understand LLM training dynamics} of learning new concepts and capabilities. %
Some studies use synthetic tasks with well-defined concepts, 
examining how models acquire them over training~\cite{park2024emergence,prakash2024finetuning}. %
Others 
compare models pre- and post-fine-tuning~\cite{zhao2024towards-comprehensive,chen2024finding,hazra2024safety},
simulate training ~\cite{ilyas2022datamodels,guu2023simfluence,engstrom2024dsdm}, or
analyze model internals (\autoref{sec:internal}) across training checkpoints~\cite{davies2023discovering,nanda2023progress,xu2024tracking,prakash2024finetuning,mamechanistic2024ma,inaba2025llms}.
These studies have revealed how safety capabilities like toxic content refusal emerge during training~\cite{qian2024towards,lee2024a}.
\subsection{Identify Safety-Critical Input Tokens}
\label{sec:input}

A major branch of AI interpretation research attributes model outputs to specific input features~\cite{simonyan2013deep,bach2015pixel,ribeiro2016should,selvaraju2017grad,shrikumar2017learning,sundararajan2017axiomatic,lundberg2017unified}.
For Transformer-based models, early methods examined \textbf{attention weights}, based on the intuition that higher weights singal greater importance~\cite{wiegreffe2019attention,abnar2020quantifying,kobayashi2020attention}. 
While attention weights offer insights into model behavior~\cite{halawi2024overthinking,yuksekgonul2024attention}, monitoring all heads in LLMs can be overwhelming.
Aggregation strategies, from simple heuristics (e.g., mean, max)~\cite{tu2021keywordmap,sarti2024quantifying} to more principled attention rollout~\cite{abnar2020quantifying}, help reduce complexity.

To improve the faithfulness by accounting for other model components like residuals and layer norms~\cite{kobayashi2021incorporating},
\textbf{vector-based methods} decompose latent vectors into vectors attributable to input tokens~\cite{kobayashi2021incorporating,modarressi2022globenc,ferrando2022measuring,ferrando2023explaining,modarressi2023decompx,yang2023local,achtibat2024attnlrp,song2024better,kobayashi2024analyzing}.
These are applied to modern LLMs~\cite{arras2025close} 
and used to analyze hallucinations~\cite{ferrando2022towards,dale2023detecting,chuang2024lookback}.
However, these methods require model-specific designs, 
limiting adaptability~\cite{abbasi2024normxlogit}.

\textbf{Perturbation-based methods} 
offer a model-agnostic approach by modifying input tokens and observing output changes~\cite{ribeiro2016should}.
Perturbations include
altering latent vectors~\cite{deiseroth2023atman,madani2025noiser},
masking or zeroing token embeddings~\cite{jacovi2021contrastive,yin2022interpreting,mohebbi2023quantifying,cohen2024contextcite},
replacing tokens~\cite{finlayson2021causal,liu2023towards,mohebbi2023quantifying,sadr2025think},
or prompting counterfactuals~\cite{bhattacharjee2023towards,yona2023surfacing,gat2024faithful}. %
Some extend Shapley value, which has been largely explored for classical models~\cite{lundberg2017unified,covert2021explaining},
to estimate the influence of specific input tokens~\cite{horovicz2024tokenshap,mohammadi2024explaining,enouen2024textgenshap}.
These methods have identified tokens triggering %
prompt poisoning ~\cite{cohen2024contextcite} and biases~\cite{mohammadi2024explaining}.
However, 
they can be costly and create unnatural, out-of-distribution inputs,
causing unfaithful interpretations~\cite{abbasi2024normxlogit,achtibat2024attnlrp}.

To address these pitfalls, 
\textbf{gradient-based methods}
compute the gradient of the model output with respect to input embeddings, quantifying how changes to each token affect output~\cite{simonyan2013deep,bach2015pixel,selvaraju2017grad,shrikumar2017learning,sundararajan2017axiomatic}.
Initially designed for smaller models~\cite{enguehard2023sequential,achtibat2024attnlrp,wang2024gradient,song2024better},
they have since been refined %
with contrastive explanations~\cite{jacovi2021contrastive,yin2022interpreting,eberle2023rather,sarti2024quantifying} and extended to LLMs~\cite{barkan2024improving,rezaei2024mambalrp,qi2024model,pan2025hidden}.

Other approaches include
\textbf{similarity-based} methods that compare the final output representation to input token embeddings,
assuming higher similarity indicates greater token importance~\cite{ferrando2022measuring,abbasi2024normxlogit}.
In parallel,
researchers have proposed 
\textbf{prompt-based approaches}~\cite{bhattacharjee2023llms,huang2023can}, 
which instruct LLMs to identify influential tokens for behaviors like jailbreaks~\cite{wang2025delman}, and
\textbf{optimization-based techniques}, 
which search for token attributions that maximize certain interpretability metrics~\cite{zhou2023solvability,barkan2024llm,barkan2024a}.

\subsection{Interpret Safety via Inference-Time Model Internals}
\label{sec:internal}

Techniques interpreting LLMs' internals during inference fall into three types: 
(1) identifying latent regions tied to safety
(\autoref{sec:probing});
(2) perturbing  components to assess impact
(\autoref{sec:causal});
and (3) deciphering latent vectors with human-understandable terms (\autoref{sec:sae}).

\subsubsection{Probe Safety Regions in Latent Space}
\label{sec:probing}

Recent work investigates whether and how safety-related concepts are encoded in LLMs' latent vectors ~\cite{zou2023representation}.
This work builds on the \textit{linear representation hypothesis}, which posits that high-level concepts like factuality or harmfulness %
are embedded as linear directions in the model's latent space~\cite{mikolov2013linguistic,elhage2022toy,park2024the}.
Under this view, researchers analyze latent vectors from individual layers---or concatenated across multiple layers---to identify which latent vectors encode safety behaviors.
A simple yet powerful approaches compute \textbf{mean latent vectors} for the data points
with and without a particular concept (e.g., hallucinated vs. factual).
These reveal directions associated with %
hallucinations~\cite{liu2024universal,chen2025attributive} and
jailbreaks~\cite{arditi2024refusal,zhao2024eeg,zhu2024locking,lin2024towards}.
Dimensionality reduction using PCA or SVD further uncover axes responsible for unsafe behaviors~\cite{duan2024llms,ball2024understanding,pan2025hidden}. %

Another widely used technique is \textbf{probing classifiers}, 
where a model is trained to predict whether a latent vector encodes a safety-related property~\cite{alain2016understanding,tenney2019you,dalvi2019one,kadavath2022language,gurnee2023finding,liu2024universal,ju2024large}.
Probing successfully detects
hallucinations~\cite{burns2022discovering,slobodkin2023curious,orgad2024llms,ashok2025language}, 
jailbreaks~\cite{zhou2024alignment,xu2024uncovering,abdelnabi2024you,ashok2025language}, 
and
bias~\cite{orgad2024llms}.
However, %
these properties are not always linearly separable~\cite{hildebrandt2025refusal}, 
introducing non-linear classifiers~\cite{azaria2023internal,ji2024llm,zhang2024prompt,su2024unsupervised,burger2024truth,he2024jailbreaklens,li2025mixhd,tan2025revprag} or contrastive learning~\cite{he2024llm,beigi2024internalinspector} 
to better capture complex boundaries of unsafe model behaviors.
A known challenge of the probing methods is poor generalization across tasks and datasets~\cite{belinkov2022probing,ch-wang2024androids,levinstein2024still}, %
which has been partially resolved by incorporating distributional differences into the loss function~\cite{burger2024truth} or training probing models on diverse datasets~\cite{liu2024universal}.
\subsubsection{Perturb to Assess Safety Impact}
\label{sec:causal}

A common way to understand how specific components affect model behavior is to perturb them and observe changes.
One approach uses \textbf{gradient-based analysis}, computing output gradients with respect to model parameters to evaluate each parameter's influence.
While useful for explaining mechanisms behind 
knowledge conflicts in RAG~\cite{jin2024cutting} and biased generations~\cite{liu2024the},
such methods may not sufficiently capture causal relations
~\cite{chattopadhyay2019neural}.
A more direct approach is
\textbf{component knockout}, which ablates 
layers, attention heads, or parameters
to identify their influence~\cite{olsson2022in,geva2023dissecting}.
This has localized components responsible for hallucinations~\cite{jin2024cutting,li2024look}, jailbreaks~\cite{zhao2024defending,wei2024assessing}, and biases~\cite{yang2023bias,ma2023deciphering}. 
Instead of full ablation, \textbf{parameter scaling} adjusts component influence
~\cite{luick2024universal}, %
pinpointing safety-critical layers~\cite{li2025safety} and heads~\cite{zhou2025on},
while \textbf{parameter perturbation} modifies model weights and evaluates how safety properties respond to the structural changes~\cite{peng2024navigating,huang2024harmful,leong2024no},
offering a broader perspective on the stability and robustness of safety alignment across the model’s parameter landscape.

\textbf{Activation patching},
inspired by causal mediation analysis~\cite{pearl2001direct,vig2020investigating}, 
replaces intermediate activations (e.g., latent vectors, attention weights) from one input with those from another input, measuring how such intervention affects the model output.
It localizes 
model components %
associated with
hallucinations~\cite{monea2024glitch,deng2025cram} and biases~\cite{vig2020investigating},
as well as general model capabilities~\cite{geiger2021causal,stolfo2023mechanistic,davies2023discovering,cabannes2024iteration} and factual knowledge~\cite{meng2022locating,nanda2023fact,ghandeharioun2024patchscopes}.
To uncover how components interact, researchers extract \textit{computational circuits},
graphs with important components as nodes and information flow as edges~\cite{geiger2021causal,elhage2021mathematical}.
\textbf{Path patching}, an extension of activation patching, 
modifies outputs along specific computational paths while freezing the rest of the network%
\footnote{Path patching differs from activation patching in that it selectively modifies only the information flowing along specific paths, whereas activation patching replaces entire activations at specific components.}~\cite{wang2023interpretability,goldowsky2023localizing,prakash2023fine,hanna2023how}.
Due to the high reliance on human inspection,
several efforts %
automate circuit discovery~\cite{conmy2023towards,ferrando2024information,bhaskar2024finding}, while
attribution patching approximates causal effect for scalability~\cite{nanda2024attribution,syed2024attribution,kramar2024atp,hanna2024have}.
However, as LLM circuit analysis is still in its early stages, most focus on simple grammatical or arithmetic tasks, with very few addressing real-world safety problems~\cite{hanna2024have}.

\subsubsection{Decipher Latent Vectors with Language} %
\label{sec:sae}

One approach to understand latent vectors through language is to analyze how 
their individual neurons respond to input data.
By identifying inputs %
that highly activate a neuron and their shared patterns, %
researchers have inferred concepts encoded by each neuron~\cite{geva2021transformer,foote2023neuron}.
However, many neurons in LLMs are \textit{polysemantic}, 
encoding multiple unrelated concepts, making interpretation challenging~\cite{arora2018linear,bricken2023towards,templeton2024scaling}.
To address this,
researchers have developed techniques to disentangle concepts superposed in the latent vectors~\cite{elhage2022toy}.
A prominent method is using Sparse dictionary learning~\cite{mairal2008supervised,makhzani2013k,elhage2021mathematical} to train \textbf{sparse autoencoders (SAEs)}~\cite{sharkey2022taking,bricken2023towards,huben2024sparse,lieberum2024gemma}.
An SAE consist of an encoder and a decoder; the encoder maps a latent vector into a sparse, high-dimensional concept vector,
where each dimension --- SAE neuron --- represents a distinct, interpretable concept,
characterized by the inputs that strongly activate it~\cite{paulo2024automatically}; the decoder reconstructs the original latent vector from the concept vector.
Since training separate SAEs for each (sub)layer in LLMs
can be computationally intensive and redundant,
later research enhances 
scalability and expressiveness through
new architectures~\cite{rajamanoharan2024improving,templeton2024scaling,dunefsky2024transcoders,mudide2025efficient}, 
activation functions~\cite{rajamanoharan2024jumping}, and 
training  strategies~\cite{kissane2024interpreting,ghilardi2024efficient,braun2024identifying,shi2025route,farnik2025jacobian}.
These advances have discovered more diverse concepts~\cite{o2024disentangling,templeton2024scaling,he2024llama},
offering insights into LLMs' %
hallucinations~\cite{ferrando2025do,theodorus2025finding}, 
jailbreaks~\cite{haerle2024scar,muhamed2025decoding,gallifant2025sparse}, biases~\cite{hegde2024effectiveness,zhou2025llm}, and
privacy leakage~\cite{frikha2025privacyscalpel}.

Beyond individual neurons,
\textbf{SAE circuits} are extracted %
to reveal how interpretable concepts interact to produce specific outputs~\cite{he2024dictionary,dunefsky2024case,marks2025sparse,balagansky2025mechanistic}. %
SAE variants, such as Crosscoder~\cite{lindsey2024sparse} and Transcoder~\cite{dunefsky2024transcoders,dunefsky2024transcoders-neurips,ameisen2025circuit}, enhance circuit interpretability and reduce redundancy, making it easier to 
isolate mechanisms behind unsafe behaviors~\cite{lindsey2025on}. %
In parallel,
\textbf{logit lens}
projects intermediate latent vectors onto the model's vocabulary space using the final projection matrix,
viewing latent vectors on the vocabulary level~\cite{nostalgebraist2020logitlens,elhage2021mathematical,geva2022transformer,dar2023analyzing}.
Further research enhances its robustness~\cite{belrose2023eliciting,din2023jump} %
and extends it to analyze training dynamics~\cite{katz2024backward}.
The logit lens has been leveraged to investigate
LLMs' knowledge store and recall mechanisms~\cite{haviv2023understanding,yu2023characterizing}
and safety issues, 
such as 
hallucinations~\cite{yu2024mechanistic,jiang2024large,halawi2024overthinking,jin2024cutting} and jailbreaks and harmfulness~\cite{zhao2024defending,feng2024unveiling,lee2024a}. 
\subsection{Self-explain with Reason Generation}
\label{sec:reasoning}

Recent work explores how LLMs can interpret their own outputs by expressing the reasoning behind them in natural language, offering insights into their internal knowledge and decision-making
~\cite{huang2023towards,zhao2024explainability,yu2024natural}.
A common approach is \textbf{in-generation reasoning}, where LLMs are prompted or trained to generate responses along with rationales~\cite{camburu2018snli,rajani2019explain,marasovic2022shot} or uncertainty estimates~\cite{kadavath2022language,chaudhry2024finetuning,amayuelas2024knowledge,xu2024sayself}.
Chain-of-thought (CoT) prompting is a notable example,
where LLMs %
generate intermediate reasoning steps to reach ananswer~\cite{wei2022chain,zhao2023selfexplain,chu2025domaino1s,cahlik2025reasoning}.
Many CoT variants 
support more complex reasoning~\cite{yao2023tree,besta2024graph} and
improve explanation faithfulness~\cite{qu2022interpretable,lyu2023faithful,tafjord2022entailer,creswell2022faithful,creswell2023selectioninference}.
However, %
such explanations can be unreliable~\cite{gao2023chatgpt,ye2022the,araya2025chains},
necessitating further verification
~\cite{ye2022the,turpin2023language,weng2023large,miao2024selfcheck}.

Alternatively,
\textbf{post-hoc explanation} methods assess and explain a response after generation~\cite{jiang2024forward,binder2025looking}.
These methods prompt LLMs to evaluate the correctness or safety of their outputs and provide rationales~\cite{li2024safetyanalyst,liu2025guardreasoner,betley2025tell}.
To explain hallucinations,
a response may be split into factual claims or questions, which the model is then asked to verify against its knowledge~\cite{dhuliawala2024chain,akbar2024hallumeasure,lee2025hudex}.
\section{Enhance Safety using Interpretation}
\label{sec:enhance}

Recent advances in LLM interpretation (\autoref{sec:interpretation}) have 
inspired techniques to enhance model safety. %
This section reviews methods that leverage interpretation to mitigate unsafe behaviors, following the stages of the LLM workflow discussed in \autoref{sec:interpretation}.

\subsection{Attend to Relevant Input Tokens (\autoref{sec:input})} 

Some methods 
prompt LLMs to attend to relevant input tokens
to reduce hallucinations~\cite{liu2025selfelicit} and improve factuality~\cite{krishna2023post}.
Others
remove jailbreak-triggering tokens~\cite{pan2025hidden} 
or manipulate attention to user-specified relevant tokens~\cite{zhang2024tell}.

\subsection{Modify Model Internals for Safety (\autoref{sec:internal})}

\subsubsection{Steer Latent Vectors To Safe Directions}
\textit{Representation engineering}
guides LLMs' latent vectors toward safe regions
by adding safety vectors identified by probing~\cite{zou2023representation} %
or training transformations that map unsafe vectors into safe regions~\cite{hernandez2024inspecting} (\autoref{sec:probing}).
These methods mitigate a range of safety concerns~\cite{qian2024towards,rimsky2024steering,singh2024representation,chu2024a}, such as
hallucinations~\cite{li2023inference,yang2024enhancing,zhang2024truthx,duan2024llms}, 
jailbreaks and harmfulness~\cite{turner2023steering,bhattacharjee2024towards,zhu2024locking,ball2024understanding,gao2024shaping,shen2024jailbreak,li2025revisiting,han2025internal}, and 
bias~\cite{hernandez2024inspecting}.

\subsubsection{Modulate (Un)Safe Neurons' Activations}

Suppressing risky neurons or amplifying safer ones guides LLMs away from unsafe behaviors.
SAEs help locate and control (un)safe SAE neuron activations ~\cite{soo2025interpretable} (\autoref{sec:sae}),
addressing risks~\cite{he2025towards} like
hallucinations~\cite{abdaljalil2025safe,bayat2025steering}, 
jailbreaks and harmfulness~\cite{o2024steering,harle2024scar,khoriaty2025don,wu2025interpreting},
biases~\cite{liu2024the,hegde2024effectiveness,marks2025sparse,zhou2025llm}, 
and privacy leaks~\cite{frikha2025privacyscalpel}.
Alternatives remove dependency on SAEs
by
using logit lens (\autoref{sec:sae}) to find and upscale safe MLP sublayers~\cite{geva2022transformer,wang2024model}
or amplify safety-critical attention weights (\autoref{sec:probing}, \autoref{sec:causal}) 
on user-specified reliable tokens~\cite{zhang2024tell,deng2025cram}.
\subsubsection{Edit Harmful Model Components}

Safety can be improved by pruning or downscaling components (e.g., attention heads or (sub)layers) linked to %
hallucinations~\cite{li2024look,yu2024mechanistic},
jailbreaks~\cite{zhao2024defending,wang2024detoxifying,li2024precision,li2025safety,wang2025delman}, and biases~\cite{ma2023deciphering}.
Other techniques identify safety directions in parameter space 
by comparing aligned and unaligned model weights, then steer critical parameters accordingly~\cite{hazra2024safety,wang2024model,zhao2024towards-comprehensive}.
\subsection{Verify Safety before Outputs (\autoref{sec:internal}, \autoref{sec:reasoning})} 
Some approaches
generate multiple candidate responses, 
evaluate their safety (\autoref{sec:internal}, \autoref{sec:reasoning}), %
and select only safe ones to construct final output~\cite{burns2022discovering,zhao2023verify,weng2023large,dale2023detecting,miao2024selfcheck,chuang2024lookback,dhuliawala2024chain,chen2025attributive}.
Others intervene during generation, resamping tokens when an unsafe sequence is detected
~\cite{li2024hallucana,cheng2025think} or
producing refusal messages~\cite{zhao2024eeg,mou2025saro}.

\smallskip
\noindent
\subsection{Output with Self-Reasoning (\autoref{sec:reasoning}).}
Building on CoT reasoning's success for performance and interpretation~\cite{ji2024chain}, 
several approaches fine-tune or prompt LLMs to  
generate intermediate reasoning steps to reinforce safety constraints during generation~\cite{kaneko2024evaluating,prahallad2024significance,li2024chain,cao2024defending,sicilia2024eliciting,moore2024reasoning,rad2025refining,zhang2025safety,mou2025saro}.
For instance, GuardReasoner~\cite{liu2025guardreasoner} prompts models to explain why a response may be harmful, enabling safer behavior through self-reflection.
\section{Tools Operationalizing Safety-Focused Interpretation}
\label{sec:toolkits}

To apply interpretation methods (\autoref{sec:interpretation}) and safety enhancement strategies (\autoref{sec:enhance}), practitioners need tools %
that support actionability 
~\cite{kaur2020interpreting,lakkaraju2022rethinking,sharkey2025open}.
Some efforts focus on developing
libraries to \textbf{ease implementation} of interpretation and safety techniques~\cite{choe2024your, kokhlikyan2020captum,sarti2023inseq, hao2024llm},
while others introduce interactive visual tools,
inspired by their effectiveness in enhancing human understanding of classical AI models~\cite{hohman2019visual,beauxis2021role,la2023state,liao2024ai,wang2025human}. %
\noindent
\textbf{5.1\hspace{1em}Training Data Attribution (TDA) Visualization} (\autoref{sec:training}) %
shows how training examples shape model behavior.
A prominent tool is LLM Attributor~\cite{lee2025llm}, which traces outputs to training data, identifying hallucination sources.
\noindent
\textbf{5.2\hspace{1em}Input Token Visualizations} (\autoref{sec:input}) 
reveal individual tokens' importance and attention patterns, showing how tokens influence one another across heads.
These 
are incorporated into LLM analysis tools~\cite{park2019sanvis,tenney2020language,wang2021dodrio,li2023visual,coscia2024iscore,yeh2024attentionviz}, %
and reveal spurious token associations indicative of bias~\cite{vig2019multiscale}. 
Many tools also support interactive perturbation, allowing users to edit tokens or attention weights and observe the effects~\cite{strobelt2019seq2seqvis,tenney2020language,coscia2024iscore,mishra2025promptaid}.

\noindent
\textbf{5.3\hspace{1em}Latent vector visualizations} (\autoref{sec:internal}) show how concepts are encoded and propagated during model inference.
Some tools project latent vectors into 2D space~\cite{tenney2020language,li2023visual,kwon2023finspector}. 
Others visualize 
semantics of latent vectors revealed by logit lens~\cite{katz2023visit,pal2023future,hernandez2024linearity} (\autoref{sec:sae}), while
some enable users to steer latent vectors for safer outputs~\cite{geva2022lm,chen2024designing} (\autoref{sec:enhance}).

\noindent
\textbf{5.4\hspace{1em}Neuron visualizations} (\autoref{sec:internal})
display data points that highly activate each neuron during inference, revealing interpretable concepts~\cite{nanda2022neuroscope,garde2023deepdecipher,bills2023language}.
(\autoref{sec:sae}). 
Similar approaches are applied to SAE neurons~\cite{lin2023neuronpedia},
helping concept identification and SAE circuit discovery~\cite{chalnev2024improving}
for multiple unsafe behaviors~\cite{lindsey2025on,ameisen2025circuit}. %
\section{Research Directions and Conclusion}
\label{sec:challenges}
\paragraph{Defense against interpretation-based attacks.}
As interpretation methods grow more powerful, they
can be misused to break guardrails or reveal model vulnerabilities~\cite{lin2024towards,li2024model,arditi2024refusal,su2024enhancing,winninger2025using}.
Future research should discern a balance between transparency and safety by 
developing 
robust defense strategies \citep{wu2025interpreting}.
\smallskip
\noindent
\textbf{Reliable Evaluation of Interpretation.}
Misleading interpretations --- such as unsafe LLM self-reasoning~\cite{shaikh2023second} --- can lead to overtrust and unsafe decisions~\cite{jacovi2020towards,ajwani2024llm}. 
To ensure safe and trustworthy use, interpretation methods must be rigorously evaluated.
Despite ongoing evaluation research~\cite{schwettmann2023find,li2024evaluating,shi2024hypothesis,makelov2024principled}, standardized benchmarks are needed to assess the reliability, robustness, and faithfulness of interpretation methods~\cite{alangari2023exploring}.

\smallskip
\noindent
\textbf{Using Training Attribution for Safety Enhancement.}
TDA 
shows promise for tracing unsafe behavior to training examples (\autoref{sec:training}), but its use in safety enhancement is limited.
Prior work on retraining after removing problematic data~\cite{kong2022resolving,mozes2023gradient}
focuses on non-safety issues on small non-generative models and cannot be scaled to LLMs.
Developing safety enhancement methods based on training attribution could open new paths for risk mitigation.  
\smallskip
\noindent
\textbf{User-centered Presentation of Safety Interpretations.}
How to present interpretation results to assist safety-critical decisions remains underexplored (\autoref{sec:toolkits}).
In particular, presentation of long, complex textual explanations from LLMs should be further investigated (\autoref{sec:reasoning});
conversational interaction helps human understanding~\cite{slack2023explaining,wang2024llmcheckup}, yet no tools apply this to safety-oriented interpretation.
Future work should explore interaction and design strategies tailored to diverse stakeholders. 

\smallskip
\noindent
\textbf{Broadening and Refining Safety Dimensions.}
Interpretation research has largely focused on 
hallucinations, jailbreaks, harmfulness, bias, and privacy leakage, 
while other risks---like out-of-distribution robustness, code security, and over-refusal---are understudied \citep{siska2025attentiondefense, yang2024robustness,xiong2024defensive, abdaljalil2025safe}.
Incorporating user intent and social impact into safety definitions may enable more nuanced and targeted interpretations \citep{sarker2024llm}. 

\smallskip
\noindent
\textbf{Conclusion.}
By bridging the gap between interpretation and safety research, our survey systematically examines interpretation methods across the LLM workflow, safety enhancement strategies, and practical tools,
while highlighting open problems and future directions.

\clearpage
\section{Limitations}

This survey provides an overview and categorization of interpretation techniques, with an emphasis on their role in improving the safety of LLMs.
Given the fast-evolving and extensive nature of the field, some latest advancements may not be included.
We focus on autoregressive Transformer-based generative LLMs, as they are among the most widely used an studied models for interpretation; therefore, the interpretation and safety enhancement techniques we discuss may not generalize to other model architectures.
Our paper selection aims to capture the breadth and diversity of existing approaches, though full technical details are omitted due to space constraints.
We also highlight tools that facilitate understanding and use of interpretation results, recognizing that notions of practicality can vary across stakeholders and that actionability of interpretation remains an actively researched open question. %
Despite its limitations, this survey introduces a taxonomy that can help newcomers quickly understand the landscape of interpretation for safety and guide future research exploring its application to other model architectures and emerging techniques.

\section{Potential Risks}

Our paper focuses on four major safety concerns addressed by interpretation research (hallucination, jailbreaks and harmfulness, bias, and privacy leakage), but this view may be too narrow, risking overlooking other safety issues such as code security and over-refusal (\autoref{sec:challenges}). 
While our survey covers a wide range of interpretation techniques, it does not include quantitative comparisons.
As a result, readers may overly rely on certain techniques or mistakenly assume that interpretation guarantees safety.

\section*{Acknowledgements}
ChatGPT was used to check grammar and spelling of this paper. %

\bibliography{references}

\begin{thebibliography}{390}
\providecommand{\natexlab}[1]{#1}

\bibitem[{Abbasi et~al.(2024)Abbasi, Modarres, and
  Pilehvar}]{abbasi2024normxlogit}
Sina Abbasi, Mohammad~Reza Modarres, and Mohammad~Taher Pilehvar. 2024.
\newblock Normxlogit: The head-on-top never lies.
\newblock \emph{arXiv preprint arXiv:2411.16252}.

\bibitem[{Abdaljalil et~al.(2025)Abdaljalil, Pallucchini, Seveso, Kurban,
  Mercorio, and Serpedin}]{abdaljalil2025safe}
Samir Abdaljalil, Filippo Pallucchini, Andrea Seveso, Hasan Kurban, Fabio
  Mercorio, and Erchin Serpedin. 2025.
\newblock {SAFE}: A sparse autoencoder-based framework for robust query
  enrichment and hallucination mitigation in llms.
\newblock \emph{arXiv preprint arXiv:2503.03032}.

\bibitem[{Abdelnabi et~al.(2024)Abdelnabi, Fay, Cherubin, Salem, Fritz, and
  Paverd}]{abdelnabi2024you}
Sahar Abdelnabi, Aideen Fay, Giovanni Cherubin, Ahmed Salem, Mario Fritz, and
  Andrew Paverd. 2024.
\newblock Are you still on track!? catching llm task drift with activations.
\newblock \emph{arXiv preprint arXiv:2406.00799}.

\bibitem[{Abnar and Zuidema(2020)}]{abnar2020quantifying}
Samira Abnar and Willem Zuidema. 2020.
\newblock \href {https://doi.org/10.18653/v1/2020.acl-main.385} {Quantifying
  attention flow in transformers}.
\newblock In \emph{Proceedings of the 58th Annual Meeting of the Association
  for Computational Linguistics}, pages 4190--4197, Online. Association for
  Computational Linguistics.

\bibitem[{Achiam et~al.(2023)Achiam, Adler, Agarwal, Ahmad, Akkaya, Aleman,
  Almeida, Altenschmidt, Altman, Anadkat et~al.}]{achiam2023gpt}
Josh Achiam, Steven Adler, Sandhini Agarwal, Lama Ahmad, Ilge Akkaya,
  Florencia~Leoni Aleman, Diogo Almeida, Janko Altenschmidt, Sam Altman,
  Shyamal Anadkat, et~al. 2023.
\newblock Gpt-4 technical report.
\newblock \emph{arXiv preprint arXiv:2303.08774}.

\bibitem[{Achtibat et~al.(2024)Achtibat, Hatefi, Dreyer, Jain, Wiegand,
  Lapuschkin, and Samek}]{achtibat2024attnlrp}
Reduan Achtibat, Sayed Mohammad~Vakilzadeh Hatefi, Maximilian Dreyer, Aakriti
  Jain, Thomas Wiegand, Sebastian Lapuschkin, and Wojciech Samek. 2024.
\newblock Attnlrp: attention-aware layer-wise relevance propagation for
  transformers.
\newblock In \emph{Proceedings of the 41st International Conference on Machine
  Learning}, ICML'24. JMLR.org.

\bibitem[{Ajwani et~al.(2024)Ajwani, Javaji, Rudzicz, and Zhu}]{ajwani2024llm}
Rohan Ajwani, Shashidhar~Reddy Javaji, Frank Rudzicz, and Zining Zhu. 2024.
\newblock Llm-generated black-box explanations can be adversarially helpful.
\newblock \emph{arXiv preprint arXiv:2405.06800}.

\bibitem[{Akbar et~al.(2024)Akbar, Hossain, Wood, Chin, Salinas, Alvarez, and
  Cornejo}]{akbar2024hallumeasure}
Shayan~Ali Akbar, Md~Mosharaf Hossain, Tess Wood, Si-Chi Chin, Erica~M Salinas,
  Victor Alvarez, and Erwin Cornejo. 2024.
\newblock \href {https://doi.org/10.18653/v1/2024.emnlp-main.837}
  {{H}allu{M}easure: Fine-grained hallucination measurement using
  chain-of-thought reasoning}.
\newblock In \emph{Proceedings of the 2024 Conference on Empirical Methods in
  Natural Language Processing}, pages 15020--15037, Miami, Florida, USA.
  Association for Computational Linguistics.

\bibitem[{Akyurek et~al.(2022)Akyurek, Bolukbasi, Liu, Xiong, Tenney, Andreas,
  and Guu}]{akyurek2022towards}
Ekin Akyurek, Tolga Bolukbasi, Frederick Liu, Binbin Xiong, Ian Tenney, Jacob
  Andreas, and Kelvin Guu. 2022.
\newblock \href {https://doi.org/10.18653/v1/2022.findings-emnlp.180} {Towards
  tracing knowledge in language models back to the training data}.
\newblock In \emph{Findings of the Association for Computational Linguistics:
  EMNLP 2022}, pages 2429--2446, Abu Dhabi, United Arab Emirates. Association
  for Computational Linguistics.

\bibitem[{Alain and Bengio(2016)}]{alain2016understanding}
Guillaume Alain and Yoshua Bengio. 2016.
\newblock Understanding intermediate layers using linear classifier probes.
\newblock \emph{arXiv preprint arXiv:1610.01644}.

\bibitem[{Alangari et~al.(2023)Alangari, El~Bachir~Menai, Mathkour, and
  Almosallam}]{alangari2023exploring}
Nourah Alangari, Mohamed El~Bachir~Menai, Hassan Mathkour, and Ibrahim
  Almosallam. 2023.
\newblock Exploring evaluation methods for interpretable machine learning: A
  survey.
\newblock \emph{Information}, 14(8):469.

\bibitem[{Amayuelas et~al.(2024)Amayuelas, Wong, Pan, Chen, and
  Wang}]{amayuelas2024knowledge}
Alfonso Amayuelas, Kyle Wong, Liangming Pan, Wenhu Chen, and William~Yang Wang.
  2024.
\newblock \href {https://doi.org/10.18653/v1/2024.findings-acl.383} {Knowledge
  of knowledge: Exploring known-unknowns uncertainty with large language
  models}.
\newblock In \emph{Findings of the Association for Computational Linguistics:
  ACL 2024}, pages 6416--6432, Bangkok, Thailand. Association for Computational
  Linguistics.

\bibitem[{Ameisen et~al.(2025)Ameisen, Lindsey, Pearce, Gurnee, Turner, Chen,
  Citro, Abrahams, Carter, Hosmer, Marcus, Sklar, Templeton, Bricken,
  McDougall, Cunningham, Henighan, Jermyn, Jones, Persic, Qi, Thompson,
  Zimmerman, Rivoire, Conerly, Olah, and Batson}]{ameisen2025circuit}
Emmanuel Ameisen, Jack Lindsey, Adam Pearce, Wes Gurnee, Nicholas~L. Turner,
  Brian Chen, Craig Citro, David Abrahams, Shan Carter, Basil Hosmer, Jonathan
  Marcus, Michael Sklar, Adly Templeton, Trenton Bricken, Callum McDougall,
  Hoagy Cunningham, Thomas Henighan, Adam Jermyn, Andy Jones, Andrew Persic,
  Zhenyi Qi, T.~Ben Thompson, Sam Zimmerman, Kelley Rivoire, Thomas Conerly,
  Chris Olah, and Joshua Batson. 2025.
\newblock Circuit tracing: Revealing computational graphs in language models.
\newblock
  \url{https://transformer-circuits.pub/2025/attribution-graphs/methods.html}.
\newblock Transformer Circuits Thread.

\bibitem[{Araya(2025)}]{araya2025chains}
Roberto Araya. 2025.
\newblock Do chains-of-thoughts of large language models suffer from
  hallucinations, cognitive biases, or phobias in bayesian reasoning?
\newblock \emph{arXiv preprint arXiv:2503.15268}.

\bibitem[{Arditi et~al.(2024)Arditi, Obeso, Syed, Paleka, Panickssery, Gurnee,
  and Nanda}]{arditi2024refusal}
Andy Arditi, Oscar Obeso, Aaquib Syed, Daniel Paleka, Nina Panickssery, Wes
  Gurnee, and Neel Nanda. 2024.
\newblock Refusal in language models is mediated by a single direction.
\newblock \emph{arXiv preprint arXiv:2406.11717}.

\bibitem[{Arora et~al.(2018)Arora, Li, Liang, Ma, and
  Risteski}]{arora2018linear}
Sanjeev Arora, Yuanzhi Li, Yingyu Liang, Tengyu Ma, and Andrej Risteski. 2018.
\newblock \href {https://doi.org/10.1162/tacl_a_00034} {Linear algebraic
  structure of word senses, with applications to polysemy}.
\newblock \emph{Transactions of the Association for Computational Linguistics},
  6:483--495.

\bibitem[{Arras et~al.(2025)Arras, Puri, Kahardipraja, Lapuschkin, and
  Samek}]{arras2025close}
Leila Arras, Bruno Puri, Patrick Kahardipraja, Sebastian Lapuschkin, and
  Wojciech Samek. 2025.
\newblock A close look at decomposition-based xai-methods for transformer
  language models.
\newblock \emph{arXiv preprint arXiv:2502.15886}.

\bibitem[{Ashok and May(2025)}]{ashok2025language}
Dhananjay Ashok and Jonathan May. 2025.
\newblock Language models can predict their own behavior.
\newblock \emph{arXiv preprint arXiv:2502.13329}.

\bibitem[{Ayyamperumal and Ge(2024)}]{ayyamperumal2024current}
Suriya~Ganesh Ayyamperumal and Limin Ge. 2024.
\newblock Current state of llm risks and ai guardrails.
\newblock \emph{arXiv preprint arXiv:2406.12934}.

\bibitem[{Azaria and Mitchell(2023)}]{azaria2023internal}
Amos Azaria and Tom Mitchell. 2023.
\newblock The internal state of an llm knows when it's lying.
\newblock \emph{arXiv preprint arXiv:2304.13734}.

\bibitem[{Bach et~al.(2015)Bach, Binder, Montavon, Klauschen, M{\"u}ller, and
  Samek}]{bach2015pixel}
Sebastian Bach, Alexander Binder, Gr{\'e}goire Montavon, Frederick Klauschen,
  Klaus-Robert M{\"u}ller, and Wojciech Samek. 2015.
\newblock On pixel-wise explanations for non-linear classifier decisions by
  layer-wise relevance propagation.
\newblock \emph{PloS one}, 10(7):e0130140.

\bibitem[{Balagansky et~al.(2025)Balagansky, Maksimov, and
  Gavrilov}]{balagansky2025mechanistic}
Nikita Balagansky, Ian Maksimov, and Daniil Gavrilov. 2025.
\newblock \href {https://openreview.net/forum?id=MDvecs7EvO} {Mechanistic
  permutability: Match features across layers}.
\newblock In \emph{The Thirteenth International Conference on Learning
  Representations}.

\bibitem[{Ball et~al.(2024)Ball, Kreuter, and
  Panickssery}]{ball2024understanding}
Sarah Ball, Frauke Kreuter, and Nina Panickssery. 2024.
\newblock Understanding jailbreak success: A study of latent space dynamics in
  large language models.
\newblock \emph{arXiv preprint arXiv:2406.09289}.

\bibitem[{Barkan et~al.(2024{\natexlab{a}})Barkan, Elisha, Toib, Weill, and
  Koenigstein}]{barkan2024improving}
Oren Barkan, Yehonatan Elisha, Yonatan Toib, Jonathan Weill, and Noam
  Koenigstein. 2024{\natexlab{a}}.
\newblock \href {https://doi.org/10.18653/v1/2024.findings-emnlp.551}
  {Improving {LLM} attributions with randomized path-integration}.
\newblock In \emph{Findings of the Association for Computational Linguistics:
  EMNLP 2024}, pages 9430--9446, Miami, Florida, USA. Association for
  Computational Linguistics.

\bibitem[{Barkan et~al.(2024{\natexlab{b}})Barkan, Toib, Elisha, and
  Koenigstein}]{barkan2024a}
Oren Barkan, Yonatan Toib, Yehonatan Elisha, and Noam Koenigstein.
  2024{\natexlab{b}}.
\newblock \href {https://doi.org/10.1145/3627673.3679548} {A learning-based
  approach for explaining language models}.
\newblock In \emph{Proceedings of the 33rd ACM International Conference on
  Information and Knowledge Management}, CIKM '24, page 98–108, New York, NY,
  USA. Association for Computing Machinery.

\bibitem[{Barkan et~al.(2024{\natexlab{c}})Barkan, Toib, Elisha, Weill, and
  Koenigstein}]{barkan2024llm}
Oren Barkan, Yonatan Toib, Yehonatan Elisha, Jonathan Weill, and Noam
  Koenigstein. 2024{\natexlab{c}}.
\newblock \href {https://doi.org/10.18653/v1/2024.findings-emnlp.556} {{LLM}
  explainability via attributive masking learning}.
\newblock In \emph{Findings of the Association for Computational Linguistics:
  EMNLP 2024}, pages 9522--9537, Miami, Florida, USA. Association for
  Computational Linguistics.

\bibitem[{Barshan et~al.(2020)Barshan, Brunet, and
  Dziugaite}]{barshan2020relatif}
Elnaz Barshan, Marc-Etienne Brunet, and Gintare~Karolina Dziugaite. 2020.
\newblock Relatif: Identifying explanatory training samples via relative
  influence.
\newblock In \emph{International Conference on Artificial Intelligence and
  Statistics}, pages 1899--1909. PMLR.

\bibitem[{Basu et~al.(2021)Basu, Pope, and Feizi}]{basu2021influence}
Samyadeep Basu, Phil Pope, and Soheil Feizi. 2021.
\newblock \href {https://openreview.net/forum?id=xHKVVHGDOEk} {Influence
  functions in deep learning are fragile}.
\newblock In \emph{International Conference on Learning Representations}.

\bibitem[{Bayat et~al.(2025)Bayat, Rahimi-Kalahroudi, Pezeshki, Chandar, and
  Vincent}]{bayat2025steering}
Reza Bayat, Ali Rahimi-Kalahroudi, Mohammad Pezeshki, Sarath Chandar, and
  Pascal Vincent. 2025.
\newblock Steering large language model activations in sparse spaces.
\newblock \emph{arXiv preprint arXiv:2503.00177}.

\bibitem[{Beauxis-Aussalet et~al.(2021)Beauxis-Aussalet, Behrisch, Borgo, Chau,
  Collins, Ebert, El-Assady, Endert, Keim, Kohlhammer et~al.}]{beauxis2021role}
Emma Beauxis-Aussalet, Michael Behrisch, Rita Borgo, Duen~Horng Chau,
  Christopher Collins, David Ebert, Mennatallah El-Assady, Alex Endert,
  Daniel~A Keim, J{\"o}rn Kohlhammer, et~al. 2021.
\newblock The role of interactive visualization in fostering trust in ai.
\newblock \emph{IEEE Computer Graphics and Applications}, 41(6):7--12.

\bibitem[{Beigi et~al.(2024)Beigi, Shen, Yang, Lin, Wang, Mohan, He, Jin, Lu,
  and Huang}]{beigi2024internalinspector}
Mohammad Beigi, Ying Shen, Runing Yang, Zihao Lin, Qifan Wang, Ankith Mohan,
  Jianfeng He, Ming Jin, Chang-Tien Lu, and Lifu Huang. 2024.
\newblock Internalinspector $i^2$: Robust confidence estimation in llms through
  internal states.
\newblock \emph{arXiv preprint arXiv:2406.12053}.

\bibitem[{Belinkov(2022)}]{belinkov2022probing}
Yonatan Belinkov. 2022.
\newblock Probing classifiers: Promises, shortcomings, and advances.
\newblock \emph{Computational Linguistics}, 48(1):207--219.

\bibitem[{Belrose et~al.(2023)Belrose, Furman, Smith, Halawi, Ostrovsky,
  McKinney, Biderman, and Steinhardt}]{belrose2023eliciting}
Nora Belrose, Zach Furman, Logan Smith, Danny Halawi, Igor Ostrovsky, Lev
  McKinney, Stella Biderman, and Jacob Steinhardt. 2023.
\newblock Eliciting latent predictions from transformers with the tuned lens.
\newblock \emph{arXiv preprint arXiv:2303.08112}.

\bibitem[{Bereska and Gavves(2024)}]{bereska2024mechanistic}
Leonard Bereska and Efstratios Gavves. 2024.
\newblock Mechanistic interpretability for ai safety--a review.
\newblock \emph{arXiv preprint arXiv:2404.14082}.

\bibitem[{Besta et~al.(2024)Besta, Blach, Kubicek, Gerstenberger, Podstawski,
  Gianinazzi, Gajda, Lehmann, Niewiadomski, Nyczyk et~al.}]{besta2024graph}
Maciej Besta, Nils Blach, Ales Kubicek, Robert Gerstenberger, Michal
  Podstawski, Lukas Gianinazzi, Joanna Gajda, Tomasz Lehmann, Hubert
  Niewiadomski, Piotr Nyczyk, et~al. 2024.
\newblock Graph of thoughts: Solving elaborate problems with large language
  models.
\newblock In \emph{Proceedings of the AAAI Conference on Artificial
  Intelligence}, volume~38, pages 17682--17690.

\bibitem[{Betley et~al.(2025)Betley, Bao, Soto, Sztyber-Betley, Chua, and
  Evans}]{betley2025tell}
Jan Betley, Xuchan Bao, Mart{\'\i}n Soto, Anna Sztyber-Betley, James Chua, and
  Owain Evans. 2025.
\newblock Tell me about yourself: Llms are aware of their learned behaviors.
\newblock \emph{arXiv preprint arXiv:2501.11120}.

\bibitem[{Bhaskar et~al.(2024)Bhaskar, Wettig, Friedman, and
  Chen}]{bhaskar2024finding}
Adithya Bhaskar, Alexander Wettig, Dan Friedman, and Danqi Chen. 2024.
\newblock Finding transformer circuits with edge pruning.
\newblock \emph{Advances in Neural Information Processing Systems},
  37:18506--18534.

\bibitem[{Bhattacharjee et~al.(2024)Bhattacharjee, Ghosh, Rebedea, and
  Parisien}]{bhattacharjee2024towards}
Amrita Bhattacharjee, Shaona Ghosh, Traian Rebedea, and Christopher Parisien.
  2024.
\newblock \href {https://openreview.net/forum?id=EkQRNLPFcn} {Towards
  inference-time category-wise safety steering for large language models}.
\newblock In \emph{Neurips Safe Generative AI Workshop 2024}.

\bibitem[{Bhattacharjee et~al.(2023{\natexlab{a}})Bhattacharjee, Moraffah,
  Garland, and Liu}]{bhattacharjee2023llms}
Amrita Bhattacharjee, Raha Moraffah, Joshua Garland, and Huan Liu.
  2023{\natexlab{a}}.
\newblock Llms as counterfactual explanation modules: can chatgpt explain
  blackbox text classifiers.
\newblock \emph{arXiv preprint arXiv:2309.13340}, 4.

\bibitem[{Bhattacharjee et~al.(2023{\natexlab{b}})Bhattacharjee, Moraffah,
  Garland, and Liu}]{bhattacharjee2023towards}
Amrita Bhattacharjee, Raha Moraffah, Joshua Garland, and Huan Liu.
  2023{\natexlab{b}}.
\newblock Towards llm-guided causal explainability for black-box text
  classifiers.
\newblock \emph{arXiv preprint arXiv:2309.13340}.

\bibitem[{Bills et~al.(2023)Bills, Cammarata, Mossing, Tillman, Gao, Goh,
  Sutskever, Leike, Wu, and Saunders}]{bills2023language}
Steven Bills, Nick Cammarata, Dan Mossing, Henk Tillman, Leo Gao, Gabriel Goh,
  Ilya Sutskever, Jan Leike, Jeff Wu, and William Saunders. 2023.
\newblock Language models can explain neurons in language models.
\newblock
  \url{https://openaipublic.blob.core.windows.net/neuron-explainer/paper/index.html}.
\newblock OpenAI.

\bibitem[{Binder et~al.(2025)Binder, Chua, Korbak, Sleight, Hughes, Long,
  Perez, Turpin, and Evans}]{binder2025looking}
Felix~Jedidja Binder, James Chua, Tomek Korbak, Henry Sleight, John Hughes,
  Robert Long, Ethan Perez, Miles Turpin, and Owain Evans. 2025.
\newblock \href {https://openreview.net/forum?id=eb5pkwIB5i} {Looking inward:
  Language models can learn about themselves by introspection}.
\newblock In \emph{The Thirteenth International Conference on Learning
  Representations}.

\bibitem[{Bommasani et~al.(2021)Bommasani, Hudson, Adeli, Altman, Arora, von
  Arx, Bernstein, Bohg, Bosselut, Brunskill
  et~al.}]{bommasani2021opportunities}
Rishi Bommasani, Drew~A Hudson, Ehsan Adeli, Russ Altman, Simran Arora, Sydney
  von Arx, Michael~S Bernstein, Jeannette Bohg, Antoine Bosselut, Emma
  Brunskill, et~al. 2021.
\newblock On the opportunities and risks of foundation models.
\newblock \emph{arXiv preprint arXiv:2108.07258}.

\bibitem[{Braun et~al.(2024)Braun, Taylor, Goldowsky-Dill, and
  Sharkey}]{braun2024identifying}
Dan Braun, Jordan Taylor, Nicholas Goldowsky-Dill, and Lee Sharkey. 2024.
\newblock Identifying functionally important features with end-to-end sparse
  dictionary learning.
\newblock \emph{Advances in Neural Information Processing Systems},
  37:107286--107325.

\bibitem[{Bricken et~al.(2023)Bricken, Templeton, Batson, Chen, Jermyn,
  Conerly, Turner, Anil, Denison, Askell, Lasenby, Wu, Kravec, Schiefer,
  Maxwell, Joseph, Tamkin, Nguyen, McLean, Burke, Hume, Carter, Henighan, and
  Olah}]{bricken2023towards}
Trenton Bricken, Adly Templeton, Joshua Batson, Brian Chen, Adam Jermyn, Tom
  Conerly, Nicholas~L Turner, Cem Anil, Carson Denison, Amanda Askell, Robert
  Lasenby, Yifan Wu, Shauna Kravec, Nicholas Schiefer, Tim Maxwell, Nicholas
  Joseph, Alex Tamkin, Karina Nguyen, Brayden McLean, Josiah~E Burke, Tristan
  Hume, Shan Carter, Tom Henighan, and Chris Olah. 2023.
\newblock \href
  {https://transformer-circuits.pub/2023/monosemantic-features/index.html}
  {Towards monosemanticity: Decomposing language models with dictionary
  learning}.

\bibitem[{Brown et~al.(2020)Brown, Mann, Ryder, Subbiah, Kaplan, Dhariwal,
  Neelakantan, Shyam, Sastry, Askell, Agarwal, Herbert-Voss, Krueger, Henighan,
  Child, Ramesh, Ziegler, Wu, Winter, Hesse, Chen, Sigler, Litwin, Gray, Chess,
  Clark, Berner, McCandlish, Radford, Sutskever, and
  Amodei}]{brown2020language}
Tom~B. Brown, Benjamin Mann, Nick Ryder, Melanie Subbiah, Jared Kaplan,
  Prafulla Dhariwal, Arvind Neelakantan, Pranav Shyam, Girish Sastry, Amanda
  Askell, Sandhini Agarwal, Ariel Herbert-Voss, Gretchen Krueger, Tom Henighan,
  Rewon Child, Aditya Ramesh, Daniel~M. Ziegler, Jeffrey Wu, Clemens Winter,
  Christopher Hesse, Mark Chen, Eric Sigler, Mateusz Litwin, Scott Gray,
  Benjamin Chess, Jack Clark, Christopher Berner, Sam McCandlish, Alec Radford,
  Ilya Sutskever, and Dario Amodei. 2020.
\newblock \href {https://arxiv.org/abs/2005.14165} {Language models are
  few-shot learners}.
\newblock \emph{Preprint}, arXiv:2005.14165.

\bibitem[{B{\"u}rger et~al.(2024)B{\"u}rger, Hamprecht, and
  Nadler}]{burger2024truth}
Lennart B{\"u}rger, Fred~A. Hamprecht, and Boaz Nadler. 2024.
\newblock \href {https://openreview.net/forum?id=1Fc2Xa2cDK} {Truth is
  universal: Robust detection of lies in {LLM}s}.
\newblock In \emph{The Thirty-eighth Annual Conference on Neural Information
  Processing Systems}.

\bibitem[{Burns et~al.(2022)Burns, Ye, Klein, and
  Steinhardt}]{burns2022discovering}
Collin Burns, Haotian Ye, Dan Klein, and Jacob Steinhardt. 2022.
\newblock Discovering latent knowledge in language models without supervision.
\newblock \emph{arXiv preprint arXiv:2212.03827}.

\bibitem[{Cabannes et~al.(2024)Cabannes, Arnal, Bouaziz, Yang, Charton, and
  Kempe}]{cabannes2024iteration}
Vivien Cabannes, Charles Arnal, Wassim Bouaziz, Xingyu Yang, Francois Charton,
  and Julia Kempe. 2024.
\newblock Iteration head: A mechanistic study of chain-of-thought.
\newblock \emph{Advances in Neural Information Processing Systems},
  37:109101--109122.

\bibitem[{Cahlik et~al.(2025)Cahlik, Alves, and Kordik}]{cahlik2025reasoning}
Vojtech Cahlik, Rodrigo Alves, and Pavel Kordik. 2025.
\newblock Reasoning-grounded natural language explanations for language models.
\newblock \emph{arXiv preprint arXiv:2503.11248}.

\bibitem[{Calderon and Reichart(2025)}]{calderon2025behalf}
Nitay Calderon and Roi Reichart. 2025.
\newblock \href {https://aclanthology.org/2025.naacl-long.29/} {On behalf of
  the stakeholders: Trends in {NLP} model interpretability in the era of
  {LLM}s}.
\newblock In \emph{Proceedings of the 2025 Conference of the Nations of the
  Americas Chapter of the Association for Computational Linguistics: Human
  Language Technologies (Volume 1: Long Papers)}, pages 656--693, Albuquerque,
  New Mexico. Association for Computational Linguistics.

\bibitem[{Camburu et~al.(2018)Camburu, Rockt{\"a}schel, Lukasiewicz, and
  Blunsom}]{camburu2018snli}
Oana-Maria Camburu, Tim Rockt{\"a}schel, Thomas Lukasiewicz, and Phil Blunsom.
  2018.
\newblock e-snli: Natural language inference with natural language
  explanations.
\newblock \emph{Advances in Neural Information Processing Systems}, 31.

\bibitem[{Cao et~al.(2024)Cao, Gu, Shen, Yang, and Zhang}]{cao2024defending}
Yanfei Cao, Naijie Gu, Xinyue Shen, Daiyuan Yang, and Xingmin Zhang. 2024.
\newblock \href {https://doi.org/10.1109/NaNA63151.2024.00028} {Defending large
  language models against jailbreak attacks through chain of thought
  prompting}.
\newblock In \emph{2024 International Conference on Networking and Network
  Applications (NaNA)}, pages 125--130.

\bibitem[{CH-Wang et~al.(2024)CH-Wang, Van~Durme, Eisner, and
  Kedzie}]{ch-wang2024androids}
Sky CH-Wang, Benjamin Van~Durme, Jason Eisner, and Chris Kedzie. 2024.
\newblock \href {https://doi.org/10.18653/v1/2024.findings-acl.260} {Do
  androids know they`re only dreaming of electric sheep?}
\newblock In \emph{Findings of the Association for Computational Linguistics:
  ACL 2024}, pages 4401--4420, Bangkok, Thailand. Association for Computational
  Linguistics.

\bibitem[{Chalnev et~al.(2024)Chalnev, Siu, and Conmy}]{chalnev2024improving}
Sviatoslav Chalnev, Matthew Siu, and Arthur Conmy. 2024.
\newblock \href {https://arxiv.org/abs/2411.02193} {Improving steering vectors
  by targeting sparse autoencoder features}.
\newblock \emph{Preprint}, arXiv:2411.02193.

\bibitem[{Chang et~al.(2025)Chang, Rajagopal, Bolukbasi, Dixon, and
  Tenney}]{chang2025scalable}
Tyler~A. Chang, Dheeraj Rajagopal, Tolga Bolukbasi, Lucas Dixon, and Ian
  Tenney. 2025.
\newblock \href {https://openreview.net/forum?id=gLa96FlWwn} {Scalable
  influence and fact tracing for large language model pretraining}.
\newblock In \emph{The Thirteenth International Conference on Learning
  Representations}.

\bibitem[{Chattopadhyay et~al.(2019)Chattopadhyay, Manupriya, Sarkar, and
  Balasubramanian}]{chattopadhyay2019neural}
Aditya Chattopadhyay, Piyushi Manupriya, Anirban Sarkar, and Vineeth~N
  Balasubramanian. 2019.
\newblock Neural network attributions: a causal perspective.
\newblock In \emph{International Conference on Machine Learning}, pages
  981--990. PMLR.

\bibitem[{Chaudhry et~al.(2024)Chaudhry, Thiagarajan, and
  Gorur}]{chaudhry2024finetuning}
Arslan Chaudhry, Sridhar Thiagarajan, and Dilan Gorur. 2024.
\newblock Finetuning language models to emit linguistic expressions of
  uncertainty.
\newblock \emph{arXiv preprint arXiv:2409.12180}.

\bibitem[{Chen et~al.(2024{\natexlab{a}})Chen, Wang, Yao, Bai, Hou, and
  Li}]{chen2024finding}
Jianhui Chen, Xiaozhi Wang, Zijun Yao, Yushi Bai, Lei Hou, and Juanzi Li.
  2024{\natexlab{a}}.
\newblock \href {https://arxiv.org/abs/2406.14144} {Finding safety neurons in
  large language models}.
\newblock \emph{Preprint}, arXiv:2406.14144.

\bibitem[{Chen et~al.(2024{\natexlab{b}})Chen, Wu, DePodesta, Yeh, Li, Marin,
  Patel, Riecke, Raval, Seow, Wattenberg, and Viégas}]{chen2024designing}
Yida Chen, Aoyu Wu, Trevor DePodesta, Catherine Yeh, Kenneth Li,
  Nicholas~Castillo Marin, Oam Patel, Jan Riecke, Shivam Raval, Olivia Seow,
  Martin Wattenberg, and Fernanda Viégas. 2024{\natexlab{b}}.
\newblock \href {https://arxiv.org/abs/2406.07882} {Designing a dashboard for
  transparency and control of conversational ai}.
\newblock \emph{Preprint}, arXiv:2406.07882.

\bibitem[{Chen et~al.(2025)Chen, Li, You, Chen, Chang, Zhang, Dai, Guo, and
  Xiao}]{chen2025attributive}
Yuyan Chen, Zehao Li, Shuangjie You, Zhengyu Chen, Jingwen Chang, Yi~Zhang,
  Weinan Dai, Qingpei Guo, and Yanghua Xiao. 2025.
\newblock Attributive reasoning for hallucination diagnosis of large language
  models.
\newblock In \emph{Proceedings of the AAAI Conference on Artificial
  Intelligence}, volume~39, pages 23660--23668.

\bibitem[{Cheng et~al.(2025{\natexlab{a}})Cheng, Bae, Bullock, and
  Kristofferson}]{cheng2025training}
Deric Cheng, Juhan Bae, Justin Bullock, and David Kristofferson.
  2025{\natexlab{a}}.
\newblock Training data attribution (tda): Examining its adoption \& use cases.
\newblock \emph{arXiv preprint arXiv:2501.12642}.

\bibitem[{Cheng et~al.(2025{\natexlab{b}})Cheng, Li, Zhao, and
  Wen}]{cheng2025think}
Xiaoxue Cheng, Junyi Li, Wayne~Xin Zhao, and Ji-Rong Wen. 2025{\natexlab{b}}.
\newblock Think more, hallucinate less: Mitigating hallucinations via dual
  process of fast and slow thinking.
\newblock \emph{arXiv preprint arXiv:2501.01306}.

\bibitem[{Choe et~al.(2024)Choe, Ahn, Bae, Zhao, Kang, Chung, Pratapa,
  Neiswanger, Strubell, Mitamura et~al.}]{choe2024your}
Sang~Keun Choe, Hwijeen Ahn, Juhan Bae, Kewen Zhao, Minsoo Kang, Youngseog
  Chung, Adithya Pratapa, Willie Neiswanger, Emma Strubell, Teruko Mitamura,
  et~al. 2024.
\newblock What is your data worth to gpt? llm-scale data valuation with
  influence functions.
\newblock \emph{arXiv preprint arXiv:2405.13954}.

\bibitem[{Chu et~al.(2025)Chu, Tan, Xue, Wang, Mo, and Li}]{chu2025domaino1s}
Xu~Chu, Zhijie Tan, Hanlin Xue, Guanyu Wang, Tong Mo, and Weiping Li. 2025.
\newblock Domaino1s: Guiding llm reasoning for explainable answers in
  high-stakes domains.
\newblock \emph{arXiv preprint arXiv:2501.14431}.

\bibitem[{Chu et~al.(2024)Chu, Wang, Li, Wang, Qin, and Ren}]{chu2024a}
Zhixuan Chu, Yan Wang, Longfei Li, Zhibo Wang, Zhan Qin, and Kui Ren. 2024.
\newblock \href {https://doi.org/10.1145/3658644.3690217} {A causal explainable
  guardrails for large language models}.
\newblock In \emph{Proceedings of the 2024 on ACM SIGSAC Conference on Computer
  and Communications Security}, CCS '24, page 1136–1150, New York, NY, USA.
  Association for Computing Machinery.

\bibitem[{Chua et~al.(2024)Chua, Li, Yang, Wang, and Yao}]{chua2024ai}
Jaymari Chua, Yun Li, Shiyi Yang, Chen Wang, and Lina Yao. 2024.
\newblock Ai safety in generative ai large language models: A survey.
\newblock \emph{arXiv preprint arXiv:2407.18369}.

\bibitem[{Chuang et~al.(2024)Chuang, Qiu, Hsieh, Krishna, Kim, and
  Glass}]{chuang2024lookback}
Yung-Sung Chuang, Linlu Qiu, Cheng-Yu Hsieh, Ranjay Krishna, Yoon Kim, and
  James~R. Glass. 2024.
\newblock \href {https://doi.org/10.18653/v1/2024.emnlp-main.84} {Lookback
  lens: Detecting and mitigating contextual hallucinations in large language
  models using only attention maps}.
\newblock In \emph{Proceedings of the 2024 Conference on Empirical Methods in
  Natural Language Processing}, pages 1419--1436, Miami, Florida, USA.
  Association for Computational Linguistics.

\bibitem[{Cohen-Wang et~al.(2024)Cohen-Wang, Shah, Georgiev, and
  Madry}]{cohen2024contextcite}
Benjamin Cohen-Wang, Harshay Shah, Kristian Georgiev, and Aleksander Madry.
  2024.
\newblock Contextcite: Attributing model generation to context.
\newblock \emph{Advances in Neural Information Processing Systems},
  37:95764--95807.

\bibitem[{Conmy et~al.(2023)Conmy, Mavor-Parker, Lynch, Heimersheim, and
  Garriga-Alonso}]{conmy2023towards}
Arthur Conmy, Augustine Mavor-Parker, Aengus Lynch, Stefan Heimersheim, and
  Adri\`{a} Garriga-Alonso. 2023.
\newblock \href
  {https://proceedings.neurips.cc/paper_files/paper/2023/file/34e1dbe95d34d7ebaf99b9bcaeb5b2be-Paper-Conference.pdf}
  {Towards automated circuit discovery for mechanistic interpretability}.
\newblock In \emph{Advances in Neural Information Processing Systems},
  volume~36, pages 16318--16352. Curran Associates, Inc.

\bibitem[{Cook and Weisberg(1980)}]{cook1980characterizations}
R~Dennis Cook and Sanford Weisberg. 1980.
\newblock Characterizations of an empirical influence function for detecting
  influential cases in regression.
\newblock \emph{Technometrics}, 22(4):495--508.

\bibitem[{Coscia et~al.(2024)Coscia, Holmes, Morris, Choi, Crossley, and
  Endert}]{coscia2024iscore}
Adam Coscia, Langdon Holmes, Wesley Morris, Joon~Suh Choi, Scott Crossley, and
  Alex Endert. 2024.
\newblock \href {https://doi.org/10.1145/3640543.3645142} {iscore: Visual
  analytics for interpreting how language models automatically score
  summaries}.
\newblock In \emph{Proceedings of the 29th International Conference on
  Intelligent User Interfaces}, IUI '24, page 787–802, New York, NY, USA.
  Association for Computing Machinery.

\bibitem[{Covert et~al.(2021)Covert, Lundberg, and Lee}]{covert2021explaining}
Ian Covert, Scott Lundberg, and Su-In Lee. 2021.
\newblock Explaining by removing: A unified framework for model explanation.
\newblock \emph{Journal of Machine Learning Research}, 22(209):1--90.

\bibitem[{Creswell and Shanahan(2022)}]{creswell2022faithful}
Antonia Creswell and Murray Shanahan. 2022.
\newblock Faithful reasoning using large language models.
\newblock \emph{arXiv preprint arXiv:2208.14271}.

\bibitem[{Creswell et~al.(2023)Creswell, Shanahan, and
  Higgins}]{creswell2023selectioninference}
Antonia Creswell, Murray Shanahan, and Irina Higgins. 2023.
\newblock \href {https://openreview.net/forum?id=3Pf3Wg6o-A4}
  {Selection-inference: Exploiting large language models for interpretable
  logical reasoning}.
\newblock In \emph{The Eleventh International Conference on Learning
  Representations}.

\bibitem[{Dale et~al.(2023)Dale, Voita, Barrault, and
  Costa-juss{\`a}}]{dale2023detecting}
David Dale, Elena Voita, Loic Barrault, and Marta~R. Costa-juss{\`a}. 2023.
\newblock \href {https://doi.org/10.18653/v1/2023.acl-long.3} {Detecting and
  mitigating hallucinations in machine translation: Model internal workings
  alone do well, sentence similarity {E}ven better}.
\newblock In \emph{Proceedings of the 61st Annual Meeting of the Association
  for Computational Linguistics (Volume 1: Long Papers)}, pages 36--50,
  Toronto, Canada. Association for Computational Linguistics.

\bibitem[{Dalvi et~al.(2019)Dalvi, Durrani, Sajjad, Belinkov, Bau, and
  Glass}]{dalvi2019one}
Fahim Dalvi, Nadir Durrani, Hassan Sajjad, Yonatan Belinkov, Anthony Bau, and
  James Glass. 2019.
\newblock What is one grain of sand in the desert? analyzing individual neurons
  in deep nlp models.
\newblock In \emph{Proceedings of the AAAI Conference on Artificial
  Intelligence}, volume~33, pages 6309--6317.

\bibitem[{Dar et~al.(2023)Dar, Geva, Gupta, and Berant}]{dar2023analyzing}
Guy Dar, Mor Geva, Ankit Gupta, and Jonathan Berant. 2023.
\newblock \href {https://doi.org/10.18653/v1/2023.acl-long.893} {Analyzing
  transformers in embedding space}.
\newblock In \emph{Proceedings of the 61st Annual Meeting of the Association
  for Computational Linguistics (Volume 1: Long Papers)}, pages 16124--16170,
  Toronto, Canada. Association for Computational Linguistics.

\bibitem[{Davies et~al.(2023)Davies, Nadeau, Prakash, Shaham, and
  Bau}]{davies2023discovering}
Xander Davies, Max Nadeau, Nikhil Prakash, Tamar~Rott Shaham, and David Bau.
  2023.
\newblock Discovering variable binding circuitry with desiderata.
\newblock \emph{arXiv preprint arXiv:2307.03637}.

\bibitem[{Deiseroth et~al.(2023)Deiseroth, Deb, Weinbach, Brack, Schramowski,
  and Kersting}]{deiseroth2023atman}
Bj\"{o}rn Deiseroth, Mayukh Deb, Samuel Weinbach, Manuel Brack, Patrick
  Schramowski, and Kristian Kersting. 2023.
\newblock Atman: understanding transformer predictions through memory efficient
  attention manipulation.
\newblock In \emph{Proceedings of the 37th International Conference on Neural
  Information Processing Systems}, NIPS '23, Red Hook, NY, USA. Curran
  Associates Inc.

\bibitem[{Deng et~al.(2025)Deng, Wang, Zhu, Wang, and Feng}]{deng2025cram}
Boyi Deng, Wenjie Wang, Fengbin Zhu, Qifan Wang, and Fuli Feng. 2025.
\newblock \href {http://arxiv.org/abs/2406.11497} {Cram: Credibility-aware
  attention modification in llms for combating misinformation in rag}.
\newblock In \emph{Proceedings of the AAAI Conference on Artificial
  Intelligence}, volume~39, pages 23760--23768.

\bibitem[{Dhuliawala et~al.(2024)Dhuliawala, Komeili, Xu, Raileanu, Li,
  Celikyilmaz, and Weston}]{dhuliawala2024chain}
Shehzaad Dhuliawala, Mojtaba Komeili, Jing Xu, Roberta Raileanu, Xian Li, Asli
  Celikyilmaz, and Jason Weston. 2024.
\newblock \href {https://doi.org/10.18653/v1/2024.findings-acl.212}
  {Chain-of-verification reduces hallucination in large language models}.
\newblock In \emph{Findings of the Association for Computational Linguistics:
  ACL 2024}, pages 3563--3578, Bangkok, Thailand. Association for Computational
  Linguistics.

\bibitem[{Din et~al.(2023)Din, Karidi, Choshen, and Geva}]{din2023jump}
Alexander~Yom Din, Taelin Karidi, Leshem Choshen, and Mor Geva. 2023.
\newblock Jump to conclusions: Short-cutting transformers with linear
  transformations.
\newblock \emph{arXiv preprint arXiv:2303.09435}.

\bibitem[{Doshi-Velez and Kim(2017)}]{doshi2017towards}
Finale Doshi-Velez and Been Kim. 2017.
\newblock Towards a rigorous science of interpretable machine learning.
\newblock \emph{arXiv preprint arXiv:1702.08608}.

\bibitem[{Duan et~al.(2024)Duan, Yang, and Tam}]{duan2024llms}
Hanyu Duan, Yi~Yang, and Kar~Yan Tam. 2024.
\newblock Do llms know about hallucination? an empirical investigation of llm's
  hidden states.
\newblock \emph{arXiv preprint arXiv:2402.09733}.

\bibitem[{Dunefsky et~al.(2024{\natexlab{a}})Dunefsky, Chlenski, and
  Nanda}]{dunefsky2024transcoders}
Jacob Dunefsky, Philippe Chlenski, and Neel Nanda. 2024{\natexlab{a}}.
\newblock Transcoders enable fine-grained interpretable circuit analysis for
  language models.
\newblock
  \url{https://www.lesswrong.com/posts/YmkjnWtZGLbHRbzrP/transcoders-enable-fine-grained-interpretable-circuit}.
\newblock LessWrong.

\bibitem[{Dunefsky et~al.(2024{\natexlab{b}})Dunefsky, Chlenski, and
  Nanda}]{dunefsky2024transcoders-neurips}
Jacob Dunefsky, Philippe Chlenski, and Neel Nanda. 2024{\natexlab{b}}.
\newblock \href {https://openreview.net/forum?id=J6zHcScAo0} {Transcoders find
  interpretable {LLM} feature circuits}.
\newblock In \emph{The Thirty-eighth Annual Conference on Neural Information
  Processing Systems}.

\bibitem[{Dunefsky et~al.(2024{\natexlab{c}})Dunefsky, Chlenski, Rajamanoharan,
  and Nanda}]{dunefsky2024case}
Jacob Dunefsky, Philippe Chlenski, Senthooran Rajamanoharan, and Neel Nanda.
  2024{\natexlab{c}}.
\newblock Case studies in reverse-engineering sparse autoencoder features by
  using mlp linearization.
\newblock
  \url{https://www.lesswrong.com/posts/93nKtsDL6YY5fRbQv/case-studies-in-reverse-engineering-sparse-autoencoder}.
\newblock LessWrong.

\bibitem[{Eberle et~al.(2023)Eberle, Chalkidis, Cabello, and
  Brandl}]{eberle2023rather}
Oliver Eberle, Ilias Chalkidis, Laura Cabello, and Stephanie Brandl. 2023.
\newblock \href {https://doi.org/10.18653/v1/2023.emnlp-main.427} {Rather a
  nurse than a physician - contrastive explanations under investigation}.
\newblock In \emph{Proceedings of the 2023 Conference on Empirical Methods in
  Natural Language Processing}, pages 6907--6920, Singapore. Association for
  Computational Linguistics.

\bibitem[{Elhage et~al.(2022)Elhage, Hume, Olsson, Schiefer, Henighan, Kravec,
  Hatfield-Dodds, Lasenby, Drain, Chen et~al.}]{elhage2022toy}
Nelson Elhage, Tristan Hume, Catherine Olsson, Nicholas Schiefer, Tom Henighan,
  Shauna Kravec, Zac Hatfield-Dodds, Robert Lasenby, Dawn Drain, Carol Chen,
  et~al. 2022.
\newblock Toy models of superposition.
\newblock \emph{arXiv preprint arXiv:2209.10652}.

\bibitem[{Elhage et~al.(2021)Elhage, Nanda, Olsson, Henighan, Joseph, Mann,
  Askell, Bai, Chen, Conerly et~al.}]{elhage2021mathematical}
Nelson Elhage, Neel Nanda, Catherine Olsson, Tom Henighan, Nicholas Joseph, Ben
  Mann, Amanda Askell, Yuntao Bai, Anna Chen, Tom Conerly, et~al. 2021.
\newblock A mathematical framework for transformer circuits.
\newblock \emph{Transformer Circuits Thread}, 1(1):12.

\bibitem[{Engstrom et~al.(2024)Engstrom, Feldmann, and
  M\k{a}dry}]{engstrom2024dsdm}
Logan Engstrom, Axel Feldmann, and Aleksander M\k{a}dry. 2024.
\newblock Dsdm: model-aware dataset selection with datamodels.
\newblock In \emph{Proceedings of the 41st International Conference on Machine
  Learning}, ICML'24. JMLR.org.

\bibitem[{Enguehard(2023)}]{enguehard2023sequential}
Joseph Enguehard. 2023.
\newblock \href {https://doi.org/10.18653/v1/2023.findings-acl.477} {Sequential
  integrated gradients: a simple but effective method for explaining language
  models}.
\newblock In \emph{Findings of the Association for Computational Linguistics:
  ACL 2023}, pages 7555--7565, Toronto, Canada. Association for Computational
  Linguistics.

\bibitem[{Enouen et~al.(2024)Enouen, Nakhost, Ebrahimi, Arik, Liu, and
  Pfister}]{enouen2024textgenshap}
James Enouen, Hootan Nakhost, Sayna Ebrahimi, Sercan Arik, Yan Liu, and Tomas
  Pfister. 2024.
\newblock \href {https://doi.org/10.18653/v1/2024.findings-acl.832}
  {{T}ext{G}en{SHAP}: Scalable post-hoc explanations in text generation with
  long documents}.
\newblock In \emph{Findings of the Association for Computational Linguistics:
  ACL 2024}, pages 13984--14011, Bangkok, Thailand. Association for
  Computational Linguistics.

\bibitem[{Farnik et~al.(2025)Farnik, Lawson, Houghton, and
  Aitchison}]{farnik2025jacobian}
Lucy Farnik, Tim Lawson, Conor Houghton, and Laurence Aitchison. 2025.
\newblock Jacobian sparse autoencoders: Sparsify computations, not just
  activations.
\newblock \emph{arXiv preprint arXiv:2502.18147}.

\bibitem[{Feldman and Zhang(2020)}]{feldman2020what}
Vitaly Feldman and Chiyuan Zhang. 2020.
\newblock \href
  {https://proceedings.neurips.cc/paper_files/paper/2020/file/1e14bfe2714193e7af5abc64ecbd6b46-Paper.pdf}
  {What neural networks memorize and why: Discovering the long tail via
  influence estimation}.
\newblock In \emph{Advances in Neural Information Processing Systems},
  volume~33, pages 2881--2891. Curran Associates, Inc.

\bibitem[{Feng et~al.(2024)Feng, Zhou, ZHU, Qian, and Mao}]{feng2024unveiling}
Zijian Feng, Hanzhang Zhou, ZIXIAO ZHU, Junlang Qian, and Kezhi Mao. 2024.
\newblock \href {https://openreview.net/forum?id=ap1ByuwQrX} {Unveiling and
  manipulating prompt influence in large language models}.
\newblock In \emph{The Twelfth International Conference on Learning
  Representations}.

\bibitem[{Ferrando et~al.(2022{\natexlab{a}})Ferrando, G{\'a}llego, Alastruey,
  Escolano, and Costa-juss{\`a}}]{ferrando2022towards}
Javier Ferrando, Gerard~I. G{\'a}llego, Belen Alastruey, Carlos Escolano, and
  Marta~R. Costa-juss{\`a}. 2022{\natexlab{a}}.
\newblock \href {https://doi.org/10.18653/v1/2022.emnlp-main.599} {Towards
  opening the black box of neural machine translation: Source and target
  interpretations of the transformer}.
\newblock In \emph{Proceedings of the 2022 Conference on Empirical Methods in
  Natural Language Processing}, pages 8756--8769, Abu Dhabi, United Arab
  Emirates. Association for Computational Linguistics.

\bibitem[{Ferrando et~al.(2022{\natexlab{b}})Ferrando, G{\'a}llego, and
  Costa-juss{\`a}}]{ferrando2022measuring}
Javier Ferrando, Gerard~I. G{\'a}llego, and Marta~R. Costa-juss{\`a}.
  2022{\natexlab{b}}.
\newblock \href {https://doi.org/10.18653/v1/2022.emnlp-main.595} {Measuring
  the mixing of contextual information in the transformer}.
\newblock In \emph{Proceedings of the 2022 Conference on Empirical Methods in
  Natural Language Processing}, pages 8698--8714, Abu Dhabi, United Arab
  Emirates. Association for Computational Linguistics.

\bibitem[{Ferrando et~al.(2023)Ferrando, G{\'a}llego, Tsiamas, and
  Costa-juss{\`a}}]{ferrando2023explaining}
Javier Ferrando, Gerard~I. G{\'a}llego, Ioannis Tsiamas, and Marta~R.
  Costa-juss{\`a}. 2023.
\newblock \href {https://doi.org/10.18653/v1/2023.acl-long.301} {Explaining how
  transformers use context to build predictions}.
\newblock In \emph{Proceedings of the 61st Annual Meeting of the Association
  for Computational Linguistics (Volume 1: Long Papers)}, pages 5486--5513,
  Toronto, Canada. Association for Computational Linguistics.

\bibitem[{Ferrando et~al.(2025)Ferrando, Obeso, Rajamanoharan, and
  Nanda}]{ferrando2025do}
Javier Ferrando, Oscar~Balcells Obeso, Senthooran Rajamanoharan, and Neel
  Nanda. 2025.
\newblock \href {https://openreview.net/forum?id=WCRQFlji2q} {Do i know this
  entity? knowledge awareness and hallucinations in language models}.
\newblock In \emph{The Thirteenth International Conference on Learning
  Representations}.

\bibitem[{Ferrando et~al.(2024)Ferrando, Sarti, Bisazza, and
  Costa-Juss{\`a}}]{ferrando2024primer}
Javier Ferrando, Gabriele Sarti, Arianna Bisazza, and Marta~R Costa-Juss{\`a}.
  2024.
\newblock A primer on the inner workings of transformer-based language models.
\newblock \emph{arXiv preprint arXiv:2405.00208}.

\bibitem[{Ferrando and Voita(2024)}]{ferrando2024information}
Javier Ferrando and Elena Voita. 2024.
\newblock \href {https://doi.org/10.18653/v1/2024.emnlp-main.965} {Information
  flow routes: Automatically interpreting language models at scale}.
\newblock In \emph{Proceedings of the 2024 Conference on Empirical Methods in
  Natural Language Processing}, pages 17432--17445, Miami, Florida, USA.
  Association for Computational Linguistics.

\bibitem[{Finlayson et~al.(2021)Finlayson, Mueller, Gehrmann, Shieber, Linzen,
  and Belinkov}]{finlayson2021causal}
Matthew Finlayson, Aaron Mueller, Sebastian Gehrmann, Stuart Shieber, Tal
  Linzen, and Yonatan Belinkov. 2021.
\newblock \href {https://doi.org/10.18653/v1/2021.acl-long.144} {Causal
  analysis of syntactic agreement mechanisms in neural language models}.
\newblock In \emph{Proceedings of the 59th Annual Meeting of the Association
  for Computational Linguistics and the 11th International Joint Conference on
  Natural Language Processing (Volume 1: Long Papers)}, pages 1828--1843,
  Online. Association for Computational Linguistics.

\bibitem[{Foote et~al.(2023)Foote, Nanda, Kran, Konstas, Cohen, and
  Barez}]{foote2023neuron}
Alex Foote, Neel Nanda, Esben Kran, Ioannis Konstas, Shay Cohen, and Fazl
  Barez. 2023.
\newblock \href {https://arxiv.org/abs/2305.19911} {Neuron to graph:
  Interpreting language model neurons at scale}.
\newblock \emph{Preprint}, arXiv:2305.19911.

\bibitem[{Frikha et~al.(2025)Frikha, Razi, Nakka, Mendes, Jiang, and
  Zhou}]{frikha2025privacyscalpel}
Ahmed Frikha, Muhammad Reza~Ar Razi, Krishna~Kanth Nakka, Ricardo Mendes, Xue
  Jiang, and Xuebing Zhou. 2025.
\newblock Privacyscalpel: Enhancing llm privacy via interpretable feature
  intervention with sparse autoencoders.
\newblock \emph{arXiv preprint arXiv:2503.11232}.

\bibitem[{Gallifant et~al.(2025)Gallifant, Chen, Sasse, Aerts, Hartvigsen, and
  Bitterman}]{gallifant2025sparse}
Jack Gallifant, Shan Chen, Kuleen Sasse, Hugo Aerts, Thomas Hartvigsen, and
  Danielle~S Bitterman. 2025.
\newblock Sparse autoencoder features for classifications and transferability.
\newblock \emph{arXiv preprint arXiv:2502.11367}.

\bibitem[{Gao et~al.(2023)Gao, Ding, Qin, and Liu}]{gao2023chatgpt}
Jinglong Gao, Xiao Ding, Bing Qin, and Ting Liu. 2023.
\newblock \href {https://doi.org/10.18653/v1/2023.findings-emnlp.743} {Is
  {C}hat{GPT} a good causal reasoner? a comprehensive evaluation}.
\newblock In \emph{Findings of the Association for Computational Linguistics:
  EMNLP 2023}, pages 11111--11126, Singapore. Association for Computational
  Linguistics.

\bibitem[{Gao et~al.(2024)Gao, Zhang, Nakov, and Chen}]{gao2024shaping}
Lang Gao, Xiangliang Zhang, Preslav Nakov, and Xiuying Chen. 2024.
\newblock Shaping the safety boundaries: Understanding and defending against
  jailbreaks in large language models.
\newblock \emph{arXiv preprint arXiv:2412.17034}.

\bibitem[{Garde et~al.(2023)Garde, Kran, and Barez}]{garde2023deepdecipher}
Albert Garde, Esben Kran, and Fazl Barez. 2023.
\newblock \href {https://arxiv.org/abs/2310.01870} {Deepdecipher: Accessing and
  investigating neuron activation in large language models}.
\newblock \emph{Preprint}, arXiv:2310.01870.

\bibitem[{Gat et~al.(2024)Gat, Calderon, Feder, Chapanin, Sharma, and
  Reichart}]{gat2024faithful}
Yair~Ori Gat, Nitay Calderon, Amir Feder, Alexander Chapanin, Amit Sharma, and
  Roi Reichart. 2024.
\newblock \href {https://openreview.net/forum?id=UMfcdRIotC} {Faithful
  explanations of black-box {NLP} models using {LLM}-generated
  counterfactuals}.
\newblock In \emph{The Twelfth International Conference on Learning
  Representations}.

\bibitem[{Geiger et~al.(2021)Geiger, Lu, Icard, and Potts}]{geiger2021causal}
Atticus Geiger, Hanson Lu, Thomas Icard, and Christopher Potts. 2021.
\newblock Causal abstractions of neural networks.
\newblock \emph{Advances in Neural Information Processing Systems},
  34:9574--9586.

\bibitem[{Geva et~al.(2023)Geva, Bastings, Filippova, and
  Globerson}]{geva2023dissecting}
Mor Geva, Jasmijn Bastings, Katja Filippova, and Amir Globerson. 2023.
\newblock \href {https://doi.org/10.18653/v1/2023.emnlp-main.751} {Dissecting
  recall of factual associations in auto-regressive language models}.
\newblock In \emph{Proceedings of the 2023 Conference on Empirical Methods in
  Natural Language Processing}, pages 12216--12235, Singapore. Association for
  Computational Linguistics.

\bibitem[{Geva et~al.(2022{\natexlab{a}})Geva, Caciularu, Dar, Roit, Sadde,
  Shlain, Tamir, and Goldberg}]{geva2022lm}
Mor Geva, Avi Caciularu, Guy Dar, Paul Roit, Shoval Sadde, Micah Shlain, Bar
  Tamir, and Yoav Goldberg. 2022{\natexlab{a}}.
\newblock \href {https://doi.org/10.18653/v1/2022.emnlp-demos.2}
  {{LM}-debugger: An interactive tool for inspection and intervention in
  transformer-based language models}.
\newblock In \emph{Proceedings of the 2022 Conference on Empirical Methods in
  Natural Language Processing: System Demonstrations}, pages 12--21, Abu Dhabi,
  UAE. Association for Computational Linguistics.

\bibitem[{Geva et~al.(2022{\natexlab{b}})Geva, Caciularu, Wang, and
  Goldberg}]{geva2022transformer}
Mor Geva, Avi Caciularu, Kevin Wang, and Yoav Goldberg. 2022{\natexlab{b}}.
\newblock \href {https://doi.org/10.18653/v1/2022.emnlp-main.3} {Transformer
  feed-forward layers build predictions by promoting concepts in the vocabulary
  space}.
\newblock In \emph{Proceedings of the 2022 Conference on Empirical Methods in
  Natural Language Processing}, pages 30--45, Abu Dhabi, United Arab Emirates.
  Association for Computational Linguistics.

\bibitem[{Geva et~al.(2021)Geva, Schuster, Berant, and
  Levy}]{geva2021transformer}
Mor Geva, Roei Schuster, Jonathan Berant, and Omer Levy. 2021.
\newblock \href {https://doi.org/10.18653/v1/2021.emnlp-main.446} {Transformer
  feed-forward layers are key-value memories}.
\newblock In \emph{Proceedings of the 2021 Conference on Empirical Methods in
  Natural Language Processing}, pages 5484--5495, Online and Punta Cana,
  Dominican Republic. Association for Computational Linguistics.

\bibitem[{Ghandeharioun et~al.(2024)Ghandeharioun, Caciularu, Pearce, Dixon,
  and Geva}]{ghandeharioun2024patchscopes}
Asma Ghandeharioun, Avi Caciularu, Adam Pearce, Lucas Dixon, and Mor Geva.
  2024.
\newblock Patchscopes: a unifying framework for inspecting hidden
  representations of language models.
\newblock In \emph{Proceedings of the 41st International Conference on Machine
  Learning}, ICML'24. JMLR.org.

\bibitem[{Ghilardi et~al.(2024)Ghilardi, Belotti, and
  Molinari}]{ghilardi2024efficient}
Davide Ghilardi, Federico Belotti, and Marco Molinari. 2024.
\newblock Efficient training of sparse autoencoders for large language models
  via layer groups.
\newblock \emph{arXiv preprint arXiv:2410.21508}.

\bibitem[{Ghorbani and Zou(2019)}]{ghorbani2019data}
Amirata Ghorbani and James Zou. 2019.
\newblock Data shapley: Equitable valuation of data for machine learning.
\newblock In \emph{International conference on machine learning}, pages
  2242--2251. PMLR.

\bibitem[{Goldowsky-Dill et~al.(2023)Goldowsky-Dill, MacLeod, Sato, and
  Arora}]{goldowsky2023localizing}
Nicholas Goldowsky-Dill, Chris MacLeod, Lucas Sato, and Aryaman Arora. 2023.
\newblock Localizing model behavior with path patching.
\newblock \emph{arXiv preprint arXiv:2304.05969}.

\bibitem[{Grosse et~al.(2023)Grosse, Bae, Anil, Elhage, Tamkin, Tajdini,
  Steiner, Li, Durmus, Perez et~al.}]{grosse2023studying}
Roger Grosse, Juhan Bae, Cem Anil, Nelson Elhage, Alex Tamkin, Amirhossein
  Tajdini, Benoit Steiner, Dustin Li, Esin Durmus, Ethan Perez, et~al. 2023.
\newblock Studying large language model generalization with influence
  functions.
\newblock \emph{arXiv preprint arXiv:2308.03296}.

\bibitem[{Guo et~al.(2021)Guo, Rajani, Hase, Bansal, and Xiong}]{guo2021fastif}
Han Guo, Nazneen Rajani, Peter Hase, Mohit Bansal, and Caiming Xiong. 2021.
\newblock \href {https://doi.org/10.18653/v1/2021.emnlp-main.808} {{F}ast{IF}:
  Scalable influence functions for efficient model interpretation and
  debugging}.
\newblock In \emph{Proceedings of the 2021 Conference on Empirical Methods in
  Natural Language Processing}, pages 10333--10350, Online and Punta Cana,
  Dominican Republic. Association for Computational Linguistics.

\bibitem[{Gurnee et~al.(2023)Gurnee, Nanda, Pauly, Harvey, Troitskii, and
  Bertsimas}]{gurnee2023finding}
Wes Gurnee, Neel Nanda, Matthew Pauly, Katherine Harvey, Dmitrii Troitskii, and
  Dimitris Bertsimas. 2023.
\newblock Finding neurons in a haystack: Case studies with sparse probing.
\newblock \emph{arXiv preprint arXiv:2305.01610}.

\bibitem[{Guu et~al.(2023)Guu, Webson, Pavlick, Dixon, Tenney, and
  Bolukbasi}]{guu2023simfluence}
Kelvin Guu, Albert Webson, Ellie Pavlick, Lucas Dixon, Ian Tenney, and Tolga
  Bolukbasi. 2023.
\newblock Simfluence: Modeling the influence of individual training examples by
  simulating training runs.
\newblock \emph{arXiv preprint arXiv:2303.08114}.

\bibitem[{Halawi et~al.(2024)Halawi, Denain, and
  Steinhardt}]{halawi2024overthinking}
Danny Halawi, Jean-Stanislas Denain, and Jacob Steinhardt. 2024.
\newblock \href {https://openreview.net/forum?id=Tigr1kMDZy} {Overthinking the
  truth: Understanding how language models process false demonstrations}.
\newblock In \emph{The Twelfth International Conference on Learning
  Representations}.

\bibitem[{Hammoudeh and Lowd(2024)}]{hammoudeh2024training}
Zayd Hammoudeh and Daniel Lowd. 2024.
\newblock \href {https://doi.org/10.1007/s10994-023-06495-7} {Training data
  influence analysis and estimation: a survey}.
\newblock \emph{Mach. Learn.}, 113(5):2351–2403.

\bibitem[{Hampel(1974)}]{hampel1974influence}
Frank~R Hampel. 1974.
\newblock The influence curve and its role in robust estimation.
\newblock \emph{Journal of the american statistical association},
  69(346):383--393.

\bibitem[{Han et~al.(2025)Han, Qian, Chen, Zhang, Zhang, and
  Ji}]{han2025internal}
Peixuan Han, Cheng Qian, Xiusi Chen, Yuji Zhang, Denghui Zhang, and Heng Ji.
  2025.
\newblock Internal activation as the polar star for steering unsafe llm
  behavior.
\newblock \emph{arXiv preprint arXiv:2502.01042}.

\bibitem[{Han and Tsvetkov(2021)}]{han2021influence}
Xiaochuang Han and Yulia Tsvetkov. 2021.
\newblock \href {https://doi.org/10.18653/v1/2021.findings-emnlp.374}
  {Influence tuning: Demoting spurious correlations via instance attribution
  and instance-driven updates}.
\newblock In \emph{Findings of the Association for Computational Linguistics:
  EMNLP 2021}, pages 4398--4409, Punta Cana, Dominican Republic. Association
  for Computational Linguistics.

\bibitem[{Han and Tsvetkov(2022)}]{han2022orca}
Xiaochuang Han and Yulia Tsvetkov. 2022.
\newblock Orca: Interpreting prompted language models via locating supporting
  data evidence in the ocean of pretraining data.
\newblock \emph{arXiv preprint arXiv:2205.12600}.

\bibitem[{Han et~al.(2020)Han, Wallace, and Tsvetkov}]{han2020explaining}
Xiaochuang Han, Byron~C. Wallace, and Yulia Tsvetkov. 2020.
\newblock \href {https://doi.org/10.18653/v1/2020.acl-main.492} {Explaining
  black box predictions and unveiling data artifacts through influence
  functions}.
\newblock In \emph{Proceedings of the 58th Annual Meeting of the Association
  for Computational Linguistics}, pages 5553--5563, Online. Association for
  Computational Linguistics.

\bibitem[{Hanna et~al.(2023)Hanna, Liu, and Variengien}]{hanna2023how}
Michael Hanna, Ollie Liu, and Alexandre Variengien. 2023.
\newblock How does gpt-2 compute greater-than? interpreting mathematical
  abilities in a pre-trained language model.
\newblock In \emph{Proceedings of the 37th International Conference on Neural
  Information Processing Systems}, NIPS '23, Red Hook, NY, USA. Curran
  Associates Inc.

\bibitem[{Hanna et~al.(2024)Hanna, Pezzelle, and Belinkov}]{hanna2024have}
Michael Hanna, Sandro Pezzelle, and Yonatan Belinkov. 2024.
\newblock Have faith in faithfulness: Going beyond circuit overlap when finding
  model mechanisms.
\newblock \emph{arXiv preprint arXiv:2403.17806}.

\bibitem[{Hao et~al.(2024)Hao, Gu, Luo, Liu, Shao, Wang, Xie, Ma, Samavedhi,
  Gao, Wang, and Hu}]{hao2024llm}
Shibo Hao, Yi~Gu, Haotian Luo, Tianyang Liu, Xiyan Shao, Xinyuan Wang, Shuhua
  Xie, Haodi Ma, Adithya Samavedhi, Qiyue Gao, Zhen Wang, and Zhiting Hu. 2024.
\newblock \href {https://openreview.net/forum?id=b0y6fbSUG0} {{LLM} reasoners:
  New evaluation, library, and analysis of step-by-step reasoning with large
  language models}.
\newblock In \emph{First Conference on Language Modeling}.

\bibitem[{H{\"a}rle et~al.(2024)H{\"a}rle, Friedrich, Brack, Deiseroth,
  Schramowski, and Kersting}]{harle2024scar}
Ruben H{\"a}rle, Felix Friedrich, Manuel Brack, Bj{\"o}rn Deiseroth, Patrick
  Schramowski, and Kristian Kersting. 2024.
\newblock Scar: Sparse conditioned autoencoders for concept detection and
  steering in llms.
\newblock \emph{arXiv preprint arXiv:2411.07122}.

\bibitem[{Haviv et~al.(2023)Haviv, Cohen, Gidron, Schuster, Goldberg, and
  Geva}]{haviv2023understanding}
Adi Haviv, Ido Cohen, Jacob Gidron, Roei Schuster, Yoav Goldberg, and Mor Geva.
  2023.
\newblock \href {https://doi.org/10.18653/v1/2023.eacl-main.19} {Understanding
  transformer memorization recall through idioms}.
\newblock In \emph{Proceedings of the 17th Conference of the European Chapter
  of the Association for Computational Linguistics}, pages 248--264, Dubrovnik,
  Croatia. Association for Computational Linguistics.

\bibitem[{Hazra et~al.(2024)Hazra, Layek, Banerjee, and
  Poria}]{hazra2024safety}
Rima Hazra, Sayan Layek, Somnath Banerjee, and Soujanya Poria. 2024.
\newblock \href {https://doi.org/10.18653/v1/2024.emnlp-main.1212} {Safety
  arithmetic: A framework for test-time safety alignment of language models by
  steering parameters and activations}.
\newblock In \emph{Proceedings of the 2024 Conference on Empirical Methods in
  Natural Language Processing}, pages 21759--21776, Miami, Florida, USA.
  Association for Computational Linguistics.

\bibitem[{He et~al.(2024{\natexlab{a}})He, Gong, Lin, Wei, Zhao, and
  Chen}]{he2024llm}
Jinwen He, Yujia Gong, Zijin Lin, Cheng{'}an Wei, Yue Zhao, and Kai Chen.
  2024{\natexlab{a}}.
\newblock \href {https://doi.org/10.18653/v1/2024.findings-acl.608} {{LLM}
  factoscope: Uncovering {LLM}s' factual discernment through measuring inner
  states}.
\newblock In \emph{Findings of the Association for Computational Linguistics:
  ACL 2024}, pages 10218--10230, Bangkok, Thailand. Association for
  Computational Linguistics.

\bibitem[{He et~al.(2024{\natexlab{b}})He, Xia, and Henderson}]{he2024what}
Luxi He, Mengzhou Xia, and Peter Henderson. 2024{\natexlab{b}}.
\newblock \href {https://openreview.net/forum?id=Hi8jKh4HE9} {What is in your
  safe data? identifying benign data that breaks safety}.
\newblock In \emph{First Conference on Language Modeling}.

\bibitem[{He et~al.(2024{\natexlab{c}})He, Wang, Chu, Xu, Zheng, Ren, and
  Chen}]{he2024jailbreaklens}
Zeqing He, Zhibo Wang, Zhixuan Chu, Huiyu Xu, Rui Zheng, Kui Ren, and Chun
  Chen. 2024{\natexlab{c}}.
\newblock Jailbreaklens: Interpreting jailbreak mechanism in the lens of
  representation and circuit.
\newblock \emph{arXiv preprint arXiv:2411.11114}.

\bibitem[{He et~al.(2025)He, Wang, Xu, and Ren}]{he2025towards}
Zeqing He, Zhibo Wang, Huiyu Xu, and Kui Ren. 2025.
\newblock Towards llm guardrails via sparse representation steering.
\newblock \emph{arXiv preprint arXiv:2503.16851}.

\bibitem[{He et~al.(2024{\natexlab{d}})He, Ge, Tang, Sun, Cheng, and
  Qiu}]{he2024dictionary}
Zhengfu He, Xuyang Ge, Qiong Tang, Tianxiang Sun, Qinyuan Cheng, and Xipeng
  Qiu. 2024{\natexlab{d}}.
\newblock Dictionary learning improves patch-free circuit discovery in
  mechanistic interpretability: A case study on othello-gpt.
\newblock \emph{arXiv preprint arXiv:2402.12201}.

\bibitem[{He et~al.(2024{\natexlab{e}})He, Shu, Ge, Chen, Wang, Zhou, Liu, Guo,
  Huang, Wu et~al.}]{he2024llama}
Zhengfu He, Wentao Shu, Xuyang Ge, Lingjie Chen, Junxuan Wang, Yunhua Zhou,
  Frances Liu, Qipeng Guo, Xuanjing Huang, Zuxuan Wu, et~al.
  2024{\natexlab{e}}.
\newblock Llama scope: Extracting millions of features from llama-3.1-8b with
  sparse autoencoders.
\newblock \emph{arXiv preprint arXiv:2410.20526}.

\bibitem[{Hegde(2024)}]{hegde2024effectiveness}
Praveen Hegde. 2024.
\newblock \href {https://openreview.net/forum?id=IxEMXN5hCg} {Effectiveness of
  sparse autoencoder for understanding and removing gender bias in {LLM}s}.
\newblock In \emph{NeurIPS 2024 Workshop on Scientific Methods for
  Understanding Deep Learning}.

\bibitem[{Hernandez et~al.(2024{\natexlab{a}})Hernandez, Li, and
  Andreas}]{hernandez2024inspecting}
Evan Hernandez, Belinda~Z. Li, and Jacob Andreas. 2024{\natexlab{a}}.
\newblock \href {https://openreview.net/forum?id=ADtL6fgNRv} {Inspecting and
  editing knowledge representations in language models}.
\newblock In \emph{First Conference on Language Modeling}.

\bibitem[{Hernandez et~al.(2024{\natexlab{b}})Hernandez, Sharma, Haklay, Meng,
  Wattenberg, Andreas, Belinkov, and Bau}]{hernandez2024linearity}
Evan Hernandez, Arnab~Sen Sharma, Tal Haklay, Kevin Meng, Martin Wattenberg,
  Jacob Andreas, Yonatan Belinkov, and David Bau. 2024{\natexlab{b}}.
\newblock \href {https://openreview.net/forum?id=w7LU2s14kE} {Linearity of
  relation decoding in transformer language models}.
\newblock In \emph{The Twelfth International Conference on Learning
  Representations}.

\bibitem[{Hildebrandt et~al.(2025)Hildebrandt, Maier, Krauss, and
  Schilling}]{hildebrandt2025refusal}
Fabian Hildebrandt, Andreas Maier, Patrick Krauss, and Achim Schilling. 2025.
\newblock Refusal behavior in large language models: A nonlinear perspective.
\newblock \emph{arXiv preprint arXiv:2501.08145}.

\bibitem[{Hohman et~al.(2019)Hohman, Kahng, Pienta, and
  Chau}]{hohman2019visual}
Fred Hohman, Minsuk Kahng, Robert Pienta, and Duen~Horng Chau. 2019.
\newblock \href {https://doi.org/10.1109/TVCG.2018.2843369} {Visual analytics
  in deep learning: An interrogative survey for the next frontiers}.
\newblock \emph{IEEE Transactions on Visualization and Computer Graphics},
  25(8):2674–2693.

\bibitem[{Horovicz and Goldshmidt(2024)}]{horovicz2024tokenshap}
Miriam Horovicz and Roni Goldshmidt. 2024.
\newblock \href {https://doi.org/10.18653/v1/2024.nlp4science-1.1}
  {{T}oken{SHAP}: Interpreting large language models with {M}onte {C}arlo
  shapley value estimation}.
\newblock In \emph{Proceedings of the 1st Workshop on NLP for Science
  (NLP4Science)}, pages 1--8, Miami, FL, USA. Association for Computational
  Linguistics.

\bibitem[{Hsieh et~al.(2024)Hsieh, Bi, Jiang, Liu, Peng, Zhang, Pan, Xu, Wang,
  Chen et~al.}]{hsieh2024comprehensive}
Weiche Hsieh, Ziqian Bi, Chuanqi Jiang, Junyu Liu, Benji Peng, Sen Zhang,
  Xuanhe Pan, Jiawei Xu, Jinlang Wang, Keyu Chen, et~al. 2024.
\newblock A comprehensive guide to explainable ai: from classical models to
  llms.
\newblock \emph{arXiv preprint arXiv:2412.00800}.

\bibitem[{Huang and Chang(2023)}]{huang2023towards}
Jie Huang and Kevin Chen-Chuan Chang. 2023.
\newblock \href {https://doi.org/10.18653/v1/2023.findings-acl.67} {Towards
  reasoning in large language models: A survey}.
\newblock In \emph{Findings of the Association for Computational Linguistics:
  ACL 2023}, pages 1049--1065, Toronto, Canada. Association for Computational
  Linguistics.

\bibitem[{Huang et~al.(2023{\natexlab{a}})Huang, Mamidanna, Jangam, Zhou, and
  Gilpin}]{huang2023can}
Shiyuan Huang, Siddarth Mamidanna, Shreedhar Jangam, Yilun Zhou, and Leilani~H
  Gilpin. 2023{\natexlab{a}}.
\newblock Can large language models explain themselves? a study of
  llm-generated self-explanations.
\newblock \emph{arXiv preprint arXiv:2310.11207}.

\bibitem[{Huang et~al.(2024)Huang, Hu, Ilhan, Tekin, and
  Liu}]{huang2024harmful}
Tiansheng Huang, Sihao Hu, Fatih Ilhan, Selim~Furkan Tekin, and Ling Liu. 2024.
\newblock Harmful fine-tuning attacks and defenses for large language models: A
  survey.
\newblock \emph{arXiv preprint arXiv:2409.18169}.

\bibitem[{Huang et~al.(2023{\natexlab{b}})Huang, Ruan, Huang, Jin, Dong, Wu,
  Bensalem, Mu, Qi, Zhao, Cai, Zhang, Wu, Xu, Wu, Freitas, and
  Mustafa}]{huang2023a}
Xiaowei Huang, Wenjie Ruan, Wei Huang, Gao Jin, Yizhen Dong, Changshun Wu,
  Saddek Bensalem, Ronghui Mu, Yi~Qi, Xingyu Zhao, Kaiwen Cai, Yanghao Zhang,
  Sihao Wu, Peipei Xu, Dengyu Wu, Andr{\'e} Freitas, and Mustafa~A. Mustafa.
  2023{\natexlab{b}}.
\newblock \href {https://api.semanticscholar.org/CorpusID:258823083} {A survey
  of safety and trustworthiness of large language models through the lens of
  verification and validation}.
\newblock \emph{Artif. Intell. Rev.}, 57:175.

\bibitem[{Huang et~al.(2023{\natexlab{c}})Huang, Xu, Lai, Jiang, Chen, Li, Yao,
  Ma, Yang, Chen et~al.}]{huang2023advancing}
Yunpeng Huang, Jingwei Xu, Junyu Lai, Zixu Jiang, Taolue Chen, Zenan Li, Yuan
  Yao, Xiaoxing Ma, Lijuan Yang, Hao Chen, et~al. 2023{\natexlab{c}}.
\newblock Advancing transformer architecture in long-context large language
  models: A comprehensive survey.
\newblock \emph{arXiv preprint arXiv:2311.12351}.

\bibitem[{Huben et~al.(2024)Huben, Cunningham, Smith, Ewart, and
  Sharkey}]{huben2024sparse}
Robert Huben, Hoagy Cunningham, Logan~Riggs Smith, Aidan Ewart, and Lee
  Sharkey. 2024.
\newblock \href {https://openreview.net/forum?id=F76bwRSLeK} {Sparse
  autoencoders find highly interpretable features in language models}.
\newblock In \emph{The Twelfth International Conference on Learning
  Representations}.

\bibitem[{Härle et~al.(2024)Härle, Friedrich, Brack, Deiseroth, Schramowski,
  and Kersting}]{haerle2024scar}
Ruben Härle, Felix Friedrich, Manuel Brack, Björn Deiseroth, Patrick
  Schramowski, and Kristian Kersting. 2024.
\newblock \href {https://arxiv.org/abs/2411.07122} {Scar: Sparse conditioned
  autoencoders for concept detection and steering in llms}.
\newblock \emph{Preprint}, arXiv:2411.07122.

\bibitem[{Ilyas et~al.(2022)Ilyas, Park, Engstrom, Leclerc, and
  Madry}]{ilyas2022datamodels}
Andrew Ilyas, Sung~Min Park, Logan Engstrom, Guillaume Leclerc, and Aleksander
  Madry. 2022.
\newblock Datamodels: Predicting predictions from training data.
\newblock In \emph{ArXiv preprint arXiv:2202.00622}.

\bibitem[{Inaba et~al.(2025)Inaba, Inui, Miyao, Oseki, Heinzerling, and
  Takagi}]{inaba2025llms}
Tatsuro Inaba, Kentaro Inui, Yusuke Miyao, Yohei Oseki, Benjamin Heinzerling,
  and Yu~Takagi. 2025.
\newblock How llms learn: Tracing internal representations with sparse
  autoencoders.
\newblock \emph{arXiv preprint arXiv:2503.06394}.

\bibitem[{Jacovi and Goldberg(2020)}]{jacovi2020towards}
Alon Jacovi and Yoav Goldberg. 2020.
\newblock \href {https://doi.org/10.18653/v1/2020.acl-main.386} {Towards
  faithfully interpretable {NLP} systems: How should we define and evaluate
  faithfulness?}
\newblock In \emph{Proceedings of the 58th Annual Meeting of the Association
  for Computational Linguistics}, pages 4198--4205, Online. Association for
  Computational Linguistics.

\bibitem[{Jacovi et~al.(2021)Jacovi, Swayamdipta, Ravfogel, Elazar, Choi, and
  Goldberg}]{jacovi2021contrastive}
Alon Jacovi, Swabha Swayamdipta, Shauli Ravfogel, Yanai Elazar, Yejin Choi, and
  Yoav Goldberg. 2021.
\newblock \href {https://doi.org/10.18653/v1/2021.emnlp-main.120} {Contrastive
  explanations for model interpretability}.
\newblock In \emph{Proceedings of the 2021 Conference on Empirical Methods in
  Natural Language Processing}, pages 1597--1611, Online and Punta Cana,
  Dominican Republic. Association for Computational Linguistics.

\bibitem[{Ji et~al.(2024{\natexlab{a}})Ji, Liu, Du, and Ng}]{ji2024chain}
Bin Ji, Huijun Liu, Mingzhe Du, and See-Kiong Ng. 2024{\natexlab{a}}.
\newblock Chain-of-thought improves text generation with citations in large
  language models.
\newblock In \emph{Proceedings of the AAAI Conference on Artificial
  Intelligence}, volume~38, pages 18345--18353.

\bibitem[{Ji et~al.(2024{\natexlab{b}})Ji, Chen, Ishii, Cahyawijaya, Bang,
  Wilie, and Fung}]{ji2024llm}
Ziwei Ji, Delong Chen, Etsuko Ishii, Samuel Cahyawijaya, Yejin Bang, Bryan
  Wilie, and Pascale Fung. 2024{\natexlab{b}}.
\newblock \href {https://doi.org/10.18653/v1/2024.blackboxnlp-1.6} {{LLM}
  internal states reveal hallucination risk faced with a query}.
\newblock In \emph{Proceedings of the 7th BlackboxNLP Workshop: Analyzing and
  Interpreting Neural Networks for NLP}, pages 88--104, Miami, Florida, US.
  Association for Computational Linguistics.

\bibitem[{Jia et~al.(2019)Jia, Dao, Wang, Hubis, Hynes, G{\"u}rel, Li, Zhang,
  Song, and Spanos}]{jia2019towards}
Ruoxi Jia, David Dao, Boxin Wang, Frances~Ann Hubis, Nick Hynes, Nezihe~Merve
  G{\"u}rel, Bo~Li, Ce~Zhang, Dawn Song, and Costas~J Spanos. 2019.
\newblock Towards efficient data valuation based on the shapley value.
\newblock In \emph{The 22nd International Conference on Artificial Intelligence
  and Statistics}, pages 1167--1176. PMLR.

\bibitem[{Jiang et~al.(2024{\natexlab{a}})Jiang, Qi, Hong, Fu, Cheng, Meng, Yu,
  Zhou, and Zhou}]{jiang2024large}
Che Jiang, Biqing Qi, Xiangyu Hong, Dayuan Fu, Yang Cheng, Fandong Meng, Mo~Yu,
  Bowen Zhou, and Jie Zhou. 2024{\natexlab{a}}.
\newblock \href {https://doi.org/10.18653/v1/2024.naacl-long.60} {On large
  language models' hallucination with regard to known facts}.
\newblock In \emph{Proceedings of the 2024 Conference of the North American
  Chapter of the Association for Computational Linguistics: Human Language
  Technologies (Volume 1: Long Papers)}, pages 1041--1053, Mexico City, Mexico.
  Association for Computational Linguistics.

\bibitem[{Jiang et~al.(2024{\natexlab{b}})Jiang, Shi, Yu, Liu, Zhang, Li, and
  Kwok}]{jiang2024forward}
Weisen Jiang, Han Shi, Longhui Yu, Zhengying Liu, Yu~Zhang, Zhenguo Li, and
  James Kwok. 2024{\natexlab{b}}.
\newblock \href {https://doi.org/10.18653/v1/2024.findings-acl.397}
  {Forward-backward reasoning in large language models for mathematical
  verification}.
\newblock In \emph{Findings of the Association for Computational Linguistics:
  ACL 2024}, pages 6647--6661, Bangkok, Thailand. Association for Computational
  Linguistics.

\bibitem[{Jiang et~al.(2025)Jiang, Chen, Yang, Li, Wang, Wu, Li, and
  Zhang}]{jiang2025comt}
Yue Jiang, Jiawei Chen, Dingkang Yang, Mingcheng Li, Shunli Wang, Tong Wu,
  Ke~Li, and Lihua Zhang. 2025.
\newblock \href {https://doi.org/10.1109/ICASSP49660.2025.10887699} {Comt:
  Chain-of-medical-thought reduces hallucination in medical report generation}.
\newblock In \emph{ICASSP 2025 - 2025 IEEE International Conference on
  Acoustics, Speech and Signal Processing (ICASSP)}, pages 1--5.

\bibitem[{Jin et~al.(2024)Jin, Cao, Yuan, Chen, Xu, Li, Jiang, Liu, and
  Zhao}]{jin2024cutting}
Zhuoran Jin, Pengfei Cao, Hongbang Yuan, Yubo Chen, Jiexin Xu, Huaijun Li,
  Xiaojian Jiang, Kang Liu, and Jun Zhao. 2024.
\newblock \href {https://doi.org/10.18653/v1/2024.findings-acl.70} {Cutting off
  the head ends the conflict: A mechanism for interpreting and mitigating
  knowledge conflicts in language models}.
\newblock In \emph{Findings of the Association for Computational Linguistics:
  ACL 2024}, pages 1193--1215, Bangkok, Thailand. Association for Computational
  Linguistics.

\bibitem[{Ju et~al.(2024)Ju, Sun, Du, Yuan, Ren, and Liu}]{ju2024large}
Tianjie Ju, Weiwei Sun, Wei Du, Xinwei Yuan, Zhaochun Ren, and Gongshen Liu.
  2024.
\newblock How large language models encode context knowledge? a layer-wise
  probing study.
\newblock \emph{arXiv preprint arXiv:2402.16061}.

\bibitem[{Kadavath et~al.(2022)Kadavath, Conerly, Askell, Henighan, Drain,
  Perez, Schiefer, Hatfield-Dodds, DasSarma, Tran-Johnson
  et~al.}]{kadavath2022language}
Saurav Kadavath, Tom Conerly, Amanda Askell, Tom Henighan, Dawn Drain, Ethan
  Perez, Nicholas Schiefer, Zac Hatfield-Dodds, Nova DasSarma, Eli
  Tran-Johnson, et~al. 2022.
\newblock Language models (mostly) know what they know.
\newblock \emph{arXiv preprint arXiv:2207.05221}.

\bibitem[{Kaneko et~al.(2024)Kaneko, Bollegala, Okazaki, and
  Baldwin}]{kaneko2024evaluating}
Masahiro Kaneko, Danushka Bollegala, Naoaki Okazaki, and Timothy Baldwin. 2024.
\newblock Evaluating gender bias in large language models via chain-of-thought
  prompting.
\newblock \emph{arXiv preprint arXiv:2401.15585}.

\bibitem[{Katz and Belinkov(2023)}]{katz2023visit}
Shahar Katz and Yonatan Belinkov. 2023.
\newblock \href {https://openreview.net/forum?id=7O9bTjLgTQ} {{VISIT}:
  Visualizing and interpreting the semantic information flow of transformers}.
\newblock In \emph{The 2023 Conference on Empirical Methods in Natural Language
  Processing}.

\bibitem[{Katz et~al.(2024)Katz, Belinkov, Geva, and Wolf}]{katz2024backward}
Shahar Katz, Yonatan Belinkov, Mor Geva, and Lior Wolf. 2024.
\newblock \href {https://doi.org/10.18653/v1/2024.emnlp-main.142} {Backward
  lens: Projecting language model gradients into the vocabulary space}.
\newblock In \emph{Proceedings of the 2024 Conference on Empirical Methods in
  Natural Language Processing}, pages 2390--2422, Miami, Florida, USA.
  Association for Computational Linguistics.

\bibitem[{Kaur et~al.(2020)Kaur, Nori, Jenkins, Caruana, Wallach, and
  Wortman~Vaughan}]{kaur2020interpreting}
Harmanpreet Kaur, Harsha Nori, Samuel Jenkins, Rich Caruana, Hanna Wallach, and
  Jennifer Wortman~Vaughan. 2020.
\newblock \href {https://doi.org/10.1145/3313831.3376219} {Interpreting
  interpretability: Understanding data scientists' use of interpretability
  tools for machine learning}.
\newblock In \emph{Proceedings of the 2020 CHI Conference on Human Factors in
  Computing Systems}, CHI '20, page 1–14, New York, NY, USA. Association for
  Computing Machinery.

\bibitem[{Khoriaty et~al.(2025)Khoriaty, Shportko, Mercier, and
  Wood-Doughty}]{khoriaty2025don}
Matthew Khoriaty, Andrii Shportko, Gustavo Mercier, and Zach Wood-Doughty.
  2025.
\newblock Don't forget it! conditional sparse autoencoder clamping works for
  unlearning.
\newblock \emph{arXiv preprint arXiv:2503.11127}.

\bibitem[{Kissane et~al.(2024)Kissane, Krzyzanowski, Bloom, Conmy, and
  Nanda}]{kissane2024interpreting}
Connor Kissane, Robert Krzyzanowski, Joseph~Isaac Bloom, Arthur Conmy, and Neel
  Nanda. 2024.
\newblock \href {https://openreview.net/forum?id=fewUBDwjji} {Interpreting
  attention layer outputs with sparse autoencoders}.
\newblock In \emph{ICML 2024 Workshop on Mechanistic Interpretability}.

\bibitem[{Kobayashi et~al.(2020)Kobayashi, Kuribayashi, Yokoi, and
  Inui}]{kobayashi2020attention}
Goro Kobayashi, Tatsuki Kuribayashi, Sho Yokoi, and Kentaro Inui. 2020.
\newblock \href {https://doi.org/10.18653/v1/2020.emnlp-main.574} {Attention is
  not only a weight: Analyzing transformers with vector norms}.
\newblock In \emph{Proceedings of the 2020 Conference on Empirical Methods in
  Natural Language Processing (EMNLP)}, pages 7057--7075, Online. Association
  for Computational Linguistics.

\bibitem[{Kobayashi et~al.(2021)Kobayashi, Kuribayashi, Yokoi, and
  Inui}]{kobayashi2021incorporating}
Goro Kobayashi, Tatsuki Kuribayashi, Sho Yokoi, and Kentaro Inui. 2021.
\newblock \href {https://doi.org/10.18653/v1/2021.emnlp-main.373}
  {{I}ncorporating {R}esidual and {N}ormalization {L}ayers into {A}nalysis of
  {M}asked {L}anguage {M}odels}.
\newblock In \emph{Proceedings of the 2021 Conference on Empirical Methods in
  Natural Language Processing}, pages 4547--4568, Online and Punta Cana,
  Dominican Republic. Association for Computational Linguistics.

\bibitem[{Kobayashi et~al.(2024)Kobayashi, Kuribayashi, Yokoi, and
  Inui}]{kobayashi2024analyzing}
Goro Kobayashi, Tatsuki Kuribayashi, Sho Yokoi, and Kentaro Inui. 2024.
\newblock \href {https://openreview.net/forum?id=mYWsyTuiRp} {Analyzing
  feed-forward blocks in transformers through the lens of attention maps}.
\newblock In \emph{The Twelfth International Conference on Learning
  Representations}.

\bibitem[{Koh and Liang(2017)}]{koh2017understanding}
Pang~Wei Koh and Percy Liang. 2017.
\newblock Understanding black-box predictions via influence functions.
\newblock In \emph{International conference on machine learning}, pages
  1885--1894. PMLR.

\bibitem[{Kokhlikyan et~al.(2020)Kokhlikyan, Miglani, Martin, Wang, Alsallakh,
  Reynolds, Melnikov, Kliushkina, Araya, Yan et~al.}]{kokhlikyan2020captum}
Narine Kokhlikyan, Vivek Miglani, Miguel Martin, Edward Wang, Bilal Alsallakh,
  Jonathan Reynolds, Alexander Melnikov, Natalia Kliushkina, Carlos Araya, Siqi
  Yan, et~al. 2020.
\newblock Captum: A unified and generic model interpretability library for
  pytorch.
\newblock \emph{arXiv preprint arXiv:2009.07896}.

\bibitem[{Kong et~al.(2022)Kong, Shen, and Huang}]{kong2022resolving}
Shuming Kong, Yanyan Shen, and Linpeng Huang. 2022.
\newblock \href {https://openreview.net/forum?id=EskfH0bwNVn} {Resolving
  training biases via influence-based data relabeling}.
\newblock In \emph{International Conference on Learning Representations}.

\bibitem[{Kram{\'a}r et~al.(2024)Kram{\'a}r, Lieberum, Shah, and
  Nanda}]{kramar2024atp}
J{\'a}nos Kram{\'a}r, Tom Lieberum, Rohin Shah, and Neel Nanda. 2024.
\newblock Atp*: An efficient and scalable method for localizing llm behaviour
  to components.
\newblock \emph{arXiv preprint arXiv:2403.00745}.

\bibitem[{Krishna et~al.(2023)Krishna, Ma, Slack, Ghandeharioun, Singh, and
  Lakkaraju}]{krishna2023post}
Satyapriya Krishna, Jiaqi Ma, Dylan Slack, Asma Ghandeharioun, Sameer Singh,
  and Himabindu Lakkaraju. 2023.
\newblock Post hoc explanations of language models can improve language models.
\newblock In \emph{Proceedings of the 37th International Conference on Neural
  Information Processing Systems}, NIPS '23, Red Hook, NY, USA. Curran
  Associates Inc.

\bibitem[{Kwon and Mihindukulasooriya(2023)}]{kwon2023finspector}
Bum~Chul Kwon and Nandana Mihindukulasooriya. 2023.
\newblock \href {https://doi.org/10.18653/v1/2023.acl-demo.4} {Finspector: A
  human-centered visual inspection tool for exploring and comparing biases
  among foundation models}.
\newblock In \emph{Proceedings of the 61st Annual Meeting of the Association
  for Computational Linguistics (Volume 3: System Demonstrations)}, pages
  42--50, Toronto, Canada. Association for Computational Linguistics.

\bibitem[{Kwon et~al.(2024)Kwon, Wu, Wu, and Zou}]{kwon2024datainf}
Yongchan Kwon, Eric Wu, Kevin Wu, and James Zou. 2024.
\newblock \href {https://openreview.net/forum?id=9m02ib92Wz} {Datainf:
  Efficiently estimating data influence in lo{RA}-tuned {LLM}s and diffusion
  models}.
\newblock In \emph{The Twelfth International Conference on Learning
  Representations}.

\bibitem[{La~Rosa et~al.(2023)La~Rosa, Blasilli, Bourqui, Auber, Santucci,
  Capobianco, Bertini, Giot, and Angelini}]{la2023state}
Biagio La~Rosa, Graziano Blasilli, Romain Bourqui, David Auber, Giuseppe
  Santucci, Roberto Capobianco, Enrico Bertini, Romain Giot, and Marco
  Angelini. 2023.
\newblock State of the art of visual analytics for explainable deep learning.
\newblock In \emph{Computer Graphics Forum}, volume~42, pages 319--355. Wiley
  Online Library.

\bibitem[{Ladhak et~al.(2023)Ladhak, Durmus, and
  Hashimoto}]{ladhak2023contrastive}
Faisal Ladhak, Esin Durmus, and Tatsunori Hashimoto. 2023.
\newblock \href {https://doi.org/10.18653/v1/2023.acl-long.643} {Contrastive
  error attribution for finetuned language models}.
\newblock In \emph{Proceedings of the 61st Annual Meeting of the Association
  for Computational Linguistics (Volume 1: Long Papers)}, pages 11482--11498,
  Toronto, Canada. Association for Computational Linguistics.

\bibitem[{Lakkaraju et~al.(2022)Lakkaraju, Slack, Chen, Tan, and
  Singh}]{lakkaraju2022rethinking}
Himabindu Lakkaraju, Dylan Slack, Yuxin Chen, Chenhao Tan, and Sameer Singh.
  2022.
\newblock \href {https://arxiv.org/abs/2202.01875} {Rethinking explainability
  as a dialogue: A practitioner's perspective}.
\newblock \emph{Preprint}, arXiv:2202.01875.

\bibitem[{Lee et~al.(2024)Lee, Bai, Pres, Wattenberg, Kummerfeld, and
  Mihalcea}]{lee2024a}
Andrew Lee, Xiaoyan Bai, Itamar Pres, Martin Wattenberg, Jonathan~K.
  Kummerfeld, and Rada Mihalcea. 2024.
\newblock A mechanistic understanding of alignment algorithms: a case study on
  dpo and toxicity.
\newblock In \emph{Proceedings of the 41st International Conference on Machine
  Learning}, ICML'24. JMLR.org.

\bibitem[{Lee et~al.(2025{\natexlab{a}})Lee, Wang, Chakravarthy, Helbling,
  Peng, Phute, Chau, and Kahng}]{lee2025llm}
Seongmin Lee, Zijie~J Wang, Aishwarya Chakravarthy, Alec Helbling, ShengYun
  Peng, Mansi Phute, Duen Horng~Polo Chau, and Minsuk Kahng.
  2025{\natexlab{a}}.
\newblock Llm attributor: Interactive visual attribution for llm generation.
\newblock In \emph{Proceedings of the AAAI Conference on Artificial
  Intelligence}, volume~39, pages 29655--29657.

\bibitem[{Lee et~al.(2025{\natexlab{b}})Lee, Lee, Heo, and Choi}]{lee2025hudex}
Sujeong Lee, Hayoung Lee, Seongsoo Heo, and Wonik Choi. 2025{\natexlab{b}}.
\newblock Hudex: Integrating hallucination detection and explainability for
  enhancing the reliability of llm responses.
\newblock \emph{arXiv preprint arXiv:2502.08109}.

\bibitem[{Leong et~al.(2024)Leong, Cheng, Xu, Wang, Wang, and Li}]{leong2024no}
Chak~Tou Leong, Yi~Cheng, Kaishuai Xu, Jian Wang, Hanlin Wang, and Wenjie Li.
  2024.
\newblock No two devils alike: Unveiling distinct mechanisms of fine-tuning
  attacks.
\newblock \emph{arXiv preprint arXiv:2405.16229}.

\bibitem[{Levinstein and Herrmann(2024)}]{levinstein2024still}
Benjamin~A Levinstein and Daniel~A Herrmann. 2024.
\newblock Still no lie detector for language models: Probing empirical and
  conceptual roadblocks.
\newblock \emph{Philosophical Studies}, pages 1--27.

\bibitem[{Li et~al.(2025{\natexlab{a}})Li, Xing, Huo, Zhou, Xu, and
  Wang}]{li2025mixhd}
Chuang Li, Bingnan Xing, Dongdong Huo, Qihui Zhou, Zhen Xu, and Yu~Wang.
  2025{\natexlab{a}}.
\newblock \href {https://doi.org/10.1109/ICASSP49660.2025.10889328} {Mixhd: A
  method for detecting hallucinations based on the internal state and output
  probability of large language models}.
\newblock In \emph{ICASSP 2025 - 2025 IEEE International Conference on
  Acoustics, Speech and Signal Processing (ICASSP)}, pages 1--5.

\bibitem[{Li et~al.(2024{\natexlab{a}})Li, Chi, Liu, and Yang}]{li2024look}
He~Li, Haoang Chi, Mingyu Liu, and Wenjing Yang. 2024{\natexlab{a}}.
\newblock Look within, why llms hallucinate: A causal perspective.
\newblock \emph{arXiv preprint arXiv:2407.10153}.

\bibitem[{Li et~al.(2024{\natexlab{b}})Li, Pyatkin, Kleiman-Weiner, Jiang,
  Dziri, Collins, Borg, Sap, Choi, and Levine}]{li2024safetyanalyst}
Jing-Jing Li, Valentina Pyatkin, Max Kleiman-Weiner, Liwei Jiang, Nouha Dziri,
  Anne~GE Collins, Jana~Schaich Borg, Maarten Sap, Yejin Choi, and Sydney
  Levine. 2024{\natexlab{b}}.
\newblock Safetyanalyst: Interpretable, transparent, and steerable llm safety
  moderation.
\newblock \emph{arXiv preprint arXiv:2410.16665}.

\bibitem[{Li et~al.(2023{\natexlab{a}})Li, Patel, Vi{\'e}gas, Pfister, and
  Wattenberg}]{li2023inference}
Kenneth Li, Oam Patel, Fernanda Vi{\'e}gas, Hanspeter Pfister, and Martin
  Wattenberg. 2023{\natexlab{a}}.
\newblock Inference-time intervention: Eliciting truthful answers from a
  language model.
\newblock \emph{Advances in Neural Information Processing Systems},
  36:41451--41530.

\bibitem[{Li et~al.(2024{\natexlab{c}})Li, Jin, Huang, Xu, Lian, Lin, Zhang,
  and Wang}]{li2024evaluating}
Meng Li, Haoran Jin, Ruixuan Huang, Zhihao Xu, Defu Lian, Zijia Lin, Di~Zhang,
  and Xiting Wang. 2024{\natexlab{c}}.
\newblock \href {https://doi.org/10.18653/v1/2024.emnlp-main.36} {Evaluating
  readability and faithfulness of concept-based explanations}.
\newblock In \emph{Proceedings of the 2024 Conference on Empirical Methods in
  Natural Language Processing}, pages 607--625, Miami, Florida, USA.
  Association for Computational Linguistics.

\bibitem[{Li et~al.(2023{\natexlab{b}})Li, Yang, Xiao, AbuRaed, Murray, and
  Carenini}]{li2023visual}
Raymond Li, Ruixin Yang, Wen Xiao, Ahmed AbuRaed, Gabriel Murray, and Giuseppe
  Carenini. 2023{\natexlab{b}}.
\newblock \href {https://arxiv.org/abs/2311.12418} {Visual analytics for
  generative transformer models}.
\newblock \emph{Preprint}, arXiv:2311.12418.

\bibitem[{Li et~al.(2025{\natexlab{b}})Li, Yao, Zhang, and Li}]{li2025safety}
Shen Li, Liuyi Yao, Lan Zhang, and Yaliang Li. 2025{\natexlab{b}}.
\newblock \href {https://openreview.net/forum?id=kUH1yPMAn7} {Safety layers in
  aligned large language models: The key to {LLM} security}.
\newblock In \emph{The Thirteenth International Conference on Learning
  Representations}.

\bibitem[{Li et~al.(2025{\natexlab{c}})Li, Wang, Liu, Wu, Dou, Lv, Wang, Zheng,
  and Huang}]{li2025revisiting}
Tianlong Li, Zhenghua Wang, Wenhao Liu, Muling Wu, Shihan Dou, Changze Lv,
  Xiaohua Wang, Xiaoqing Zheng, and Xuanjing Huang. 2025{\natexlab{c}}.
\newblock \href {https://aclanthology.org/2025.coling-main.212/} {Revisiting
  jailbreaking for large language models: A representation engineering
  perspective}.
\newblock In \emph{Proceedings of the 31st International Conference on
  Computational Linguistics}, pages 3158--3178, Abu Dhabi, UAE. Association for
  Computational Linguistics.

\bibitem[{Li et~al.(2024{\natexlab{d}})Li, Dayanik, Tyagi, and
  Pierleoni}]{li2024hallucana}
Tianyi Li, Erenay Dayanik, Shubhi Tyagi, and Andrea Pierleoni.
  2024{\natexlab{d}}.
\newblock Hallucana: Fixing llm hallucination with a canary lookahead.
\newblock \emph{arXiv preprint arXiv:2412.07965}.

\bibitem[{Li et~al.(2024{\natexlab{e}})Li, Zhang, Lou, Wu, and
  Wang}]{li2024chain}
Xi~Li, Yusen Zhang, Renze Lou, Chen Wu, and Jiaqi Wang. 2024{\natexlab{e}}.
\newblock Chain-of-scrutiny: Detecting backdoor attacks for large language
  models.
\newblock \emph{arXiv preprint arXiv:2406.05948}.

\bibitem[{Li et~al.(2024{\natexlab{f}})Li, Li, Kosuga, Yoshida, and
  Bian}]{li2024precision}
Xuying Li, Zhuo Li, Yuji Kosuga, Yasuhiro Yoshida, and Victor Bian.
  2024{\natexlab{f}}.
\newblock Precision knowledge editing: Enhancing safety in large language
  models.
\newblock \emph{arXiv preprint arXiv:2410.03772}.

\bibitem[{Li et~al.(2024{\natexlab{g}})Li, Zhang, Wang, Shi, and
  Wang}]{li2024model}
Yuxi Li, Zhibo Zhang, Kailong Wang, Ling Shi, and Haoyu Wang.
  2024{\natexlab{g}}.
\newblock \href {https://arxiv.org/abs/2412.08201} {Model-editing-based
  jailbreak against safety-aligned large language models}.
\newblock \emph{Preprint}, arXiv:2412.08201.

\bibitem[{Li et~al.(2024{\natexlab{h}})Li, Zhao, Li, and Sun}]{li2024influence}
Zhe Li, Wei Zhao, Yige Li, and Jun Sun. 2024{\natexlab{h}}.
\newblock Do influence functions work on large language models?
\newblock \emph{arXiv preprint arXiv:2409.19998}.

\bibitem[{Liao and Wortman~Vaughan(2024)}]{liao2024ai}
Q.~Vera Liao and Jennifer Wortman~Vaughan. 2024.
\newblock {AI} {Transparency} in the {Age} of {LLMs}: A {Human}-{Centered}
  {Research} {Roadmap}.
\newblock \emph{Harvard Data Science Review}, (Special Issue 5).
\newblock Https://hdsr.mitpress.mit.edu/pub/aelql9qy.

\bibitem[{Lieberum et~al.(2024)Lieberum, Rajamanoharan, Conmy, Smith, Sonnerat,
  Varma, Kram{\'a}r, Dragan, Shah, and Nanda}]{lieberum2024gemma}
Tom Lieberum, Senthooran Rajamanoharan, Arthur Conmy, Lewis Smith, Nicolas
  Sonnerat, Vikrant Varma, J{\'a}nos Kram{\'a}r, Anca Dragan, Rohin Shah, and
  Neel Nanda. 2024.
\newblock Gemma scope: Open sparse autoencoders everywhere all at once on gemma
  2.
\newblock \emph{arXiv preprint arXiv:2408.05147}.

\bibitem[{Lin(2023)}]{lin2023neuronpedia}
Johnny Lin. 2023.
\newblock \href {https://www.neuronpedia.org} {Neuronpedia: Interactive
  reference and tooling for analyzing neural networks}.
\newblock Software available from neuronpedia.org.

\bibitem[{Lin et~al.(2024)Lin, He, Xu, Xing, Yamada, Liu, and
  Tang}]{lin2024towards}
Yuping Lin, Pengfei He, Han Xu, Yue Xing, Makoto Yamada, Hui Liu, and Jiliang
  Tang. 2024.
\newblock \href {https://doi.org/10.18653/v1/2024.emnlp-main.401} {Towards
  understanding jailbreak attacks in {LLM}s: A representation space analysis}.
\newblock In \emph{Proceedings of the 2024 Conference on Empirical Methods in
  Natural Language Processing}, pages 7067--7085, Miami, Florida, USA.
  Association for Computational Linguistics.

\bibitem[{Lindsey et~al.(2025)Lindsey, Gurnee, Ameisen, Chen, Pearce, Turner,
  Citro, Abrahams, Carter, Hosmer, Marcus, Sklar, Templeton, Bricken,
  McDougall, Cunningham, Henighan, Jermyn, Jones, Persic, Qi, Thompson,
  Zimmerman, Rivoire, Conerly, Olah, and Batson}]{lindsey2025on}
Jack Lindsey, Wes Gurnee, Emmanuel Ameisen, Brian Chen, Adam Pearce,
  Nicholas~L. Turner, Craig Citro, David Abrahams, Shan Carter, Basil Hosmer,
  Jonathan Marcus, Michael Sklar, Adly Templeton, Trenton Bricken, Callum
  McDougall, Hoagy Cunningham, Thomas Henighan, Adam Jermyn, Andy Jones, Andrew
  Persic, Zhenyi Qi, T.~Ben Thompson, Sam Zimmerman, Kelley Rivoire, Thomas
  Conerly, Chris Olah, and Joshua Batson. 2025.
\newblock On the biology of a large language model.
\newblock
  \url{https://transformer-circuits.pub/2025/attribution-graphs/biology.html}.
\newblock Transformer Circuits Thread.

\bibitem[{Lindsey et~al.(2024)Lindsey, Templeton, Marcus, Conerly, Batson, and
  Olah}]{lindsey2024sparse}
Jack Lindsey, Adly Templeton, Jonathan Marcus, Thomas Conerly, Joshua Batson,
  and Christopher Olah. 2024.
\newblock Sparse crosscoders for cross-layer features and model diffing.
\newblock \url{https://transformer-circuits.pub/2024/crosscoders/index.html}.
\newblock Transformer Circuits Thread.

\bibitem[{Liu et~al.(2023)Liu, Xu, Li, Feng, and Song}]{liu2023towards}
Fuxiao Liu, Paiheng Xu, Zongxia Li, Yue Feng, and Hyemi Song. 2023.
\newblock Towards understanding in-context learning with contrastive
  demonstrations and saliency maps.
\newblock \emph{arXiv preprint arXiv:2307.05052}.

\bibitem[{Liu et~al.(2024{\natexlab{a}})Liu, Chen, Cheng, and
  He}]{liu2024universal}
Junteng Liu, Shiqi Chen, Yu~Cheng, and Junxian He. 2024{\natexlab{a}}.
\newblock \href {https://doi.org/10.18653/v1/2024.emnlp-main.1012} {On the
  universal truthfulness hyperplane inside {LLM}s}.
\newblock In \emph{Proceedings of the 2024 Conference on Empirical Methods in
  Natural Language Processing}, pages 18199--18224, Miami, Florida, USA.
  Association for Computational Linguistics.

\bibitem[{Liu et~al.(2024{\natexlab{b}})Liu, Liu, Chen, Chen, Zan, Kan, and
  Ho}]{liu2024the}
Yan Liu, Yu~Liu, Xiaokang Chen, Pin-Yu Chen, Daoguang Zan, Min-Yen Kan, and
  Tsung-Yi Ho. 2024{\natexlab{b}}.
\newblock \href {https://openreview.net/forum?id=SQGUDc9tC8} {The devil is in
  the neurons: Interpreting and mitigating social biases in language models}.
\newblock In \emph{ICLR}.

\bibitem[{Liu et~al.(2025{\natexlab{a}})Liu, Gao, Zhai, Xia, Wu, Xue, Chen,
  Kawaguchi, Zhang, and Hooi}]{liu2025guardreasoner}
Yue Liu, Hongcheng Gao, Shengfang Zhai, Jun Xia, Tianyi Wu, Zhiwei Xue, Yulin
  Chen, Kenji Kawaguchi, Jiaheng Zhang, and Bryan Hooi. 2025{\natexlab{a}}.
\newblock Guardreasoner: Towards reasoning-based llm safeguards.
\newblock \emph{arXiv preprint arXiv:2501.18492}.

\bibitem[{Liu et~al.(2025{\natexlab{b}})Liu, Amjad, Adkathimar, Wei, and
  Tong}]{liu2025selfelicit}
Zhining Liu, Rana~Ali Amjad, Ravinarayana Adkathimar, Tianxin Wei, and Hanghang
  Tong. 2025{\natexlab{b}}.
\newblock Selfelicit: Your language model secretly knows where is the relevant
  evidence.
\newblock \emph{arXiv preprint arXiv:2502.08767}.

\bibitem[{Ludan et~al.(2023)Ludan, Lyu, Yang, Dugan, Yatskar, and
  Callison-Burch}]{ludan2023interpretable}
Josh~Magnus Ludan, Qing Lyu, Yue Yang, Liam Dugan, Mark Yatskar, and Chris
  Callison-Burch. 2023.
\newblock Interpretable-by-design text understanding with iteratively generated
  concept bottleneck.
\newblock \emph{arXiv preprint arXiv:2310.19660}.

\bibitem[{Luick(2024)}]{luick2024universal}
Niclas Luick. 2024.
\newblock Universal response and emergence of induction in llms.
\newblock \emph{arXiv preprint arXiv:2411.07071}.

\bibitem[{Lundberg and Lee(2017)}]{lundberg2017unified}
Scott~M Lundberg and Su-In Lee. 2017.
\newblock A unified approach to interpreting model predictions.
\newblock \emph{Advances in neural information processing systems}, 30.

\bibitem[{Lyu et~al.(2023)Lyu, Havaldar, Stein, Zhang, Rao, Wong, Apidianaki,
  and Callison-Burch}]{lyu2023faithful}
Qing Lyu, Shreya Havaldar, Adam Stein, Li~Zhang, Delip Rao, Eric Wong, Marianna
  Apidianaki, and Chris Callison-Burch. 2023.
\newblock \href {https://doi.org/10.18653/v1/2023.ijcnlp-main.20} {Faithful
  chain-of-thought reasoning}.
\newblock In \emph{Proceedings of the 13th International Joint Conference on
  Natural Language Processing and the 3rd Conference of the Asia-Pacific
  Chapter of the Association for Computational Linguistics (Volume 1: Long
  Papers)}, pages 305--329, Nusa Dua, Bali. Association for Computational
  Linguistics.

\bibitem[{Ma et~al.(2024)Ma, Shi, Liu, Liang, Gan, Cheng, Neiswanger, and
  Vosoughi}]{mamechanistic2024ma}
Chiyu Ma, Lin Shi, Ollie Liu, Wenhua Liang, Jiaqi Gan, Ming Cheng, Willie
  Neiswanger, and Soroush Vosoughi. 2024.
\newblock Mechanistic insights: Circuit transformations across input and
  fine-tuning landscapes.
\newblock \emph{OpenReview}.

\bibitem[{Ma et~al.(2023)Ma, Scheible, Wang, Veeramachaneni, Chowdhary, Sun,
  Koulogeorge, Wang, Yang, and Vosoughi}]{ma2023deciphering}
Weicheng Ma, Henry Scheible, Brian Wang, Goutham Veeramachaneni, Pratim
  Chowdhary, Alan Sun, Andrew Koulogeorge, Lili Wang, Diyi Yang, and Soroush
  Vosoughi. 2023.
\newblock \href {https://doi.org/10.18653/v1/2023.emnlp-main.697} {Deciphering
  stereotypes in pre-trained language models}.
\newblock In \emph{Proceedings of the 2023 Conference on Empirical Methods in
  Natural Language Processing}, pages 11328--11345, Singapore. Association for
  Computational Linguistics.

\bibitem[{Ma et~al.(2025)Ma, Gao, Wang, Wang, Wang, Sun, Ding, Xu, Chen, Zhao,
  Huang, Li, Zhang, Zheng, Bai, Wu, Qiu, Zhang, Li, Han, Li, Sun, Wang, Gu, Wu,
  Chen, Zhang, Liu, Gong, Liu, Pan, Xie, Pang, Dong, Jia, Zhang, Ma, Zhang,
  Gong, Xiao, Erfani, Baldwin, Li, Sugiyama, Tao, Bailey, and
  Jiang}]{ma2025safety}
Xingjun Ma, Yifeng Gao, Yixu Wang, Ruofan Wang, Xin Wang, Ye~Sun, Yifan Ding,
  Hengyuan Xu, Yunhao Chen, Yunhan Zhao, Hanxun Huang, Yige Li, Jiaming Zhang,
  Xiang Zheng, Yang Bai, Zuxuan Wu, Xipeng Qiu, Jingfeng Zhang, Yiming Li,
  Xudong Han, Haonan Li, Jun Sun, Cong Wang, Jindong Gu, Baoyuan Wu, Siheng
  Chen, Tianwei Zhang, Yang Liu, Mingming Gong, Tongliang Liu, Shirui Pan,
  Cihang Xie, Tianyu Pang, Yinpeng Dong, Ruoxi Jia, Yang Zhang, Shiqing Ma,
  Xiangyu Zhang, Neil Gong, Chaowei Xiao, Sarah Erfani, Tim Baldwin, Bo~Li,
  Masashi Sugiyama, Dacheng Tao, James Bailey, and Yu-Gang Jiang. 2025.
\newblock \href {https://arxiv.org/abs/2502.05206} {Safety at scale: A
  comprehensive survey of large model safety}.
\newblock \emph{Preprint}, arXiv:2502.05206.

\bibitem[{Madani et~al.(2025)Madani, Gema, Sarti, Zhao, Minervini, and
  Passerini}]{madani2025noiser}
Mohammad Reza~Ghasemi Madani, Aryo~Pradipta Gema, Gabriele Sarti, Yu~Zhao,
  Pasquale Minervini, and Andrea Passerini. 2025.
\newblock Noiser: Bounded input perturbations for attributing large language
  models.
\newblock \emph{arXiv preprint arXiv:2504.02911}.

\bibitem[{Madsen et~al.(2022)Madsen, Reddy, and Chandar}]{madsen2022post}
Andreas Madsen, Siva Reddy, and Sarath Chandar. 2022.
\newblock \href {https://doi.org/10.1145/3546577} {Post-hoc interpretability
  for neural nlp: A survey}.
\newblock \emph{ACM Comput. Surv.}, 55(8).

\bibitem[{Mairal et~al.(2008)Mairal, Ponce, Sapiro, Zisserman, and
  Bach}]{mairal2008supervised}
Julien Mairal, Jean Ponce, Guillermo Sapiro, Andrew Zisserman, and Francis
  Bach. 2008.
\newblock Supervised dictionary learning.
\newblock \emph{Advances in neural information processing systems}, 21.

\bibitem[{Makelov et~al.(2024)Makelov, Lange, and
  Nanda}]{makelov2024principled}
Aleksandar Makelov, George Lange, and Neel Nanda. 2024.
\newblock \href {https://arxiv.org/abs/2405.08366} {Towards principled
  evaluations of sparse autoencoders for interpretability and control}.
\newblock \emph{Preprint}, arXiv:2405.08366.

\bibitem[{Makhzani and Frey(2013)}]{makhzani2013k}
Alireza Makhzani and Brendan Frey. 2013.
\newblock K-sparse autoencoders.
\newblock \emph{arXiv preprint arXiv:1312.5663}.

\bibitem[{Marasovic et~al.(2022)Marasovic, Beltagy, Downey, and
  Peters}]{marasovic2022shot}
Ana Marasovic, Iz~Beltagy, Doug Downey, and Matthew Peters. 2022.
\newblock \href {https://doi.org/10.18653/v1/2022.findings-naacl.31} {Few-shot
  self-rationalization with natural language prompts}.
\newblock In \emph{Findings of the Association for Computational Linguistics:
  NAACL 2022}, pages 410--424, Seattle, United States. Association for
  Computational Linguistics.

\bibitem[{Marks et~al.(2025)Marks, Rager, Michaud, Belinkov, Bau, and
  Mueller}]{marks2025sparse}
Samuel Marks, Can Rager, Eric~J Michaud, Yonatan Belinkov, David Bau, and Aaron
  Mueller. 2025.
\newblock \href {https://openreview.net/forum?id=I4e82CIDxv} {Sparse feature
  circuits: Discovering and editing interpretable causal graphs in language
  models}.
\newblock In \emph{The Thirteenth International Conference on Learning
  Representations}.

\bibitem[{McGrath and Jonker(2024)}]{mcgrath2024what}
Amanda McGrath and Alexandra Jonker. 2024.
\newblock \href {https://www.ibm.com/think/topics/interpretability} {What is ai
  interpretability?}

\bibitem[{Meng et~al.(2022)Meng, Bau, Andonian, and
  Belinkov}]{meng2022locating}
Kevin Meng, David Bau, Alex~J Andonian, and Yonatan Belinkov. 2022.
\newblock \href {https://openreview.net/forum?id=-h6WAS6eE4} {Locating and
  editing factual associations in {GPT}}.
\newblock In \emph{Advances in Neural Information Processing Systems}.

\bibitem[{Miao et~al.(2024)Miao, Teh, and Rainforth}]{miao2024selfcheck}
Ning Miao, Yee~Whye Teh, and Tom Rainforth. 2024.
\newblock \href {https://openreview.net/forum?id=pTHfApDakA} {Selfcheck: Using
  {LLM}s to zero-shot check their own step-by-step reasoning}.
\newblock In \emph{The Twelfth International Conference on Learning
  Representations}.

\bibitem[{Mikolov et~al.(2013)Mikolov, Yih, and Zweig}]{mikolov2013linguistic}
Tom{\'a}{\v{s}} Mikolov, Wen-tau Yih, and Geoffrey Zweig. 2013.
\newblock Linguistic regularities in continuous space word representations.
\newblock In \emph{Proceedings of the 2013 conference of the north american
  chapter of the association for computational linguistics: Human language
  technologies}, pages 746--751.

\bibitem[{Mishra et~al.(2025)Mishra, Danzy, Soni, Arunkumar, Huang, Kwon, and
  Bryan}]{mishra2025promptaid}
Aditi Mishra, Bretho Danzy, Utkarsh Soni, Anjana Arunkumar, Jinbin Huang,
  Bum~Chul Kwon, and Chris Bryan. 2025.
\newblock \href {https://doi.org/10.1109/TVCG.2025.3535332} {Promptaid: Visual
  prompt exploration, perturbation, testing and iteration for large language
  models}.
\newblock \emph{IEEE Transactions on Visualization and Computer Graphics},
  pages 1--14.

\bibitem[{Modarressi et~al.(2023)Modarressi, Fayyaz, Aghazadeh, Yaghoobzadeh,
  and Pilehvar}]{modarressi2023decompx}
Ali Modarressi, Mohsen Fayyaz, Ehsan Aghazadeh, Yadollah Yaghoobzadeh, and
  Mohammad~Taher Pilehvar. 2023.
\newblock \href {https://doi.org/10.18653/v1/2023.acl-long.149} {{D}ecomp{X}:
  Explaining transformers decisions by propagating token decomposition}.
\newblock In \emph{Proceedings of the 61st Annual Meeting of the Association
  for Computational Linguistics (Volume 1: Long Papers)}, pages 2649--2664,
  Toronto, Canada. Association for Computational Linguistics.

\bibitem[{Modarressi et~al.(2022)Modarressi, Fayyaz, Yaghoobzadeh, and
  Pilehvar}]{modarressi2022globenc}
Ali Modarressi, Mohsen Fayyaz, Yadollah Yaghoobzadeh, and Mohammad~Taher
  Pilehvar. 2022.
\newblock \href {https://doi.org/10.18653/v1/2022.naacl-main.19} {{G}lob{E}nc:
  Quantifying global token attribution by incorporating the whole encoder layer
  in transformers}.
\newblock In \emph{Proceedings of the 2022 Conference of the North American
  Chapter of the Association for Computational Linguistics: Human Language
  Technologies}, pages 258--271, Seattle, United States. Association for
  Computational Linguistics.

\bibitem[{Mohammadi(2024)}]{mohammadi2024explaining}
Behnam Mohammadi. 2024.
\newblock \href {https://ssrn.com/abstract=4759713} {Explaining large language
  models decisions using shapley values}.
\newblock \emph{SSRN}.

\bibitem[{Mohebbi et~al.(2023)Mohebbi, Zuidema, Chrupa{\l}a, and
  Alishahi}]{mohebbi2023quantifying}
Hosein Mohebbi, Willem Zuidema, Grzegorz Chrupa{\l}a, and Afra Alishahi. 2023.
\newblock \href {https://doi.org/10.18653/v1/2023.eacl-main.245} {Quantifying
  context mixing in transformers}.
\newblock In \emph{Proceedings of the 17th Conference of the European Chapter
  of the Association for Computational Linguistics}, pages 3378--3400,
  Dubrovnik, Croatia. Association for Computational Linguistics.

\bibitem[{Monea et~al.(2024)Monea, Peyrard, Josifoski, Chaudhary, Eisner,
  Kiciman, Palangi, Patra, and West}]{monea2024glitch}
Giovanni Monea, Maxime Peyrard, Martin Josifoski, Vishrav Chaudhary, Jason
  Eisner, Emre Kiciman, Hamid Palangi, Barun Patra, and Robert West. 2024.
\newblock \href {https://doi.org/10.18653/v1/2024.acl-long.369} {A glitch in
  the matrix? locating and detecting language model grounding with fakepedia}.
\newblock In \emph{Proceedings of the 62nd Annual Meeting of the Association
  for Computational Linguistics (Volume 1: Long Papers)}, pages 6828--6844,
  Bangkok, Thailand. Association for Computational Linguistics.

\bibitem[{Moore et~al.(2024)Moore, Roberts, Pham, and
  Fisher}]{moore2024reasoning}
Kyle Moore, Jesse Roberts, Thao Pham, and Douglas Fisher. 2024.
\newblock Reasoning beyond bias: A study on counterfactual prompting and chain
  of thought reasoning.
\newblock \emph{arXiv preprint arXiv:2408.08651}.

\bibitem[{Mou et~al.(2025)Mou, Luo, Zhang, and Ye}]{mou2025saro}
Yutao Mou, Yuxiao Luo, Shikun Zhang, and Wei Ye. 2025.
\newblock Saro: Enhancing llm safety through reasoning-based alignment.
\newblock \emph{arXiv preprint arXiv:2504.09420}.

\bibitem[{Mozes et~al.(2023)Mozes, Bolukbasi, Yuan, Liu, Thain, and
  Dixon}]{mozes2023gradient}
Maximilian Mozes, Tolga Bolukbasi, Ann Yuan, Frederick Liu, Nithum Thain, and
  Lucas Dixon. 2023.
\newblock \href {https://arxiv.org/abs/2302.06598} {Gradient-based automated
  iterative recovery for parameter-efficient tuning}.
\newblock \emph{Preprint}, arXiv:2302.06598.

\bibitem[{Mudide et~al.(2025)Mudide, Engels, Michaud, Tegmark, and
  de~Witt}]{mudide2025efficient}
Anish Mudide, Joshua Engels, Eric~J Michaud, Max Tegmark, and
  Christian~Schroeder de~Witt. 2025.
\newblock \href {https://openreview.net/forum?id=k2ZVAzVeMP} {Efficient
  dictionary learning with switch sparse autoencoders}.
\newblock In \emph{The Thirteenth International Conference on Learning
  Representations}.

\bibitem[{Muhamed et~al.(2025)Muhamed, Diab, and Smith}]{muhamed2025decoding}
Aashiq Muhamed, Mona~T. Diab, and Virginia Smith. 2025.
\newblock \href {https://aclanthology.org/2025.findings-naacl.87/} {Decoding
  dark matter: Specialized sparse autoencoders for interpreting rare concepts
  in foundation models}.
\newblock In \emph{Findings of the Association for Computational Linguistics:
  NAACL 2025}, pages 1604--1635, Albuquerque, New Mexico. Association for
  Computational Linguistics.

\bibitem[{Nanda(2022)}]{nanda2022neuroscope}
Neel Nanda. 2022.
\newblock Neuroscope: A website for mechanistic interpretability of language
  models.
\newblock \url{https://neuroscope.io}.

\bibitem[{Nanda(2024)}]{nanda2024attribution}
Neel Nanda. 2024.
\newblock Attribution patching: Activation patching at industrial scale.
\newblock
  \url{https://www.neelnanda.io/mechanistic-interpretability/attribution-patching}.
\newblock Neel Nanda.

\bibitem[{Nanda et~al.(2023{\natexlab{a}})Nanda, Chan, Lieberum, Smith, and
  Steinhardt}]{nanda2023progress}
Neel Nanda, Lawrence Chan, Tom Lieberum, Jess Smith, and Jacob Steinhardt.
  2023{\natexlab{a}}.
\newblock \href {https://openreview.net/forum?id=9XFSbDPmdW} {Progress measures
  for grokking via mechanistic interpretability}.
\newblock In \emph{The Eleventh International Conference on Learning
  Representations}.

\bibitem[{Nanda et~al.(2023{\natexlab{b}})Nanda, Rajamanoharan, Kramár, and
  Shah}]{nanda2023fact}
Neel Nanda, Senthooran Rajamanoharan, János Kramár, and Rohin Shah.
  2023{\natexlab{b}}.
\newblock Fact finding: Attempting to reverse-engineer factual recall on the
  neuron level (post 1).
\newblock
  \url{https://www.lesswrong.com/posts/iGuwZTHWb6DFY3sKB/fact-finding-attempting-to-reverse-engineer-factual-recall}.
\newblock LessWrong.

\bibitem[{nostalgebraist(2020)}]{nostalgebraist2020logitlens}
nostalgebraist. 2020.
\newblock Interpreting gpt: The logit lens.
\newblock
  \url{https://www.lesswrong.com/posts/AcKRB8wDpdaN6v6ru/interpreting-gpt-the-logit-lens}.
\newblock LessWrong.

\bibitem[{O'Brien et~al.(2024)O'Brien, Majercak, Fernandes, Edgar, Chen, Nori,
  Carignan, Horvitz, and Poursabzi-Sangde}]{o2024steering}
Kyle O'Brien, David Majercak, Xavier Fernandes, Richard Edgar, Jingya Chen,
  Harsha Nori, Dean Carignan, Eric Horvitz, and Forough Poursabzi-Sangde. 2024.
\newblock Steering language model refusal with sparse autoencoders.
\newblock \emph{arXiv preprint arXiv:2411.11296}.

\bibitem[{Olsson et~al.(2022)Olsson, Elhage, Nanda, Joseph, DasSarma, Henighan,
  Mann, Askell, Bai, Chen, Conerly, Drain, Ganguli, Hatfield-Dodds, Hernandez,
  Johnston, Jones, Kernion, Lovitt, Ndousse, Amodei, Brown, Clark, Kaplan,
  McCandlish, and Olah}]{olsson2022in}
Catherine Olsson, Nelson Elhage, Neel Nanda, Nicholas Joseph, Nova DasSarma,
  Tom Henighan, Ben Mann, Amanda Askell, Yuntao Bai, Anna Chen, Tom Conerly,
  Dawn Drain, Deep Ganguli, Zac Hatfield-Dodds, Danny Hernandez, Scott
  Johnston, Andy Jones, Jackson Kernion, Liane Lovitt, Kamal Ndousse, Dario
  Amodei, Tom Brown, Jack Clark, Jared Kaplan, Sam McCandlish, and Chris Olah.
  2022.
\newblock In-context learning and induction heads.
\newblock
  \url{https://transformer-circuits.pub/2022/in-context-learning-and-induction-heads/index.html}.

\bibitem[{O'Neill et~al.(2024)O'Neill, Ye, Iyer, and Wu}]{o2024disentangling}
Charles O'Neill, Christine Ye, Kartheik Iyer, and John~F Wu. 2024.
\newblock Disentangling dense embeddings with sparse autoencoders.
\newblock \emph{arXiv preprint arXiv:2408.00657}.

\bibitem[{Orgad et~al.(2024)Orgad, Toker, Gekhman, Reichart, Szpektor, Kotek,
  and Belinkov}]{orgad2024llms}
Hadas Orgad, Michael Toker, Zorik Gekhman, Roi Reichart, Idan Szpektor, Hadas
  Kotek, and Yonatan Belinkov. 2024.
\newblock Llms know more than they show: On the intrinsic representation of llm
  hallucinations.
\newblock \emph{arXiv preprint arXiv:2410.02707}.

\bibitem[{Pal et~al.(2023)Pal, Sun, Yuan, Wallace, and Bau}]{pal2023future}
Koyena Pal, Jiuding Sun, Andrew Yuan, Byron Wallace, and David Bau. 2023.
\newblock \href {https://doi.org/10.18653/v1/2023.conll-1.37} {Future lens:
  Anticipating subsequent tokens from a single hidden state}.
\newblock In \emph{Proceedings of the 27th Conference on Computational Natural
  Language Learning (CoNLL)}, pages 548--560, Singapore. Association for
  Computational Linguistics.

\bibitem[{Pan et~al.(2025{\natexlab{a}})Pan, Liu, Chen, Zhou, Yu, and
  Jia}]{pan2025hidden}
Wenbo Pan, Zhichao Liu, Qiguang Chen, Xiangyang Zhou, Haining Yu, and Xiaohua
  Jia. 2025{\natexlab{a}}.
\newblock The hidden dimensions of llm alignment: A multi-dimensional safety
  analysis.
\newblock \emph{arXiv preprint arXiv:2502.09674}.

\bibitem[{Pan et~al.(2025{\natexlab{b}})Pan, Shi, Zhao, and
  Ma}]{pan2025detecting}
Yijun Pan, Taiwei Shi, Jieyu Zhao, and Jiaqi~W Ma. 2025{\natexlab{b}}.
\newblock Detecting and filtering unsafe training data via data attribution.
\newblock \emph{arXiv preprint arXiv:2502.11411}.

\bibitem[{Park et~al.(2019)Park, Na, Jo, Shin, Yoo, Kwon, Zhao, Noh, Lee, and
  Choo}]{park2019sanvis}
Cheonbok Park, Inyoup Na, Yongjang Jo, Sungbok Shin, Jaehyo Yoo, Bum~Chul Kwon,
  Jian Zhao, Hyungjong Noh, Yeonsoo Lee, and Jaegul Choo. 2019.
\newblock \href {https://doi.org/10.1109/VISUAL.2019.8933677} {Sanvis: Visual
  analytics for understanding self-attention networks}.
\newblock In \emph{2019 IEEE Visualization Conference (VIS)}, pages 146--150.

\bibitem[{Park et~al.(2024{\natexlab{a}})Park, Okawa, Lee, Lubana, and
  Tanaka}]{park2024emergence}
Core~Francisco Park, Maya Okawa, Andrew Lee, Ekdeep~S Lubana, and Hidenori
  Tanaka. 2024{\natexlab{a}}.
\newblock Emergence of hidden capabilities: Exploring learning dynamics in
  concept space.
\newblock \emph{Advances in Neural Information Processing Systems},
  37:84698--84729.

\bibitem[{Park et~al.(2024{\natexlab{b}})Park, Choe, and Veitch}]{park2024the}
Kiho Park, Yo~Joong Choe, and Victor Veitch. 2024{\natexlab{b}}.
\newblock The linear representation hypothesis and the geometry of large
  language models.
\newblock In \emph{Proceedings of the 41st International Conference on Machine
  Learning}, ICML'24. JMLR.org.

\bibitem[{Park et~al.(2023)Park, Georgiev, Ilyas, Leclerc, and
  M\k{a}dry}]{park2023trak}
Sung~Min Park, Kristian Georgiev, Andrew Ilyas, Guillaume Leclerc, and
  Aleksander M\k{a}dry. 2023.
\newblock Trak: attributing model behavior at scale.
\newblock In \emph{Proceedings of the 40th International Conference on Machine
  Learning}, ICML'23. JMLR.org.

\bibitem[{Paulo et~al.(2024)Paulo, Mallen, Juang, and
  Belrose}]{paulo2024automatically}
Gon{\c{c}}alo Paulo, Alex Mallen, Caden Juang, and Nora Belrose. 2024.
\newblock Automatically interpreting millions of features in large language
  models.
\newblock \emph{arXiv preprint arXiv:2410.13928}.

\bibitem[{Pearl(2001)}]{pearl2001direct}
Judea Pearl. 2001.
\newblock Direct and indirect effects.
\newblock In \emph{Proceedings of the Seventeenth conference on Uncertainty in
  artificial intelligence}, pages 411--420.

\bibitem[{Peng et~al.(2024)Peng, Chen, Hull, and Chau}]{peng2024navigating}
Sheng~Y Peng, Pin-Yu Chen, Matthew Hull, and Duen~H Chau. 2024.
\newblock Navigating the safety landscape: Measuring risks in finetuning large
  language models.
\newblock \emph{Advances in Neural Information Processing Systems},
  37:95692--95715.

\bibitem[{Prahallad and Mamidi(2024)}]{prahallad2024significance}
Lavanya Prahallad and Radhika Mamidi. 2024.
\newblock Significance of chain of thought in gender bias mitigation for
  english-dravidian machine translation.
\newblock \emph{arXiv preprint arXiv:2405.19701}.

\bibitem[{Prakash et~al.(2024{\natexlab{a}})Prakash, Shaham, Haklay, Belinkov,
  and Bau}]{prakash2024finetuning}
Nikhil Prakash, Tamar~Rott Shaham, Tal Haklay, Yonatan Belinkov, and David Bau.
  2024{\natexlab{a}}.
\newblock \href {https://openreview.net/forum?id=8sKcAWOf2D} {Fine-tuning
  enhances existing mechanisms: A case study on entity tracking}.
\newblock In \emph{ICLR}.

\bibitem[{Prakash et~al.(2024{\natexlab{b}})Prakash, Shaham, Haklay, Belinkov,
  and Bau}]{prakash2023fine}
Nikhil Prakash, Tamar~Rott Shaham, Tal Haklay, Yonatan Belinkov, and David Bau.
  2024{\natexlab{b}}.
\newblock Fine-tuning enhances existing mechanisms: A case study on entity
  tracking.
\newblock In \emph{Proceedings of the 2024 International Conference on Learning
  Representations}.
\newblock ArXiv:2402.14811.

\bibitem[{Pruthi et~al.(2020)Pruthi, Liu, Kale, and
  Sundararajan}]{pruthi2020estimating}
Garima Pruthi, Frederick Liu, Satyen Kale, and Mukund Sundararajan. 2020.
\newblock Estimating training data influence by tracing gradient descent.
\newblock \emph{Advances in Neural Information Processing Systems},
  33:19920--19930.

\bibitem[{Qi et~al.(2024)Qi, Sarti, Fern{\'a}ndez, and Bisazza}]{qi2024model}
Jirui Qi, Gabriele Sarti, Raquel Fern{\'a}ndez, and Arianna Bisazza. 2024.
\newblock \href {https://doi.org/10.18653/v1/2024.emnlp-main.347} {Model
  internals-based answer attribution for trustworthy retrieval-augmented
  generation}.
\newblock In \emph{Proceedings of the 2024 Conference on Empirical Methods in
  Natural Language Processing}, pages 6037--6053, Miami, Florida, USA.
  Association for Computational Linguistics.

\bibitem[{Qian et~al.(2024)Qian, Zhang, Yao, Liu, Yin, Qiao, Liu, and
  Shao}]{qian2024towards}
Chen Qian, Jie Zhang, Wei Yao, Dongrui Liu, Zhenfei Yin, Yu~Qiao, Yong Liu, and
  Jing Shao. 2024.
\newblock \href {https://doi.org/10.18653/v1/2024.findings-acl.290} {Towards
  tracing trustworthiness dynamics: Revisiting pre-training period of large
  language models}.
\newblock In \emph{Findings of the Association for Computational Linguistics:
  ACL 2024}, pages 4864--4888, Bangkok, Thailand. Association for Computational
  Linguistics.

\bibitem[{Qu et~al.(2022)Qu, Cao, Gao, Ding, and Xu}]{qu2022interpretable}
Hanhao Qu, Yu~Cao, Jun Gao, Liang Ding, and Ruifeng Xu. 2022.
\newblock \href {https://doi.org/10.18653/v1/2022.naacl-main.216}
  {Interpretable proof generation via iterative backward reasoning}.
\newblock In \emph{Proceedings of the 2022 Conference of the North American
  Chapter of the Association for Computational Linguistics: Human Language
  Technologies}, pages 2968--2981, Seattle, United States. Association for
  Computational Linguistics.

\bibitem[{Rad et~al.(2025)Rad, Nghiem, Luo, Wadhwa, Sorower, and
  Rawls}]{rad2025refining}
Melissa~Kazemi Rad, Huy Nghiem, Andy Luo, Sahil Wadhwa, Mohammad Sorower, and
  Stephen Rawls. 2025.
\newblock Refining input guardrails: Enhancing llm-as-a-judge efficiency
  through chain-of-thought fine-tuning and alignment.
\newblock \emph{arXiv preprint arXiv:2501.13080}.

\bibitem[{Rajamanoharan et~al.(2024{\natexlab{a}})Rajamanoharan, Conmy, Smith,
  Lieberum, Varma, Kramar, Shah, and Nanda}]{rajamanoharan2024improving}
Senthooran Rajamanoharan, Arthur Conmy, Lewis Smith, Tom Lieberum, Vikrant
  Varma, Janos Kramar, Rohin Shah, and Neel Nanda. 2024{\natexlab{a}}.
\newblock Improving sparse decomposition of language model activations with
  gated sparse autoencoders.
\newblock In \emph{The Thirty-eighth Annual Conference on Neural Information
  Processing Systems}.

\bibitem[{Rajamanoharan et~al.(2024{\natexlab{b}})Rajamanoharan, Lieberum,
  Sonnerat, Conmy, Varma, Kram{\'a}r, and Nanda}]{rajamanoharan2024jumping}
Senthooran Rajamanoharan, Tom Lieberum, Nicolas Sonnerat, Arthur Conmy, Vikrant
  Varma, J{\'a}nos Kram{\'a}r, and Neel Nanda. 2024{\natexlab{b}}.
\newblock Jumping ahead: Improving reconstruction fidelity with jumprelu sparse
  autoencoders.
\newblock \emph{arXiv preprint arXiv:2407.14435}.

\bibitem[{Rajani et~al.(2019)Rajani, McCann, Xiong, and
  Socher}]{rajani2019explain}
Nazneen~Fatema Rajani, Bryan McCann, Caiming Xiong, and Richard Socher. 2019.
\newblock \href {https://doi.org/10.18653/v1/P19-1487} {Explain yourself!
  leveraging language models for commonsense reasoning}.
\newblock In \emph{Proceedings of the 57th Annual Meeting of the Association
  for Computational Linguistics}, pages 4932--4942, Florence, Italy.
  Association for Computational Linguistics.

\bibitem[{R{\"a}uker et~al.(2023)R{\"a}uker, Ho, Casper, and
  Hadfield-Menell}]{rauker2023toward}
Tilman R{\"a}uker, Anson Ho, Stephen Casper, and Dylan Hadfield-Menell. 2023.
\newblock Toward transparent ai: A survey on interpreting the inner structures
  of deep neural networks.
\newblock In \emph{2023 ieee conference on secure and trustworthy machine
  learning (satml)}, pages 464--483. IEEE.

\bibitem[{Ren et~al.(2020)Ren, Yeh, and Schwing}]{ren2020not}
Zhongzheng Ren, Raymond Yeh, and Alexander Schwing. 2020.
\newblock Not all unlabeled data are equal: Learning to weight data in
  semi-supervised learning.
\newblock \emph{Advances in Neural Information Processing Systems},
  33:21786--21797.

\bibitem[{Rezaei~Jafari et~al.(2024)Rezaei~Jafari, Montavon, M{\"u}ller, and
  Eberle}]{rezaei2024mambalrp}
Farnoush Rezaei~Jafari, Gr{\'e}goire Montavon, Klaus-Robert M{\"u}ller, and
  Oliver Eberle. 2024.
\newblock Mambalrp: Explaining selective state space sequence models.
\newblock \emph{Advances in Neural Information Processing Systems},
  37:118540--118570.

\bibitem[{Ribeiro et~al.(2016)Ribeiro, Singh, and Guestrin}]{ribeiro2016should}
Marco~Tulio Ribeiro, Sameer Singh, and Carlos Guestrin. 2016.
\newblock " why should i trust you?" explaining the predictions of any
  classifier.
\newblock In \emph{Proceedings of the 22nd ACM SIGKDD international conference
  on knowledge discovery and data mining}, pages 1135--1144.

\bibitem[{Rimsky et~al.(2024)Rimsky, Gabrieli, Schulz, Tong, Hubinger, and
  Turner}]{rimsky2024steering}
Nina Rimsky, Nick Gabrieli, Julian Schulz, Meg Tong, Evan Hubinger, and
  Alexander Turner. 2024.
\newblock \href {https://doi.org/10.18653/v1/2024.acl-long.828} {Steering llama
  2 via contrastive activation addition}.
\newblock In \emph{Proceedings of the 62nd Annual Meeting of the Association
  for Computational Linguistics (Volume 1: Long Papers)}, pages 15504--15522,
  Bangkok, Thailand. Association for Computational Linguistics.

\bibitem[{Sadr et~al.(2025)Sadr, Madhusudan, and Emami}]{sadr2025think}
Nikta~Gohari Sadr, Sangmitra Madhusudan, and Ali Emami. 2025.
\newblock Think or step-by-step? unzipping the black box in zero-shot prompts.
\newblock \emph{arXiv preprint arXiv:2502.03418}.

\bibitem[{Sadybekov and Katritch(2023)}]{sadybekov2023computational}
Anastasiia~V Sadybekov and Vsevolod Katritch. 2023.
\newblock Computational approaches streamlining drug discovery.
\newblock \emph{Nature}, 616(7958):673--685.

\bibitem[{Sarker(2024)}]{sarker2024llm}
Iqbal~H Sarker. 2024.
\newblock Llm potentiality and awareness: a position paper from the perspective
  of trustworthy and responsible ai modeling.
\newblock \emph{Discover Artificial Intelligence}, 4(1):40.

\bibitem[{Sarti et~al.(2024)Sarti, Chrupa{\l}a, Nissim, and
  Bisazza}]{sarti2024quantifying}
Gabriele Sarti, Grzegorz Chrupa{\l}a, Malvina Nissim, and Arianna Bisazza.
  2024.
\newblock \href {https://openreview.net/forum?id=XTHfNGI3zT} {Quantifying the
  plausibility of context reliance in neural machine translation}.
\newblock In \emph{The Twelfth International Conference on Learning
  Representations}.

\bibitem[{Sarti et~al.(2023)Sarti, Feldhus, Sickert, and van~der
  Wal}]{sarti2023inseq}
Gabriele Sarti, Nils Feldhus, Ludwig Sickert, and Oskar van~der Wal. 2023.
\newblock \href {https://doi.org/10.18653/v1/2023.acl-demo.40} {Inseq: An
  interpretability toolkit for sequence generation models}.
\newblock In \emph{Proceedings of the 61st Annual Meeting of the Association
  for Computational Linguistics (Volume 3: System Demonstrations)}, pages
  421--435, Toronto, Canada. Association for Computational Linguistics.

\bibitem[{Schioppa et~al.(2022)Schioppa, Zablotskaia, Vilar, and
  Sokolov}]{schioppa2022scaling}
Andrea Schioppa, Polina Zablotskaia, David Vilar, and Artem Sokolov. 2022.
\newblock Scaling up influence functions.
\newblock In \emph{Proceedings of the AAAI Conference on Artificial
  Intelligence}, volume~36, pages 8179--8186.

\bibitem[{Schwettmann et~al.(2023)Schwettmann, Shaham, Materzynska, Chowdhury,
  Li, Andreas, Bau, and Torralba}]{schwettmann2023find}
Sarah Schwettmann, Tamar Shaham, Joanna Materzynska, Neil Chowdhury, Shuang Li,
  Jacob Andreas, David Bau, and Antonio Torralba. 2023.
\newblock Find: A function description benchmark for evaluating
  interpretability methods.
\newblock \emph{Advances in Neural Information Processing Systems},
  36:75688--75715.

\bibitem[{Selvaraju et~al.(2017)Selvaraju, Cogswell, Das, Vedantam, Parikh, and
  Batra}]{selvaraju2017grad}
Ramprasaath~R Selvaraju, Michael Cogswell, Abhishek Das, Ramakrishna Vedantam,
  Devi Parikh, and Dhruv Batra. 2017.
\newblock Grad-cam: Visual explanations from deep networks via gradient-based
  localization.
\newblock In \emph{Proceedings of the IEEE international conference on computer
  vision}, pages 618--626.

\bibitem[{Shaikh et~al.(2023)Shaikh, Zhang, Held, Bernstein, and
  Yang}]{shaikh2023second}
Omar Shaikh, Hongxin Zhang, William Held, Michael Bernstein, and Diyi Yang.
  2023.
\newblock \href {https://doi.org/10.18653/v1/2023.acl-long.244} {On second
  thought, let`s not think step by step! bias and toxicity in zero-shot
  reasoning}.
\newblock In \emph{Proceedings of the 61st Annual Meeting of the Association
  for Computational Linguistics (Volume 1: Long Papers)}, pages 4454--4470,
  Toronto, Canada. Association for Computational Linguistics.

\bibitem[{Sharkey et~al.(2022)Sharkey, Braun, and Millidge}]{sharkey2022taking}
Lee Sharkey, Dan Braun, and Beren Millidge. 2022.
\newblock Taking features out of superposition with sparse autoencoders.
\newblock In \emph{AI Alignment Forum}, volume~8, pages 15--16.

\bibitem[{Sharkey et~al.(2025)Sharkey, Chughtai, Batson, Lindsey, Wu, Bushnaq,
  Goldowsky-Dill, Heimersheim, Ortega, Bloom, Biderman, Garriga-Alonso, Conmy,
  Nanda, Rumbelow, Wattenberg, Schoots, Miller, Michaud, Casper, Tegmark,
  Saunders, Bau, Todd, Geiger, Geva, Hoogland, Murfet, and
  McGrath}]{sharkey2025open}
Lee Sharkey, Bilal Chughtai, Joshua Batson, Jack Lindsey, Jeff Wu, Lucius
  Bushnaq, Nicholas Goldowsky-Dill, Stefan Heimersheim, Alejandro Ortega,
  Joseph Bloom, Stella Biderman, Adria Garriga-Alonso, Arthur Conmy, Neel
  Nanda, Jessica Rumbelow, Martin Wattenberg, Nandi Schoots, Joseph Miller,
  Eric~J. Michaud, Stephen Casper, Max Tegmark, William Saunders, David Bau,
  Eric Todd, Atticus Geiger, Mor Geva, Jesse Hoogland, Daniel Murfet, and Tom
  McGrath. 2025.
\newblock \href {https://arxiv.org/abs/2501.16496} {Open problems in
  mechanistic interpretability}.
\newblock \emph{Preprint}, arXiv:2501.16496.

\bibitem[{Shen et~al.(2024)Shen, Zhao, Dong, He, and Zeng}]{shen2024jailbreak}
Guobin Shen, Dongcheng Zhao, Yiting Dong, Xiang He, and Yi~Zeng. 2024.
\newblock Jailbreak antidote: Runtime safety-utility balance via sparse
  representation adjustment in large language models.
\newblock \emph{arXiv preprint arXiv:2410.02298}.

\bibitem[{Shi et~al.(2024{\natexlab{a}})Shi, Beltran~Velez, Nazaret, Zheng,
  Garriga-Alonso, Jesson, Makar, and Blei}]{shi2024hypothesis}
Claudia Shi, Nicolas Beltran~Velez, Achille Nazaret, Carolina Zheng, Adri{\`a}
  Garriga-Alonso, Andrew Jesson, Maggie Makar, and David Blei.
  2024{\natexlab{a}}.
\newblock Hypothesis testing the circuit hypothesis in llms.
\newblock \emph{Advances in Neural Information Processing Systems},
  37:94539--94567.

\bibitem[{Shi et~al.(2024{\natexlab{b}})Shi, Shen, Huang, Li, Leng, Jin, Liu,
  Wu, Guo, Yu et~al.}]{shi2024large}
Dan Shi, Tianhao Shen, Yufei Huang, Zhigen Li, Yongqi Leng, Renren Jin, Chuang
  Liu, Xinwei Wu, Zishan Guo, Linhao Yu, et~al. 2024{\natexlab{b}}.
\newblock Large language model safety: A holistic survey.
\newblock \emph{arXiv preprint arXiv:2412.17686}.

\bibitem[{Shi et~al.(2025)Shi, Li, Liang, Wan, Ma, Wang, and He}]{shi2025route}
Wei Shi, Sihang Li, Tao Liang, Mingyang Wan, Guojun Ma, Xiang Wang, and
  Xiangnan He. 2025.
\newblock Route sparse autoencoder to interpret large language models.
\newblock \emph{arXiv preprint arXiv:2503.08200}.

\bibitem[{Shrikumar et~al.(2017)Shrikumar, Greenside, and
  Kundaje}]{shrikumar2017learning}
Avanti Shrikumar, Peyton Greenside, and Anshul Kundaje. 2017.
\newblock Learning important features through propagating activation
  differences.
\newblock In \emph{Proceedings of the 34th International Conference on Machine
  Learning - Volume 70}, ICML'17, page 3145–3153. JMLR.org.

\bibitem[{Sicilia and Alikhani(2024)}]{sicilia2024eliciting}
Anthony Sicilia and Malihe Alikhani. 2024.
\newblock \href {https://doi.org/10.18653/v1/2024.nlp4pi-1.19} {Eliciting
  uncertainty in chain-of-thought to mitigate bias against forecasting harmful
  user behaviors}.
\newblock In \emph{Proceedings of the Third Workshop on NLP for Positive
  Impact}, pages 211--223, Miami, Florida, USA. Association for Computational
  Linguistics.

\bibitem[{Simonyan et~al.(2014)Simonyan, Vedaldi, and
  Zisserman}]{simonyan2013deep}
Karen Simonyan, Andrea Vedaldi, and Andrew Zisserman. 2014.
\newblock \href {http://arxiv.org/abs/1312.6034} {Deep inside convolutional
  networks: Visualising image classification models and saliency maps}.
\newblock In \emph{2nd International Conference on Learning Representations,
  {ICLR} 2014, Banff, AB, Canada, April 14-16, 2014, Workshop Track
  Proceedings}.

\bibitem[{Singh et~al.(2024{\natexlab{a}})Singh, Inala, Galley, Caruana, and
  Gao}]{singh2024rethinking}
Chandan Singh, Jeevana~Priya Inala, Michel Galley, Rich Caruana, and Jianfeng
  Gao. 2024{\natexlab{a}}.
\newblock Rethinking interpretability in the era of large language models.
\newblock \emph{arXiv preprint arXiv:2402.01761}.

\bibitem[{Singh et~al.(2024{\natexlab{b}})Singh, Ravfogel, Herzig, Aharoni,
  Cotterell, and Kumaraguru}]{singh2024representation}
Shashwat Singh, Shauli Ravfogel, Jonathan Herzig, Roee Aharoni, Ryan Cotterell,
  and Ponnurangam Kumaraguru. 2024{\natexlab{b}}.
\newblock Representation surgery: theory and practice of affine steering.
\newblock In \emph{Proceedings of the 41st International Conference on Machine
  Learning}, ICML'24. JMLR.org.

\bibitem[{Singhal et~al.(2023)Singhal, Azizi, Tu, Mahdavi, Wei, Chung, Scales,
  Tanwani, Cole-Lewis, Pfohl et~al.}]{singhal2023large}
Karan Singhal, Shekoofeh Azizi, Tao Tu, S~Sara Mahdavi, Jason Wei, Hyung~Won
  Chung, Nathan Scales, Ajay Tanwani, Heather Cole-Lewis, Stephen Pfohl, et~al.
  2023.
\newblock Large language models encode clinical knowledge.
\newblock \emph{Nature}, 620(7972):172--180.

\bibitem[{Siska and Sankaran(2025)}]{siska2025attentiondefense}
Charlotte Siska and Anush Sankaran. 2025.
\newblock Attentiondefense: Leveraging system prompt attention for explainable
  defense against novel jailbreaks.
\newblock \emph{arXiv preprint arXiv:2504.12321}.

\bibitem[{Slack et~al.(2023)Slack, Krishna, Lakkaraju, and
  Singh}]{slack2023explaining}
Dylan Slack, Satyapriya Krishna, Himabindu Lakkaraju, and Sameer Singh. 2023.
\newblock Explaining machine learning models with interactive natural language
  conversations using talktomodel.
\newblock \emph{Nature Machine Intelligence}, 5(8):873--883.

\bibitem[{Slobodkin et~al.(2023)Slobodkin, Goldman, Caciularu, Dagan, and
  Ravfogel}]{slobodkin2023curious}
Aviv Slobodkin, Omer Goldman, Avi Caciularu, Ido Dagan, and Shauli Ravfogel.
  2023.
\newblock \href {https://doi.org/10.18653/v1/2023.emnlp-main.220} {The curious
  case of hallucinatory (un)answerability: Finding truths in the hidden states
  of over-confident large language models}.
\newblock In \emph{Proceedings of the 2023 Conference on Empirical Methods in
  Natural Language Processing}, pages 3607--3625, Singapore. Association for
  Computational Linguistics.

\bibitem[{Song et~al.(2024)Song, Cui, Luo, Lecue, and Li}]{song2024better}
Linxin Song, Yan Cui, Ao~Luo, Freddy Lecue, and Irene Li. 2024.
\newblock \href {https://aclanthology.org/2024.findings-eacl.138/} {Better
  explain transformers by illuminating important information}.
\newblock In \emph{Findings of the Association for Computational Linguistics:
  EACL 2024}, pages 2048--2062, St. Julian{'}s, Malta. Association for
  Computational Linguistics.

\bibitem[{Soo et~al.(2025)Soo, Teng, Balaganesh, Guoxian, and
  YAN}]{soo2025interpretable}
Samuel Soo, Wesley Teng, Chandrasekaran Balaganesh, Tan Guoxian, and Ming YAN.
  2025.
\newblock \href {https://openreview.net/forum?id=swRxS7s4rB} {Interpretable
  steering of large language models with feature guided activation additions}.
\newblock In \emph{ICLR 2025 Workshop on Building Trust in Language Models and
  Applications}.

\bibitem[{Stolfo et~al.(2023)Stolfo, Belinkov, and
  Sachan}]{stolfo2023mechanistic}
Alessandro Stolfo, Yonatan Belinkov, and Mrinmaya Sachan. 2023.
\newblock \href {https://doi.org/10.18653/v1/2023.emnlp-main.435} {A
  mechanistic interpretation of arithmetic reasoning in language models using
  causal mediation analysis}.
\newblock In \emph{Proceedings of the 2023 Conference on Empirical Methods in
  Natural Language Processing}, pages 7035--7052, Singapore. Association for
  Computational Linguistics.

\bibitem[{Strobelt et~al.(2019)Strobelt, Gehrmann, Behrisch, Perer, Pfister,
  and Rush}]{strobelt2019seq2seqvis}
Hendrik Strobelt, Sebastian Gehrmann, Michael Behrisch, Adam Perer, Hanspeter
  Pfister, and Alexander~M. Rush. 2019.
\newblock \href {https://doi.org/10.1109/TVCG.2018.2865044} {Seq2seq-vis: A
  visual debugging tool for sequence-to-sequence models}.
\newblock \emph{IEEE Transactions on Visualization and Computer Graphics},
  25(1):353--363.

\bibitem[{Su(2024)}]{su2024enhancing}
Jingbo Su. 2024.
\newblock Enhancing adversarial attacks through chain of thought.
\newblock \emph{arXiv preprint arXiv:2410.21791}.

\bibitem[{Su et~al.(2024{\natexlab{a}})Su, Wang, Ai, Hu, Wu, Zhou, and
  Liu}]{su2024unsupervised}
Weihang Su, Changyue Wang, Qingyao Ai, Yiran Hu, Zhijing Wu, Yujia Zhou, and
  Yiqun Liu. 2024{\natexlab{a}}.
\newblock Unsupervised real-time hallucination detection based on the internal
  states of large language models.
\newblock \emph{arXiv preprint arXiv:2403.06448}.

\bibitem[{Su et~al.(2024{\natexlab{b}})Su, Li, and Lease}]{su2024wrapper}
Yiheng Su, Junyi~Jessy Li, and Matthew Lease. 2024{\natexlab{b}}.
\newblock \href {https://doi.org/10.18653/v1/2024.blackboxnlp-1.33} {Wrapper
  boxes for faithful attribution of model predictions to training data}.
\newblock In \emph{Proceedings of the 7th BlackboxNLP Workshop: Analyzing and
  Interpreting Neural Networks for NLP}, pages 551--576, Miami, Florida, US.
  Association for Computational Linguistics.

\bibitem[{Sun et~al.(2025)Sun, Oikarinen, Ustun, and Weng}]{sun2025concept}
Chung-En Sun, Tuomas Oikarinen, Berk Ustun, and Tsui-Wei Weng. 2025.
\newblock \href {https://openreview.net/forum?id=RC5FPYVQaH} {Concept
  bottleneck large language models}.
\newblock In \emph{The Thirteenth International Conference on Learning
  Representations}.

\bibitem[{Sundararajan et~al.(2017)Sundararajan, Taly, and
  Yan}]{sundararajan2017axiomatic}
Mukund Sundararajan, Ankur Taly, and Qiqi Yan. 2017.
\newblock Axiomatic attribution for deep networks.
\newblock In \emph{International conference on machine learning}, pages
  3319--3328. PMLR.

\bibitem[{Syed et~al.(2024)Syed, Rager, and Conmy}]{syed2024attribution}
Aaquib Syed, Can Rager, and Arthur Conmy. 2024.
\newblock \href {https://doi.org/10.18653/v1/2024.blackboxnlp-1.25}
  {Attribution patching outperforms automated circuit discovery}.
\newblock In \emph{Proceedings of the 7th BlackboxNLP Workshop: Analyzing and
  Interpreting Neural Networks for NLP}, pages 407--416, Miami, Florida, US.
  Association for Computational Linguistics.

\bibitem[{Tafjord et~al.(2022)Tafjord, Dalvi~Mishra, and
  Clark}]{tafjord2022entailer}
Oyvind Tafjord, Bhavana Dalvi~Mishra, and Peter Clark. 2022.
\newblock \href {https://doi.org/10.18653/v1/2022.emnlp-main.134} {Entailer:
  Answering questions with faithful and truthful chains of reasoning}.
\newblock In \emph{Proceedings of the 2022 Conference on Empirical Methods in
  Natural Language Processing}, pages 2078--2093, Abu Dhabi, United Arab
  Emirates. Association for Computational Linguistics.

\bibitem[{Tan et~al.(2025)Tan, Luan, Luo, Sun, Chen, and Dai}]{tan2025revprag}
Xue Tan, Hao Luan, Mingyu Luo, Xiaoyan Sun, Ping Chen, and Jun Dai. 2025.
\newblock \href {https://arxiv.org/abs/2411.18948} {Revprag: Revealing
  poisoning attacks in retrieval-augmented generation through llm activation
  analysis}.
\newblock \emph{arXiv preprint arXiv:2411.18948}.

\bibitem[{Tan et~al.(2024)Tan, Cheng, Wang, Yuan, Li, and
  Liu}]{tan2024interpreting}
Zhen Tan, Lu~Cheng, Song Wang, Bo~Yuan, Jundong Li, and Huan Liu. 2024.
\newblock Interpreting pretrained language models via concept bottlenecks.
\newblock In \emph{Pacific-Asia Conference on Knowledge Discovery and Data
  Mining}, pages 56--74. Springer.

\bibitem[{Tang and Li(2025)}]{tang2025investigation}
Qiyao Tang and Xiangyang Li. 2025.
\newblock An investigation of large language models and their vulnerabilities
  in spam detection.
\newblock \emph{arXiv preprint arXiv:2504.09776}.

\bibitem[{Templeton et~al.(2024)Templeton, Conerly, Marcus, Lindsey, Bricken,
  Chen, Pearce, Citro, Ameisen, Jones, Cunningham, Turner, McDougall,
  MacDiarmid, Tamkin, Durmus, Hume, Mosconi, Freeman, Sumers, Rees, Batson,
  Jermyn, Carter, Olah, and Henighan}]{templeton2024scaling}
Adly Templeton, Tom Conerly, Jonathan Marcus, Jack Lindsey, Trenton Bricken,
  Brian Chen, Adam Pearce, Craig Citro, Emmanuel Ameisen, Andy Jones, Hoagy
  Cunningham, Nicholas~L Turner, Callum McDougall, Monte MacDiarmid, Alex
  Tamkin, Esin Durmus, Tristan Hume, Francesco Mosconi, C.~Daniel Freeman,
  Theodore~R. Sumers, Edward Rees, Joshua Batson, Adam Jermyn, Shan Carter,
  Chris Olah, and Tom Henighan. 2024.
\newblock Scaling monosemanticity: Extracting interpretable features from
  claude 3 sonnet.
\newblock
  \url{https://transformer-circuits.pub/2024/scaling-monosemanticity/index.html}.
\newblock Transformer Circuits Thread.

\bibitem[{Tenney et~al.(2020)Tenney, Wexler, Bastings, Bolukbasi, Coenen,
  Gehrmann, Jiang, Pushkarna, Radebaugh, Reif, and Yuan}]{tenney2020language}
Ian Tenney, James Wexler, Jasmijn Bastings, Tolga Bolukbasi, Andy Coenen,
  Sebastian Gehrmann, Ellen Jiang, Mahima Pushkarna, Carey Radebaugh, Emily
  Reif, and Ann Yuan. 2020.
\newblock \href {https://doi.org/10.18653/v1/2020.emnlp-demos.15} {The language
  interpretability tool: Extensible, interactive visualizations and analysis
  for {NLP} models}.
\newblock In \emph{Proceedings of the 2020 Conference on Empirical Methods in
  Natural Language Processing: System Demonstrations}, pages 107--118, Online.
  Association for Computational Linguistics.

\bibitem[{Tenney et~al.(2019)Tenney, Xia, Chen, Wang, Poliak, McCoy, Kim,
  Van~Durme, Bowman, Das et~al.}]{tenney2019you}
Ian Tenney, Patrick Xia, Berlin Chen, Alex Wang, Adam Poliak, R~Thomas McCoy,
  Najoung Kim, Benjamin Van~Durme, Samuel~R Bowman, Dipanjan Das, et~al. 2019.
\newblock What do you learn from context? probing for sentence structure in
  contextualized word representations.
\newblock \emph{arXiv preprint arXiv:1905.06316}.

\bibitem[{Theodorus et~al.(2025)Theodorus, Swaytha, Gautam, Ward, Shah,
  Blondin, and Zhu}]{theodorus2025finding}
Justin Theodorus, V~Swaytha, Shivani Gautam, Adam Ward, Mahir Shah, Cole
  Blondin, and Kevin Zhu. 2025.
\newblock \href {https://openreview.net/forum?id=oCprwPRqwW} {Finding sparse
  autoencoder representations of errors in cot prompting}.
\newblock In \emph{ICLR 2025 Workshop on Building Trust in Language Models and
  Applications}.

\bibitem[{Tsai et~al.(2023)Tsai, Yeh, and Ravikumar}]{tsai2023sample}
Che-Ping Tsai, Chih-Kuan Yeh, and Pradeep Ravikumar. 2023.
\newblock Sample based explanations via generalized representers.
\newblock In \emph{Proceedings of the 37th International Conference on Neural
  Information Processing Systems}, NIPS '23, Red Hook, NY, USA. Curran
  Associates Inc.

\bibitem[{Tu et~al.(2021)Tu, Xu, and Shen}]{tu2021keywordmap}
Yamei Tu, Jiayi Xu, and Han-Wei Shen. 2021.
\newblock Keywordmap: Attention-based visual exploration for keyword analysis.
\newblock In \emph{2021 IEEE 14th Pacific Visualization Symposium
  (PacificVis)}, pages 206--215. IEEE.

\bibitem[{Turner et~al.(2023)Turner, Thiergart, Leech, Udell, Vazquez, Mini,
  and MacDiarmid}]{turner2023steering}
Alexander~Matt Turner, Lisa Thiergart, Gavin Leech, David Udell, Juan~J
  Vazquez, Ulisse Mini, and Monte MacDiarmid. 2023.
\newblock Steering language models with activation engineering.
\newblock \emph{arXiv preprint arXiv:2308.10248}.

\bibitem[{Turpin et~al.(2023)Turpin, Michael, Perez, and
  Bowman}]{turpin2023language}
Miles Turpin, Julian Michael, Ethan Perez, and Samuel~R. Bowman. 2023.
\newblock \href {https://openreview.net/forum?id=bzs4uPLXvi} {Language models
  don't always say what they think: Unfaithful explanations in chain-of-thought
  prompting}.
\newblock In \emph{Thirty-seventh Conference on Neural Information Processing
  Systems}.

\bibitem[{Vaswani et~al.(2017)Vaswani, Shazeer, Parmar, Uszkoreit, Jones,
  Gomez, Kaiser, and Polosukhin}]{vaswani2017attention}
Ashish Vaswani, Noam Shazeer, Niki Parmar, Jakob Uszkoreit, Llion Jones,
  Aidan~N Gomez, {\L}ukasz Kaiser, and Illia Polosukhin. 2017.
\newblock Attention is all you need.
\newblock \emph{Advances in neural information processing systems}, 30.

\bibitem[{Veeramachaneni(2025)}]{veeramachaneni2025large}
Vinod Veeramachaneni. 2025.
\newblock Large language models: A comprehensive survey on architectures,
  applications, and challenges.
\newblock \emph{Advanced Innovations in Computer Programming Languages},
  7(1):20--39.

\bibitem[{Vig(2019)}]{vig2019multiscale}
Jesse Vig. 2019.
\newblock \href {https://doi.org/10.18653/v1/P19-3007} {A multiscale
  visualization of attention in the transformer model}.
\newblock In \emph{Proceedings of the 57th Annual Meeting of the Association
  for Computational Linguistics: System Demonstrations}, pages 37--42,
  Florence, Italy. Association for Computational Linguistics.

\bibitem[{Vig et~al.(2020)Vig, Gehrmann, Belinkov, Qian, Nevo, Singer, and
  Shieber}]{vig2020investigating}
Jesse Vig, Sebastian Gehrmann, Yonatan Belinkov, Sharon Qian, Daniel Nevo,
  Yaron Singer, and Stuart Shieber. 2020.
\newblock Investigating gender bias in language models using causal mediation
  analysis.
\newblock \emph{Advances in neural information processing systems},
  33:12388--12401.

\bibitem[{Wang et~al.(2024{\natexlab{a}})Wang, Yue, Lu, Shi, Zhao, Wang, Song,
  and Huang}]{wang2024model}
Huanqian Wang, Yang Yue, Rui Lu, Jingxin Shi, Andrew Zhao, Shenzhi Wang, Shiji
  Song, and Gao Huang. 2024{\natexlab{a}}.
\newblock Model surgery: Modulating llm's behavior via simple parameter
  editing.
\newblock \emph{arXiv preprint arXiv:2407.08770}.

\bibitem[{Wang et~al.(2025{\natexlab{a}})Wang, Mittal, Song, and
  Jia}]{wang2025data}
Jiachen~T. Wang, Prateek Mittal, Dawn Song, and Ruoxi Jia. 2025{\natexlab{a}}.
\newblock \href {https://openreview.net/forum?id=HD6bWcj87Y} {Data shapley in
  one training run}.
\newblock In \emph{The Thirteenth International Conference on Learning
  Representations}.

\bibitem[{Wang et~al.(2024{\natexlab{b}})Wang, Lin, Qiao, Foo, and
  Low}]{wang2024helpful}
Jingtan Wang, Xiaoqiang Lin, Rui Qiao, Chuan-Sheng Foo, and Bryan Kian~Hsiang
  Low. 2024{\natexlab{b}}.
\newblock \href {https://openreview.net/forum?id=WSpPC1Jm0p} {Helpful or
  harmful data? fine-tuning-free shapley attribution for explaining language
  model predictions}.
\newblock In \emph{ICML}.

\bibitem[{Wang et~al.(2023)Wang, Variengien, Conmy, Shlegeris, and
  Steinhardt}]{wang2023interpretability}
Kevin~Ro Wang, Alexandre Variengien, Arthur Conmy, Buck Shlegeris, and Jacob
  Steinhardt. 2023.
\newblock \href {https://openreview.net/forum?id=NpsVSN6o4ul} {Interpretability
  in the wild: a circuit for indirect object identification in {GPT}-2 small}.
\newblock In \emph{The Eleventh International Conference on Learning
  Representations}.

\bibitem[{Wang et~al.(2024{\natexlab{c}})Wang, Zhang, Xu, Xi, Deng, Yao, Zhang,
  Yang, Wang, and Chen}]{wang2024detoxifying}
Mengru Wang, Ningyu Zhang, Ziwen Xu, Zekun Xi, Shumin Deng, Yunzhi Yao, Qishen
  Zhang, Linyi Yang, Jindong Wang, and Huajun Chen. 2024{\natexlab{c}}.
\newblock \href {https://doi.org/10.18653/v1/2024.acl-long.171} {Detoxifying
  large language models via knowledge editing}.
\newblock In \emph{Proceedings of the 62nd Annual Meeting of the Association
  for Computational Linguistics (Volume 1: Long Papers)}, pages 3093--3118,
  Bangkok, Thailand. Association for Computational Linguistics.

\bibitem[{Wang et~al.(2024{\natexlab{d}})Wang, Anikina, Feldhus, Genabith,
  Hennig, and M{\"o}ller}]{wang2024llmcheckup}
Qianli Wang, Tatiana Anikina, Nils Feldhus, Josef Genabith, Leonhard Hennig,
  and Sebastian M{\"o}ller. 2024{\natexlab{d}}.
\newblock \href {https://doi.org/10.18653/v1/2024.hcinlp-1.9} {{LLMC}heckup:
  Conversational examination of large language models via interpretability
  tools and self-explanations}.
\newblock In \emph{Proceedings of the Third Workshop on Bridging
  Human--Computer Interaction and Natural Language Processing}, pages 89--104,
  Mexico City, Mexico. Association for Computational Linguistics.

\bibitem[{Wang et~al.(2024{\natexlab{e}})Wang, Zhao, Ouyang, Liu, Wang, and
  Shen}]{wang2024interactive}
Sheng Wang, Zihao Zhao, Xi~Ouyang, Tianming Liu, Qian Wang, and Dinggang Shen.
  2024{\natexlab{e}}.
\newblock Interactive computer-aided diagnosis on medical image using large
  language models.
\newblock \emph{Communications Engineering}, 3(1):133.

\bibitem[{Wang et~al.(2025{\natexlab{b}})Wang, Weng, Yang, Qin, Huang, and
  Wang}]{wang2025delman}
Yi~Wang, Fenghua Weng, Sibei Yang, Zhan Qin, Minlie Huang, and Wenjie Wang.
  2025{\natexlab{b}}.
\newblock Delman: Dynamic defense against large language model jailbreaking
  with model editing.
\newblock \emph{arXiv preprint arXiv:2502.11647}.

\bibitem[{Wang et~al.(2024{\natexlab{f}})Wang, Zhang, Guo, and
  Shen}]{wang2024gradient}
Yongjie Wang, Tong Zhang, Xu~Guo, and Zhiqi Shen. 2024{\natexlab{f}}.
\newblock Gradient based feature attribution in explainable ai: A technical
  review.
\newblock \emph{arXiv preprint arXiv:2403.10415}.

\bibitem[{Wang et~al.(2025{\natexlab{c}})Wang, Sun, Fu, and
  Liang}]{wang2025human}
Yunchao Wang, Guodao Sun, Zihang Fu, and Ronghua Liang. 2025{\natexlab{c}}.
\newblock \href {https://arxiv.org/abs/2410.08723} {Human-computer interaction
  and visualization in natural language generation models: Applications,
  challenges, and opportunities}.
\newblock \emph{Preprint}, arXiv:2410.08723.

\bibitem[{Wang et~al.(2021)Wang, Turko, and Chau}]{wang2021dodrio}
Zijie~J. Wang, Robert Turko, and Duen~Horng Chau. 2021.
\newblock \href {https://doi.org/10.18653/v1/2021.acl-demo.16} {Dodrio:
  Exploring transformer models with interactive visualization}.
\newblock In \emph{Proceedings of the 59th Annual Meeting of the Association
  for Computational Linguistics and the 11th International Joint Conference on
  Natural Language Processing: System Demonstrations}, pages 132--141, Online.
  Association for Computational Linguistics.

\bibitem[{Wei et~al.(2024)Wei, Huang, Huang, Xie, Qi, Xia, Mittal, Wang, and
  Henderson}]{wei2024assessing}
Boyi Wei, Kaixuan Huang, Yangsibo Huang, Tinghao Xie, Xiangyu Qi, Mengzhou Xia,
  Prateek Mittal, Mengdi Wang, and Peter Henderson. 2024.
\newblock Assessing the brittleness of safety alignment via pruning and
  low-rank modifications.
\newblock In \emph{Proceedings of the 41st International Conference on Machine
  Learning}, ICML'24. JMLR.org.

\bibitem[{Wei et~al.(2022)Wei, Wang, Schuurmans, Bosma, Ichter, Xia, Chi, Le,
  and Zhou}]{wei2022chain}
Jason Wei, Xuezhi Wang, Dale Schuurmans, Maarten Bosma, Brian Ichter, Fei Xia,
  Ed~H. Chi, Quoc~V. Le, and Denny Zhou. 2022.
\newblock Chain-of-thought prompting elicits reasoning in large language
  models.
\newblock In \emph{Proceedings of the 36th International Conference on Neural
  Information Processing Systems}, NIPS '22, Red Hook, NY, USA. Curran
  Associates Inc.

\bibitem[{Weng et~al.(2023)Weng, Zhu, Xia, Li, He, Liu, Sun, Liu, and
  Zhao}]{weng2023large}
Yixuan Weng, Minjun Zhu, Fei Xia, Bin Li, Shizhu He, Shengping Liu, Bin Sun,
  Kang Liu, and Jun Zhao. 2023.
\newblock \href {https://doi.org/10.18653/v1/2023.findings-emnlp.167} {Large
  language models are better reasoners with self-verification}.
\newblock In \emph{Findings of the Association for Computational Linguistics:
  EMNLP 2023}, pages 2550--2575, Singapore. Association for Computational
  Linguistics.

\bibitem[{Wiegreffe and Pinter(2019)}]{wiegreffe2019attention}
Sarah Wiegreffe and Yuval Pinter. 2019.
\newblock \href {https://doi.org/10.18653/v1/D19-1002} {Attention is not not
  explanation}.
\newblock In \emph{Proceedings of the 2019 Conference on Empirical Methods in
  Natural Language Processing and the 9th International Joint Conference on
  Natural Language Processing (EMNLP-IJCNLP)}, pages 11--20, Hong Kong, China.
  Association for Computational Linguistics.

\bibitem[{Winninger et~al.(2025)Winninger, Addad, and
  Kapusta}]{winninger2025using}
Thomas Winninger, Boussad Addad, and Katarzyna Kapusta. 2025.
\newblock \href {https://arxiv.org/abs/2503.06269} {Using mechanistic
  interpretability to craft adversarial attacks against large language models}.
\newblock \emph{Preprint}, arXiv:2503.06269.

\bibitem[{Wu et~al.(2024{\natexlab{a}})Wu, Pang, Shen, and
  Cheng}]{wu2024enhancing}
Kangxi Wu, Liang Pang, Huawei Shen, and Xueqi Cheng. 2024{\natexlab{a}}.
\newblock \href {https://doi.org/10.18653/v1/2024.emnlp-main.782} {Enhancing
  training data attribution for large language models with fitting error
  consideration}.
\newblock In \emph{Proceedings of the 2024 Conference on Empirical Methods in
  Natural Language Processing}, pages 14131--14143, Miami, Florida, USA.
  Association for Computational Linguistics.

\bibitem[{Wu et~al.(2025)Wu, Yuan, Yao, Zhai, and Liu}]{wu2025interpreting}
Xuansheng Wu, Jiayi Yuan, Wenlin Yao, Xiaoming Zhai, and Ninghao Liu. 2025.
\newblock Interpreting and steering llms with mutual information-based
  explanations on sparse autoencoders.
\newblock \emph{arXiv preprint arXiv:2502.15576}.

\bibitem[{Wu et~al.(2024{\natexlab{b}})Wu, Zhao, Zhu, Shi, Yang, Liu, Zhai,
  Yao, Li, Du, and Liu}]{wu2024usable}
Xuansheng Wu, Haiyan Zhao, Yaochen Zhu, Yucheng Shi, Fan Yang, Tianming Liu,
  Xiaoming Zhai, Wenlin Yao, Jundong Li, Mengnan Du, and Ninghao Liu.
  2024{\natexlab{b}}.
\newblock \href {https://arxiv.org/abs/2403.08946} {Usable xai: 10 strategies
  towards exploiting explainability in the llm era}.
\newblock \emph{Preprint}, arXiv:2403.08946.

\bibitem[{Wu et~al.(2022)Wu, Shu, and Low}]{wu2022davinz}
Zhaoxuan Wu, Yao Shu, and Bryan Kian~Hsiang Low. 2022.
\newblock Davinz: Data valuation using deep neural networks at initialization.
\newblock In \emph{International Conference on Machine Learning}, pages
  24150--24176. PMLR.

\bibitem[{Xia et~al.(2024)Xia, Malladi, Gururangan, Arora, and
  Chen}]{xia2024less}
Mengzhou Xia, Sadhika Malladi, Suchin Gururangan, Sanjeev Arora, and Danqi
  Chen. 2024.
\newblock {LESS}: Selecting influential data for targeted instruction tuning.
\newblock In \emph{International Conference on Machine Learning (ICML)}.

\bibitem[{Xiong et~al.(2024)Xiong, Qi, Chen, and Ho}]{xiong2024defensive}
Chen Xiong, Xiangyu Qi, Pin-Yu Chen, and Tsung-Yi Ho. 2024.
\newblock Defensive prompt patch: A robust and interpretable defense of llms
  against jailbreak attacks.
\newblock \emph{arXiv preprint arXiv:2405.20099}.

\bibitem[{Xu et~al.(2024{\natexlab{a}})Xu, Wu, Diao, Liu, Wang, Chen, and
  Gao}]{xu2024sayself}
Tianyang Xu, Shujin Wu, Shizhe Diao, Xiaoze Liu, Xingyao Wang, Yangyi Chen, and
  Jing Gao. 2024{\natexlab{a}}.
\newblock \href {https://doi.org/10.18653/v1/2024.emnlp-main.343} {{S}ay{S}elf:
  Teaching {LLM}s to express confidence with self-reflective rationales}.
\newblock In \emph{Proceedings of the 2024 Conference on Empirical Methods in
  Natural Language Processing}, pages 5985--5998, Miami, Florida, USA.
  Association for Computational Linguistics.

\bibitem[{Xu et~al.(2024{\natexlab{b}})Xu, Wang, and Wang}]{xu2024tracking}
Yang Xu, Yi~Wang, and Hao Wang. 2024{\natexlab{b}}.
\newblock Tracking the feature dynamics in llm training: A mechanistic study.
\newblock \emph{arXiv preprint arXiv:2412.17626}.

\bibitem[{Xu et~al.(2024{\natexlab{c}})Xu, HUANG, Chen, and
  Wang}]{xu2024uncovering}
Zhihao Xu, Ruixuan HUANG, Changyu Chen, and Xiting Wang. 2024{\natexlab{c}}.
\newblock \href {https://openreview.net/forum?id=Uymv9ThB50} {Uncovering safety
  risks of large language models through concept activation vector}.
\newblock In \emph{The Thirty-eighth Annual Conference on Neural Information
  Processing Systems}.

\bibitem[{Yang et~al.(2024{\natexlab{a}})Yang, Chen, Sun, Li, Feng, and
  Peng}]{yang2024enhancing}
Jingyuan Yang, Dapeng Chen, Yajing Sun, Rongjun Li, Zhiyong Feng, and Wei Peng.
  2024{\natexlab{a}}.
\newblock \href {https://doi.org/10.18653/v1/2024.findings-acl.199} {Enhancing
  semantic consistency of large language models through model editing: An
  interpretability-oriented approach}.
\newblock In \emph{Findings of the Association for Computational Linguistics:
  ACL 2024}, pages 3343--3353, Bangkok, Thailand. Association for Computational
  Linguistics.

\bibitem[{Yang et~al.(2023{\natexlab{a}})Yang, Huang, Zou, Zhang, Dai, and
  Chen}]{yang2023local}
Sen Yang, Shujian Huang, Wei Zou, Jianbing Zhang, Xinyu Dai, and Jiajun Chen.
  2023{\natexlab{a}}.
\newblock \href {https://doi.org/10.18653/v1/2023.acl-long.572} {Local
  interpretation of transformer based on linear decomposition}.
\newblock In \emph{Proceedings of the 61st Annual Meeting of the Association
  for Computational Linguistics (Volume 1: Long Papers)}, pages 10270--10287,
  Toronto, Canada. Association for Computational Linguistics.

\bibitem[{Yang et~al.(2023{\natexlab{b}})Yang, Duan, Abbasi, Lalor, and
  Tam}]{yang2023bias}
Yi~Yang, Hanyu Duan, Ahmed Abbasi, John~P Lalor, and Kar~Yan Tam.
  2023{\natexlab{b}}.
\newblock Bias a-head? analyzing bias in transformer-based language model
  attention heads.
\newblock \emph{arXiv preprint arXiv:2311.10395}.

\bibitem[{Yang et~al.(2024{\natexlab{b}})Yang, Sun, Yue, Devanbu, and
  Lo}]{yang2024robustness}
Zhou Yang, Zhensu Sun, Terry~Zhuo Yue, Premkumar Devanbu, and David Lo.
  2024{\natexlab{b}}.
\newblock Robustness, security, privacy, explainability, efficiency, and
  usability of large language models for code.
\newblock \emph{arXiv preprint arXiv:2403.07506}.

\bibitem[{Yao et~al.(2023)Yao, Yu, Zhao, Shafran, Griffiths, Cao, and
  Narasimhan}]{yao2023tree}
Shunyu Yao, Dian Yu, Jeffrey Zhao, Izhak Shafran, Thomas~L. Griffiths, Yuan
  Cao, and Karthik Narasimhan. 2023.
\newblock Tree of thoughts: deliberate problem solving with large language
  models.
\newblock In \emph{Proceedings of the 37th International Conference on Neural
  Information Processing Systems}, NIPS '23, Red Hook, NY, USA. Curran
  Associates Inc.

\bibitem[{Ye and Durrett(2022)}]{ye2022the}
Xi~Ye and Greg Durrett. 2022.
\newblock The unreliability of explanations in few-shot prompting for textual
  reasoning.
\newblock In \emph{Proceedings of the 36th International Conference on Neural
  Information Processing Systems}, NIPS '22, Red Hook, NY, USA. Curran
  Associates Inc.

\bibitem[{Yeh et~al.(2024)Yeh, Chen, Wu, Chen, Viégas, and
  Wattenberg}]{yeh2024attentionviz}
Catherine Yeh, Yida Chen, Aoyu Wu, Cynthia Chen, Fernanda Viégas, and Martin
  Wattenberg. 2024.
\newblock \href {https://doi.org/10.1109/TVCG.2023.3327163} {Attentionviz: A
  global view of transformer attention}.
\newblock \emph{IEEE Transactions on Visualization and Computer Graphics},
  30(1):262--272.

\bibitem[{Yeh et~al.(2018)Yeh, Kim, Yen, and Ravikumar}]{yeh2018representer}
Chih-Kuan Yeh, Joon Kim, Ian En-Hsu Yen, and Pradeep~K Ravikumar. 2018.
\newblock \href
  {https://proceedings.neurips.cc/paper_files/paper/2018/file/8a7129b8f3edd95b7d969dfc2c8e9d9d-Paper.pdf}
  {Representer point selection for explaining deep neural networks}.
\newblock In \emph{Advances in Neural Information Processing Systems},
  volume~31. Curran Associates, Inc.

\bibitem[{Yeh et~al.(2022)Yeh, Taly, Sundararajan, Liu, and
  Ravikumar}]{yeh2022first}
Chih-Kuan Yeh, Ankur Taly, Mukund Sundararajan, Frederick Liu, and Pradeep
  Ravikumar. 2022.
\newblock First is better than last for language data influence.
\newblock \emph{Advances in Neural Information Processing Systems},
  35:32285--32298.

\bibitem[{Yin and Neubig(2022)}]{yin2022interpreting}
Kayo Yin and Graham Neubig. 2022.
\newblock \href {https://doi.org/10.18653/v1/2022.emnlp-main.14} {Interpreting
  language models with contrastive explanations}.
\newblock In \emph{Proceedings of the 2022 Conference on Empirical Methods in
  Natural Language Processing}, pages 184--198, Abu Dhabi, United Arab
  Emirates. Association for Computational Linguistics.

\bibitem[{Yona et~al.(2023)Yona, Honovich, Laish, and
  Aharoni}]{yona2023surfacing}
Gal Yona, Or~Honovich, Itay Laish, and Roee Aharoni. 2023.
\newblock Surfacing biases in large language models using contrastive input
  decoding.
\newblock \emph{arXiv preprint arXiv:2305.07378}.

\bibitem[{Yu et~al.(2022)Yu, Quartey, and Schilder}]{yu2022legal}
Fangyi Yu, Lee Quartey, and Frank Schilder. 2022.
\newblock Legal prompting: Teaching a language model to think like a lawyer.
\newblock \emph{arXiv preprint arXiv:2212.01326}.

\bibitem[{Yu et~al.(2024{\natexlab{a}})Yu, Zhang, Tiwari, and
  Wang}]{yu2024natural}
Fei Yu, Hongbo Zhang, Prayag Tiwari, and Benyou Wang. 2024{\natexlab{a}}.
\newblock \href {https://doi.org/10.1145/3664194} {Natural language reasoning,
  a survey}.
\newblock \emph{ACM Comput. Surv.}, 56(12).

\bibitem[{Yu et~al.(2024{\natexlab{b}})Yu, Cao, Cheung, and
  Dong}]{yu2024mechanistic}
Lei Yu, Meng Cao, Jackie~CK Cheung, and Yue Dong. 2024{\natexlab{b}}.
\newblock \href {https://doi.org/10.18653/v1/2024.findings-emnlp.466}
  {Mechanistic understanding and mitigation of language model non-factual
  hallucinations}.
\newblock In \emph{Findings of the Association for Computational Linguistics:
  EMNLP 2024}, pages 7943--7956, Miami, Florida, USA. Association for
  Computational Linguistics.

\bibitem[{Yu et~al.(2023)Yu, Merullo, and Pavlick}]{yu2023characterizing}
Qinan Yu, Jack Merullo, and Ellie Pavlick. 2023.
\newblock \href {https://doi.org/10.18653/v1/2023.emnlp-main.615}
  {Characterizing mechanisms for factual recall in language models}.
\newblock In \emph{Proceedings of the 2023 Conference on Empirical Methods in
  Natural Language Processing}, pages 9924--9959, Singapore. Association for
  Computational Linguistics.

\bibitem[{Yuksekgonul et~al.(2024)Yuksekgonul, Chandrasekaran, Jones,
  Gunasekar, Naik, Palangi, Kamar, and Nushi}]{yuksekgonul2024attention}
Mert Yuksekgonul, Varun Chandrasekaran, Erik Jones, Suriya Gunasekar, Ranjita
  Naik, Hamid Palangi, Ece Kamar, and Besmira Nushi. 2024.
\newblock \href {https://openreview.net/forum?id=gfFVATffPd} {Attention
  satisfies: A constraint-satisfaction lens on factual errors of language
  models}.
\newblock In \emph{The Twelfth International Conference on Learning
  Representations}.

\bibitem[{Zhang et~al.(2024{\natexlab{a}})Zhang, Yu, Yi, Zhang, Li, and
  Liu}]{zhang2024prompt}
Fujie Zhang, Peiqi Yu, Biao Yi, Baolei Zhang, Tong Li, and Zheli Liu.
  2024{\natexlab{a}}.
\newblock Prompt-guided internal states for hallucination detection of large
  language models.
\newblock \emph{arXiv preprint arXiv:2411.04847}.

\bibitem[{Zhang et~al.(2024{\natexlab{b}})Zhang, Singh, Liu, Liu, Yu, Gao, and
  Zhao}]{zhang2024tell}
Qingru Zhang, Chandan Singh, Liyuan Liu, Xiaodong Liu, Bin Yu, Jianfeng Gao,
  and Tuo Zhao. 2024{\natexlab{b}}.
\newblock \href {https://openreview.net/forum?id=xZDWO0oejD} {Tell your model
  where to attend: Post-hoc attention steering for {LLM}s}.
\newblock In \emph{The Twelfth International Conference on Learning
  Representations}.

\bibitem[{Zhang et~al.(2024{\natexlab{c}})Zhang, Yu, and
  Feng}]{zhang2024truthx}
Shaolei Zhang, Tian Yu, and Yang Feng. 2024{\natexlab{c}}.
\newblock \href {https://doi.org/10.18653/v1/2024.acl-long.483} {{T}ruth{X}:
  Alleviating hallucinations by editing large language models in truthful
  space}.
\newblock In \emph{Proceedings of the 62nd Annual Meeting of the Association
  for Computational Linguistics (Volume 1: Long Papers)}, pages 8908--8949,
  Bangkok, Thailand. Association for Computational Linguistics.

\bibitem[{Zhang et~al.(2025)Zhang, Li, Han, Yao, Cen, and
  Zhao}]{zhang2025safety}
Yuyou Zhang, Miao Li, William Han, Yihang Yao, Zhepeng Cen, and Ding Zhao.
  2025.
\newblock Safety is not only about refusal: Reasoning-enhanced fine-tuning for
  interpretable llm safety.
\newblock \emph{arXiv preprint arXiv:2503.05021}.

\bibitem[{Zhao et~al.(2024{\natexlab{a}})Zhao, Dou, and Huang}]{zhao2024eeg}
Chongwen Zhao, Zhihao Dou, and Kaizhu Huang. 2024{\natexlab{a}}.
\newblock Eeg-defender: Defending against jailbreak through early exit
  generation of large language models.
\newblock \emph{arXiv preprint arXiv:2408.11308}.

\bibitem[{Zhao et~al.(2024{\natexlab{b}})Zhao, Chen, Yang, Liu, Deng, Cai,
  Wang, Yin, and Du}]{zhao2024explainability}
Haiyan Zhao, Hanjie Chen, Fan Yang, Ninghao Liu, Huiqi Deng, Hengyi Cai,
  Shuaiqiang Wang, Dawei Yin, and Mengnan Du. 2024{\natexlab{b}}.
\newblock Explainability for large language models: A survey.
\newblock \emph{ACM Transactions on Intelligent Systems and Technology},
  15(2):1--38.

\bibitem[{Zhao et~al.(2024{\natexlab{c}})Zhao, Yang, Shen, Lakkaraju, and
  Du}]{zhao2024towards}
Haiyan Zhao, Fan Yang, Bo~Shen, Himabindu Lakkaraju, and Mengnan Du.
  2024{\natexlab{c}}.
\newblock Towards uncovering how large language model works: An explainability
  perspective.
\newblock \emph{arXiv preprint arXiv:2402.10688}.

\bibitem[{Zhao et~al.(2023{\natexlab{a}})Zhao, Yao, zhichao Yang, and hong
  yu}]{zhao2023selfexplain}
Jiachen Zhao, Zonghai Yao, zhichao Yang, and hong yu. 2023{\natexlab{a}}.
\newblock \href {https://openreview.net/forum?id=nN8pCTVQZD} {{SELF}-{EXPLAIN}:
  Teaching large language models to reason complex questions by themselves}.
\newblock In \emph{R0-FoMo:Robustness of Few-shot and Zero-shot Learning in
  Large Foundation Models}.

\bibitem[{Zhao et~al.(2023{\natexlab{b}})Zhao, Li, Joty, Qin, and
  Bing}]{zhao2023verify}
Ruochen Zhao, Xingxuan Li, Shafiq Joty, Chengwei Qin, and Lidong Bing.
  2023{\natexlab{b}}.
\newblock \href {https://doi.org/10.18653/v1/2023.acl-long.320}
  {Verify-and-edit: A knowledge-enhanced chain-of-thought framework}.
\newblock In \emph{Proceedings of the 61st Annual Meeting of the Association
  for Computational Linguistics (Volume 1: Long Papers)}, pages 5823--5840,
  Toronto, Canada. Association for Computational Linguistics.

\bibitem[{Zhao et~al.(2024{\natexlab{d}})Zhao, Li, Li, Zhang, and
  Sun}]{zhao2024defending}
Wei Zhao, Zhe Li, Yige Li, Ye~Zhang, and Jun Sun. 2024{\natexlab{d}}.
\newblock \href {https://doi.org/10.18653/v1/2024.findings-emnlp.293}
  {Defending large language models against jailbreak attacks via layer-specific
  editing}.
\newblock In \emph{Findings of the Association for Computational Linguistics:
  EMNLP 2024}, pages 5094--5109, Miami, Florida, USA. Association for
  Computational Linguistics.

\bibitem[{Zhao et~al.(2024{\natexlab{e}})Zhao, Hu, Li, Deng, Guo, Sui, Zhao,
  Qin, Chua, and Liu}]{zhao2024towards-comprehensive}
Weixiang Zhao, Yulin Hu, Zhuojun Li, Yang Deng, Jiahe Guo, Xingyu Sui, Yanyan
  Zhao, Bing Qin, Tat-Seng Chua, and Ting Liu. 2024{\natexlab{e}}.
\newblock \href {https://arxiv.org/abs/2405.13820} {Towards comprehensive post
  safety alignment of large language models via safety patching}.
\newblock \emph{Preprint}, arXiv:2405.13820.

\bibitem[{Zhou et~al.(2025{\natexlab{a}})Zhou, Patil, Sun, lakshmanan,
  Rajamanoharan, and Conmy}]{zhou2025llm}
Dylan Zhou, Kunal Patil, Yifan Sun, Karthik lakshmanan, Senthooran
  Rajamanoharan, and Arthur Conmy. 2025{\natexlab{a}}.
\newblock \href {https://openreview.net/forum?id=aeQeXlG2Pw} {{LLM}
  neurosurgeon: Targeted knowledge removal in {LLM}s using sparse
  autoencoders}.
\newblock In \emph{ICLR 2025 Workshop on Building Trust in Language Models and
  Applications}.

\bibitem[{Zhou and Shah(2023)}]{zhou2023solvability}
Yilun Zhou and Julie Shah. 2023.
\newblock \href {https://doi.org/10.18653/v1/2023.findings-eacl.182} {The
  solvability of interpretability evaluation metrics}.
\newblock In \emph{Findings of the Association for Computational Linguistics:
  EACL 2023}, pages 2399--2415, Dubrovnik, Croatia. Association for
  Computational Linguistics.

\bibitem[{Zhou et~al.(2024)Zhou, Yu, Zhang, Xu, Huang, and
  Li}]{zhou2024alignment}
Zhenhong Zhou, Haiyang Yu, Xinghua Zhang, Rongwu Xu, Fei Huang, and Yongbin Li.
  2024.
\newblock \href {https://doi.org/10.18653/v1/2024.findings-emnlp.139} {How
  alignment and jailbreak work: Explain {LLM} safety through intermediate
  hidden states}.
\newblock In \emph{Findings of the Association for Computational Linguistics:
  EMNLP 2024}, pages 2461--2488, Miami, Florida, USA. Association for
  Computational Linguistics.

\bibitem[{Zhou et~al.(2025{\natexlab{b}})Zhou, Yu, Zhang, Xu, Huang, Wang, Liu,
  Fang, and Li}]{zhou2025on}
Zhenhong Zhou, Haiyang Yu, Xinghua Zhang, Rongwu Xu, Fei Huang, Kun Wang, Yang
  Liu, Junfeng Fang, and Yongbin Li. 2025{\natexlab{b}}.
\newblock \href {https://openreview.net/forum?id=h0Ak8A5yqw} {On the role of
  attention heads in large language model safety}.
\newblock In \emph{The Thirteenth International Conference on Learning
  Representations}.

\bibitem[{Zhu et~al.(2024)Zhu, Yang, Wei, Zhang, and Zhang}]{zhu2024locking}
Minjun Zhu, Linyi Yang, Yifan Wei, Ningyu Zhang, and Yue Zhang. 2024.
\newblock Locking down the finetuned llms safety.
\newblock \emph{arXiv preprint arXiv:2410.10343}.

\bibitem[{Zou et~al.(2023)Zou, Phan, Chen, Campbell, Guo, Ren, Pan, Yin,
  Mazeika, Dombrowski et~al.}]{zou2023representation}
Andy Zou, Long Phan, Sarah Chen, James Campbell, Phillip Guo, Richard Ren,
  Alexander Pan, Xuwang Yin, Mantas Mazeika, Ann-Kathrin Dombrowski, et~al.
  2023.
\newblock Representation engineering: A top-down approach to ai transparency.
\newblock \emph{arXiv preprint arXiv:2310.01405}.

\end{thebibliography}
\clearpage
\appendix
\section*{Appendix}\label{sec:appendix}

\autoref{tab:overview-all-1} and \autoref{tab:overview-all-2} provide an overview of interpretation-informed safety enhancement techniques (\autoref{sec:enhance}) and tools that facilitate understanding and application of interpretation (\autoref{sec:toolkits}). This table extends \autoref{tab:overview} in the main text to include safety-oriented interpretation methods not yet leveraged for safety enhancements or tool use.
    Each row is one work; each column corresponds to a technique or tool.
    Safety issues, techniques, and tools addressed by a work are indicated by a colored cell.

\setlength{\colwidth}{0.52cm}

\begin{table*}[t]
    \centering
    \caption{
    Overview of representative works at the intersections of safety-focused interpretation (\autoref{sec:interpretation}), safety enhancements they inform (\autoref{sec:enhance}), and tools operationalizing them (\autoref{sec:toolkits}),
    extending \autoref{tab:overview}. Continued in \autoref{tab:overview-all-2}.
    }
    \sffamily
    \mosttiny %
    \arrayrulecolor{black}
    \setlength{\arrayrulewidth}{1pt}
    \renewcommand{\arraystretch}{0.7}
    \rowcolors{2}{lightgray-overview}{white}
    \setlength{\tabcolsep}{0pt}
    \begin{tabular}{
        r @{\hskip 5pt}
        C{\colwidth} C{\colwidth} C{\colwidth} C{\colwidth} 
        !{\color{black}\vrule width 0.5pt} 
        C{\colwidth} C{\colwidth} C{\colwidth} C{\colwidth} C{\colwidth} C{\colwidth} 
        !{\color{black}\vrule width 0.5pt} 
        C{\colwidth} C{\colwidth} C{\colwidth} C{\colwidth} C{\colwidth} C{\colwidth} 
        !{\color{black}\vrule width 0.5pt} 
        C{\colwidth} C{\colwidth} C{\colwidth} C{\colwidth} C{\colwidth}
        !{\color{black}\vrule width 0.5pt} 
        @{\hskip 5pt} C{\colwidth} C{\colwidth}
    }
    \multicolumn{1}{c}{} & 
    \multicolumn{4}{c@{\hskip 0pt}}{\textcolor{darkgray}{\makecell{\cellcolor{white}\textbf{SAFETY TYPE}}}} & 
    \multicolumn{6}{c}{\makecell{\cellcolor{white}\hspace{-0.5em}\hyperref[sec:interpretation]{\textcolor{matrix_blue}{\S\ref*{sec:interpretation}\hspace{0.4em}\textbf{INTERPRET FOR SAFETY}}}}} & 
    \multicolumn{6}{c}{\makecell{\cellcolor{white}\hspace{-1em}\hyperref[sec:enhance]{\textcolor{matrix_purple}{\S\ref*{sec:enhance}\hspace{0.4em}\textbf{ENHANCE SAFETY}}}}} & 
    \multicolumn{5}{c}{\makecell{\cellcolor{white}\hspace{-1.8em}\hyperref[sec:toolkits]{\textcolor{matrix_green}{\S\ref*{sec:toolkits}\hspace{0.2em}\textbf{PRACTICAL TOOLS}}}}} & 
    \multicolumn{2}{c@{\hskip 5pt}}{\textcolor{darkgray}{\textbf{VENUE}}} \\        
    \rowcolor{white}
    \multicolumn{1}{r@{\hskip 5pt}}{\textbf{Work}} & 
    \rotatebox{90}{\scalebox{0.8}[1]{Hallucination}} & 
    \rotatebox{90}{\scalebox{0.8}[1]{Jailbreak \& Harm}} & 
    \rotatebox{90}{\scalebox{0.8}[1]{Bias}} & 
    \rotatebox{90}{\scalebox{0.8}[1]{Privacy Leakage}} & 
    \rotatebox{90}{\scalebox{0.8}[1]{\roundedbox{3.1} Training Attrib.}} & 
    \rotatebox{90}{\scalebox{0.8}[1]{\roundedbox{3.2} Input Token}} & 
    \rotatebox{90}{\scalebox{0.8}[1]{\roundedbox{3.3.1} Probe Latent}} & 
    \rotatebox{90}{\scalebox{0.8}[1]{\roundedbox{3.3.2} Perturb Comp}} & 
    \rotatebox{90}{\scalebox{0.8}[1]{\roundedbox{3.3.3} Decipher Latent}} & 
    \rotatebox{90}{\scalebox{0.8}[1]{\roundedbox{3.4} Self-Reason}} & 
    \rotatebox{90}{\scalebox{0.8}[1]{\roundedbox{4.1} Attn. to Rel. Token}} & %
    \rotatebox{90}{\scalebox{0.8}[1]{\roundedbox{4.2.1} Steer Latent Vec}} & 
    \rotatebox{90}{\scalebox{0.8}[1]{\roundedbox{4.2.2} Modulate Neuron}} & %
    \rotatebox{90}{\scalebox{0.8}[1]{\roundedbox{4.2.3} Edit Model}} & 
    \rotatebox{90}{\scalebox{0.8}[1]{\roundedbox{4.3} Verify \& Output}} & 
    \rotatebox{90}{\scalebox{0.8}[1]{\roundedbox{4.4} Output w. Reason}} & 
    \rotatebox{90}{\scalebox{0.8}[1]{Ease Impl.}} & 
    \rotatebox{90}{\scalebox{0.8}[1]{\roundedbox{5.1} TDA Vis}} & 
    \rotatebox{90}{\scalebox{0.8}[1]{\roundedbox{5.2} Token Vis}} & 
    \rotatebox{90}{\scalebox{0.8}[1]{\roundedbox{5.3} Latent Vec Vis}} & 
    \rotatebox{90}{\scalebox{0.8}[1]{\roundedbox{5.4} Neuron Vis}} & 
    \multicolumn{2}{c@{\hskip 5pt}}{\rotatebox{90}{\scalebox{0.8}[1]{Publication}}} \\  
    \midrule 
    \citet{su2024wrapper}           
    &    & \g &    &    
    & \g &    &    &    &    &    
    &    &    &    &    &    &    
    &    &    &    &    &    
    & \multicolumn{2}{l}{BlackboxNLP} \\
    \citet{he2024what}          
    &    & \g &    &    
    & \g &    &    &    &    &    
    &    &    &    &    &    &    
    &    &    &    &    &    
    & \multicolumn{2}{l}{COLM} \\
    \citet{pan2025detecting}         
    &    & \g & \g &      
    & \g &    &    &    &    &    
    &    &    &    &    &    &    
    &    &    &    &    &    
    & \multicolumn{2}{l}{ArXiv} \\
    \citet{wu2024enhancing}         
    & \g &    &    &      
    & \g &    &    &    &    &    
    &    &    &    &    &    &    
    &    &    &    &    &    
    & \multicolumn{2}{l}{EMNLP} \\
    \citet{chen2024finding}
    &    & \g &    &    
    & \g &    &    &    &    &    
    &    &    &    &    &    &    
    &    &    &    &    &    
    & \multicolumn{2}{l}{ArXiv} \\
    \citet{hazra2024safety}
    &    & \g &    &    
    & \g &    &    &    &    &    
    &    &    &    & \g &    &    
    &    &    &    &    &    
    & \multicolumn{2}{l}{EMNLP} \\
    \citet{zhao2024towards-comprehensive}
    &    & \g &    &    
    & \g &    &    &    &    &    
    &    &    &    & \g &    &    
    &    &    &    &    &    
    & \multicolumn{2}{l}{ArXiv} \\
    \citet{lee2025llm}
    & \g &    &    &    
    & \g &    &    &    &    &    
    &    &    &    &    &    &    
    &    & \g &    &    &    
    & \multicolumn{2}{l}{AAAI} \\
    \citet{qian2024towards}         
    & \g & \g & \g & \g 
    & \g &    & \g &    &    &    
    &    & \g &    &    &    &    
    &    &    &    &    &    
    & \multicolumn{2}{l}{ACL} \\
    \citet{lee2024a}           
    &    & \g &    &    
    & \g &    & \g &    & \g &    
    &    &    &    &    &    &    
    &    &    &    &    &    
    & \multicolumn{2}{l}{ICML} \\
    \citet{ferrando2022towards}           
    & \g &    &    &    
    &    & \g &    &    &    &    
    &    &    &    &    &    &    
    &    &    &    &    &    
    & \multicolumn{2}{l}{EMNLP} \\
    \citet{yuksekgonul2024attention}           
    & \g &    &    &    
    &    & \g &    &    &    &    
    &    &    &    &    &    &    
    &    &    &    &    &    
    & \multicolumn{2}{l}{ICLR} \\
    \citet{mohammadi2024explaining}           
    &    &    & \g &    
    &    & \g &    &    &    &    
    &    &    &    &    &    &    
    &    &    &    &    &    
    & \multicolumn{2}{l}{SSRN} \\
    \citet{cohen2024contextcite}
    &    & \g &    &  
    &    & \g &    &    &    &    
    &    &    &    &    &    &    
    &    &    &    &    &    
    & \multicolumn{2}{l}{NeurIPS} \\
    \citet{yona2023surfacing}           
    &    &    & \g &    
    &    & \g &    &    &    &    
    &    &    &    &    &    &    
    &    &    &    &    &    
    & \multicolumn{2}{l}{ArXiv} \\
    \citet{eberle2023rather}           
    &    &    & \g &    
    &    & \g &    &    &    &    
    &    &    &    &    &    &    
    &    &    &    &    &    
    & \multicolumn{2}{l}{EMNLP} \\
    \citet{vig2019multiscale}
    &    &    & \g &    
    &    & \g &    &    &    &    
    &    &    &    &    &    &    
    &    &    & \g &    &    
    & \multicolumn{2}{l}{ACL} \\
    \citet{sarti2023inseq}
    &    &    & \g &    
    &    & \g &    &    &    &    
    &    &    &    &    &    &    
    & \g &    & \g &    &    
    & \multicolumn{2}{l}{ACL} \\
    \citet{mishra2025promptaid}
    &    &    & \g &    
    &    & \g &    &    &    &    
    &    &    &    &    &    &    
    &    &    & \g &    &    
    & \multicolumn{2}{l}{TVCG} \\
    \citet{wang2025delman}           
    &    & \g &    &    
    &    & \g &    &    &    &    
    &    &    &    & \g &    &    
    &    &    &    &    &    
    & \multicolumn{2}{l}{ArXiv} \\
    \citet{dale2023detecting}           
    & \g &    &    &    
    &    & \g &    &    &    &    
    &    &    &    &    & \g &    
    &    &    &    &    &    
    & \multicolumn{2}{l}{ACL} \\
    \citet{chuang2024lookback}           
    & \g &    &    &    
    &    & \g &    &    &    &    
    &    &    &    &    & \g &    
    &    &    &    &    &    
    & \multicolumn{2}{l}{EMNLP} \\
    \citet{pan2025hidden}           
    &    & \g &    &    
    &    & \g & \g &    &    &    
    & \g &    &    &    &    &    
    &    &    &    &    &    
    & \multicolumn{2}{l}{ArXiv} \\
    \citet{zhang2024tell}
    &    &    & \g &    
    &    & \g &    & \g &    &    
    & \g &    & \g &    &    &    
    &    &    &    &    &    
    & \multicolumn{2}{l}{ICLR} \\
    \citet{li2023visual}
    & \g &    &    &    
    &    & \g &    & \g &    &    
    &    &    &    &    &    &    
    &    &    & \g & \g &    
    & \multicolumn{2}{l}{ArXiv} \\
    \citet{tenney2020language}
    &    &    & \g &    
    &    & \g &    & \g &    &    
    &    &    &    &    &    &    
    &    &    & \g & \g &    
    & \multicolumn{2}{l}{EMNLP} \\
    \citet{feng2024unveiling}           
    &    & \g &    &    
    &    & \g &    &    & \g &    
    &    &    &    &    &    &    
    &    &    &    &    &    
    & \multicolumn{2}{l}{ICLR} \\
    \citet{halawi2024overthinking}
    & \g &    &    &    
    &    & \g &    &    & \g &    
    &    &    &    &    &    &    
    &    &    &    &    &    
    & \multicolumn{2}{l}{ICLR} \\
    \citet{liu2024universal}           
    & \g &    &    &    
    &    &    & \g &    &    &    
    &    &    &    &    &    &    
    &    &    &    &    &    
    & \multicolumn{2}{l}{EMNLP} \\
    \citet{arditi2024refusal}
    &    & \g &    &    
    &    &    & \g &    &    &    
    &    &    &    &    &    &    
    &    &    &    &    &    
    & \multicolumn{2}{l}{ArXiv} \\
    \citet{lin2024towards}
    &    & \g &    &    
    &    &    & \g &    &    &    
    &    &    &    &    &    &    
    &    &    &    &    &    
    & \multicolumn{2}{l}{EMNLP} \\
    \citet{abdelnabi2024you}
    &    & \g &    &    
    &    &    & \g &    &    &    
    &    &    &    &    &    &    
    &    &    &    &    &    
    & \multicolumn{2}{l}{ArXiv} \\
    \citet{slobodkin2023curious}
    & \g &    &    &    
    &    &    & \g &    &    &    
    &    &    &    &    &    &    
    &    &    &    &    &    
    & \multicolumn{2}{l}{EMNLP} \\
    \citet{ashok2025language}
    & \g & \g &    &    
    &    &    & \g &    &    &    
    &    &    &    &    &    &    
    &    &    &    &    &    
    & \multicolumn{2}{l}{ArXiv} \\
    \citet{orgad2024llms}
    & \g &    & \g &    
    &    &    & \g &    &    &    
    &    &    &    &    &    &    
    &    &    &    &    &    
    & \multicolumn{2}{l}{ICLR} \\
    \citet{li2025mixhd}
    & \g &    &    &    
    &    &    & \g &    &    &    
    &    &    &    &    &    &    
    &    &    &    &    &    
    & \multicolumn{2}{l}{ICASSP} \\
    \citet{zhang2024prompt}
    & \g &    &    &    
    &    &    & \g &    &    &    
    &    &    &    &    &    &    
    &    &    &    &    &    
    & \multicolumn{2}{l}{ArXiv} \\
    \citet{su2024unsupervised}
    & \g &    &    &    
    &    &    & \g &    &    &    
    &    &    &    &    &    &    
    &    &    &    &    &    
    & \multicolumn{2}{l}{ACL} \\
    \citet{azaria2023internal}
    & \g &    &    &    
    &    &    & \g &    &    &    
    &    &    &    &    &    &    
    &    &    &    &    &    
    & \multicolumn{2}{l}{EMNLP} \\
    \citet{ji2024llm}
    & \g &    &    &    
    &    &    & \g &    &    &    
    &    &    &    &    &    &    
    &    &    &    &    &    
    & \multicolumn{2}{l}{BlackboxNLP} \\
    \citet{burger2024truth}
    & \g &    &    &    
    &    &    & \g &    &    &    
    &    &    &    &    &    &    
    &    &    &    &    &    
    & \multicolumn{2}{l}{NeurIPS} \\
    \citet{zhou2024alignment}
    &    & \g &    &    
    &    &    & \g &    &    &    
    &    &    &    &    &    &    
    &    &    &    &    &    
    & \multicolumn{2}{l}{EMNLP} \\
    \citet{xu2024uncovering}
    &    & \g &    &    
    &    &    & \g &    &    &    
    &    &    &    &    &    &    
    &    &    &    &    &    
    & \multicolumn{2}{l}{NeurIPS} \\
    \citet{he2024jailbreaklens}
    &    & \g &    &    
    &    &    & \g &    &    &    
    &    &    &    &    &    &    
    &    &    &    &    &    
    & \multicolumn{2}{l}{ArXiv} \\
    \citet{tan2025revprag}
    & \g &    &    &    
    &    &    & \g &    &    &    
    &    &    &    &    &    &    
    &    &    &    &    &    
    & \multicolumn{2}{l}{ArXiv} \\
    \citet{he2024llm}
    & \g &    &    &    
    &    &    & \g &    &    &    
    &    &    &    &    &    &    
    &    &    &    &    &    
    & \multicolumn{2}{l}{ACL} \\
    \citet{beigi2024internalinspector}
    & \g &    &    &    
    &    &    & \g &    &    &    
    &    &    &    &    &    &    
    &    &    &    &    &    
    & \multicolumn{2}{l}{EMNLP} \\
    \citet{ch-wang2024androids}
    & \g &    &    &    
    &    &    & \g &    &    &    
    &    &    &    &    &    &    
    &    &    &    &    &    
    & \multicolumn{2}{l}{ACL} \\
    \citet{winninger2025using}
    &    & \g &    &    
    &    &    & \g &    &    &    
    &    &    &    &    &    &    
    &    &    &    &    &    
    & \multicolumn{2}{l}{ArXiv} \\
    \citet{arditi2024refusal}
    &    & \g &    &    
    &    &    & \g &    &    &    
    &    &    &    &    &    &    
    &    &    &    &    &    
    & \multicolumn{2}{l}{ArXiv} \\
    \citet{li2024model}
    &    & \g &    &    
    &    &    & \g &    &    &    
    &    &    &    &    &    &    
    &    &    &    &    &    
    & \multicolumn{2}{l}{ArXiv} \\
    \citet{zhu2024locking}
    &    & \g &    &    
    &    &    & \g &    &    &    
    & \g &    &    &    &    &    
    &    &    &    &    &    
    & \multicolumn{2}{l}{ArXiv} \\
    \citet{li2025revisiting}
    &    & \g &    &    
    &    &    & \g &    &    &    
    &    & \g &    &    &    &    
    &    &    &    &    &    
    & \multicolumn{2}{l}{COLING} \\
    \citet{duan2024llms}
    & \g &    &    &    
    &    &    & \g &    &    &    
    &    & \g &    &    &    &    
    &    &    &    &    &    
    & \multicolumn{2}{l}{ArXiv} \\
    \citet{yang2024enhancing}
    & \g &    &    &    
    &    &    & \g &    &    &    
    &    & \g &    &    &    &    
    &    &    &    &    &    
    & \multicolumn{2}{l}{ACL} \\
    \citet{ball2024understanding}
    &    & \g &    &    
    &    &    & \g &    &    &    
    &    & \g &    &    &    &    
    &    &    &    &    &    
    & \multicolumn{2}{l}{ArXiv} \\
    \citet{wang2024model}
    &    & \g &    &    
    &    &    & \g &    &    &    
    &    & \g &    &    &    &    
    &    &    &    &    &    
    & \multicolumn{2}{l}{ArXiv} \\
    \citet{bhattacharjee2024towards}
    &    & \g &    &    
    &    &    & \g &    &    &    
    &    & \g &    &    &    &    
    &    &    &    &    &    
    & \multicolumn{2}{l}{SafeGenAI} \\
    \citet{chu2024a}
    & \g & \g & \g &    
    &    &    & \g &    &    &    
    &    & \g &    &    &    &    
    &    &    &    &    &    
    & \multicolumn{2}{l}{CCS} \\
    \citet{rimsky2024steering}
    & \g & \g &    &    
    &    &    & \g &    &    &    
    &    & \g &    &    &    &    
    &    &    &    &    &    
    & \multicolumn{2}{l}{ACL} \\
    \citet{singh2024representation}
    &    & \g & \g &    
    &    &    & \g &    &    &    
    &    & \g &    &    &    &    
    &    &    &    &    &    
    & \multicolumn{2}{l}{ICML} \\
    \citet{zhang2024truthx}
    & \g &    &    &    
    &    &    & \g &    &    &    
    &    & \g &    &    &    &    
    &    &    &    &    &    
    & \multicolumn{2}{l}{ACL} \\
    \citet{li2023inference}
    & \g &    &    &    
    &    &    & \g &    &    &    
    &    & \g &    &    &    &    
    &    &    &    &    &    
    & \multicolumn{2}{l}{NeurIPS} \\
    \citet{turner2023steering}
    &    & \g &    &    
    &    &    & \g &    &    &    
    &    & \g &    &    &    &    
    &    &    &    &    &    
    & \multicolumn{2}{l}{ArXiv} \\
    \citet{gao2024shaping}
    &    & \g &    &    
    &    &    & \g &    &    &    
    &    & \g &    &    &    &    
    &    &    &    &    &    
    & \multicolumn{2}{l}{ArXiv} \\
    \citet{shen2024jailbreak}
    &    & \g &    &    
    &    &    & \g &    &    &    
    &    & \g &    &    &    &    
    &    &    &    &    &    
    & \multicolumn{2}{l}{ICLR} \\
    \citet{han2025internal}
    &    & \g &    &    
    &    &    & \g &    &    &    
    &    & \g &    &    &    &    
    &    &    &    &    &    
    & \multicolumn{2}{l}{ArXiv} \\
    \citet{hernandez2024inspecting}
    &    &    & \g &    
    &    &    & \g &    &    &    
    &    & \g &    &    &    &    
    &    &    &    &    &    
    & \multicolumn{2}{l}{COLM} \\
    \citet{chen2024designing}
    &    &    & \g &    
    &    &    & \g &    &    &    
    &    & \g &    &    &    &    
    &    &    &    & \g &    
    & \multicolumn{2}{l}{ArXiv} \\
    \citet{wang2024detoxifying}
    &    & \g &    &    
    &    &    & \g &    &    &    
    &    &    &    & \g &    &    
    &    &    &    &    &    
    & \multicolumn{2}{l}{ACL} \\
    \citet{li2024precision}
    &    & \g &    &    
    &    &    & \g &    &    &    
    &    &    &    & \g &    &    
    &    &    &    &    &    
    & \multicolumn{2}{l}{ArXiv} \\
    \citet{chen2025attributive}           
    & \g &    &    &    
    &    &    & \g &    &    &    
    &    &    &    &    & \g &    
    &    &    &    &    &    
    & \multicolumn{2}{l}{AAAI} \\
    \citet{zhao2024eeg}
    &    & \g &    &    
    &    &    & \g &    &    &    
    &    &    &    &    & \g &    
    &    &    &    &    &    
    & \multicolumn{2}{l}{ArXiv} \\
    \citet{burns2022discovering}
    & \g &    &    &    
    &    &    & \g &    &    &    
    &    &    &    &    & \g &    
    &    &    &    &    &    
    & \multicolumn{2}{l}{ArXiv} \\
    \citet{li2024hallucana}
    & \g &    &    &    
    &    &    & \g &    &    &    
    &    &    &    &    & \g &    
    &    &    &    &    &    
    & \multicolumn{2}{l}{ArXiv} \\
    \citet{monea2024glitch}
    & \g &    &    &    
    &    &    &    & \g &    &    
    &    &    &    &    &    &    
    &    &    &    &    &    
    & \multicolumn{2}{l}{ACL} \\
    \citet{vig2020investigating}
    &    &    & \g &    
    &    &    &    & \g &    &    
    &    &    &    &    &    &    
    &    &    &    &    &    
    & \multicolumn{2}{l}{NeurIPS} \\
    \citet{wei2024assessing}
    &    & \g &    &    
    &    &    &    & \g &    &    
    &    &    &    &    &    &    
    &    &    &    &    &    
    & \multicolumn{2}{l}{ICML} \\
    \citet{yang2023bias}
    &    &    & \g &    
    &    &    &    & \g &    &    
    &    &    &    &    &    &    
    &    &    &    &    &    
    & \multicolumn{2}{l}{ArXiv} \\
    \citet{zhou2025on}
    &    & \g &    &    
    &    &    &    & \g &    &    
    &    &    &    &    &    &    
    &    &    &    &    &    
    & \multicolumn{2}{l}{ICLR} \\
    \citet{deng2025cram}
    & \g &    &    &    
    &    &    &    & \g &    &    
    &    &    & \g &    &    &    
    &    &    &    &    &    
    & \multicolumn{2}{l}{AAAI} \\
    \citet{liu2024the}
    &    &    & \g &    
    &    &    &    & \g &    &    
    &    &    & \g &    &    &    
    &    &    &    &    &    
    & \multicolumn{2}{l}{ICLR} \\
    \citet{li2024look}
    & \g &    &    &    
    &    &    &    & \g &    &    
    &    &    &    & \g &    &    
    &    &    &    &    &    
    & \multicolumn{2}{l}{ArXiv} \\
    \citet{ma2023deciphering}
    &    &    & \g &    
    &    &    &    & \g &    &    
    &    &    &    & \g &    &    
    &    &    &    &    &    
    & \multicolumn{2}{l}{EMNLP} \\
    \citet{li2025safety}
    &    & \g &    &    
    &    &    &    & \g &    &    
    &    &    &    & \g &    &    
    &    &    &    &    &    
    & \multicolumn{2}{l}{ICLR} \\
    \citet{jin2024cutting}
    & \g &    &    &    
    &    &    &    & \g & \g &    
    &    &    &    &    &    &    
    &    &    &    &    &    
    & \multicolumn{2}{l}{ACL} \\
    \citet{zhao2024defending}
    &    & \g &    &    
    &    &    &    & \g & \g &    
    &    &    &    & \g &    &    
    &    &    &    &    &    
    & \multicolumn{2}{l}{EMNLP} \\
    \bottomrule
\end{tabular}
\rmfamily
\normalsize
\centering
\label{tab:overview-all-1}
\vspace{-5pt}
\end{table*}

\setlength{\colwidth}{0.52cm}

\begin{table*}[t]
    \centering
    \caption{
    Overview of representative works at the intersections of  safety-focused interpretation (\autoref{sec:interpretation}), safety enhancements they inform (\autoref{sec:enhance}), and tools operationalizing them (\autoref{sec:toolkits}),
    extending \autoref{tab:overview} and continuing \autoref{tab:overview-all-1}.
    }
\vspace{-5pt}
    \sffamily
    \mosttiny %
    \arrayrulecolor{black}
    \setlength{\arrayrulewidth}{1pt}
    \renewcommand{\arraystretch}{0.7}
    \rowcolors{2}{lightgray-overview}{white}
    \setlength{\tabcolsep}{0pt}
    \begin{tabular}{
        r @{\hskip 5pt}
        C{\colwidth} C{\colwidth} C{\colwidth} C{\colwidth} 
        !{\color{black}\vrule width 0.5pt} 
        C{\colwidth} C{\colwidth} C{\colwidth} C{\colwidth} C{\colwidth} C{\colwidth} 
        !{\color{black}\vrule width 0.5pt} 
        C{\colwidth} C{\colwidth} C{\colwidth} C{\colwidth} C{\colwidth} C{\colwidth} 
        !{\color{black}\vrule width 0.5pt} 
        C{\colwidth} C{\colwidth} C{\colwidth} C{\colwidth} C{\colwidth}
        !{\color{black}\vrule width 0.5pt} 
        @{\hskip 5pt} C{\colwidth} C{\colwidth}
    }
    \multicolumn{1}{c}{} & 
    \multicolumn{4}{c@{\hskip 0pt}}{\textcolor{darkgray}{\makecell{\cellcolor{white}\textbf{SAFETY TYPE}}}} & 
    \multicolumn{6}{c}{\makecell{\cellcolor{white}\hspace{-0.5em}\hyperref[sec:interpretation]{\textcolor{matrix_blue}{\S\ref*{sec:interpretation}\hspace{0.4em}\textbf{INTERPRET FOR SAFETY}}}}} & 
    \multicolumn{6}{c}{\makecell{\cellcolor{white}\hspace{-1em}\hyperref[sec:enhance]{\textcolor{matrix_purple}{\S\ref*{sec:enhance}\hspace{0.4em}\textbf{ENHANCE SAFETY}}}}} & 
    \multicolumn{5}{c}{\makecell{\cellcolor{white}\hspace{-1.8em}\hyperref[sec:toolkits]{\textcolor{matrix_green}{\S\ref*{sec:toolkits}\hspace{0.2em}\textbf{PRACTICAL TOOLS}}}}} & 
    \multicolumn{2}{c@{\hskip 5pt}}{\textcolor{darkgray}{\textbf{VENUE}}} \\        
    \rowcolor{white}
    \multicolumn{1}{r@{\hskip 5pt}}{\textbf{Work}} & 
    \rotatebox{90}{\scalebox{0.8}[1]{Hallucination}} & 
    \rotatebox{90}{\scalebox{0.8}[1]{Jailbreak \& Harm}} & 
    \rotatebox{90}{\scalebox{0.8}[1]{Bias}} & 
    \rotatebox{90}{\scalebox{0.8}[1]{Privacy Leakage}} & 
    \rotatebox{90}{\scalebox{0.8}[1]{\roundedbox{3.1} Training Attrib.}} & 
    \rotatebox{90}{\scalebox{0.8}[1]{\roundedbox{3.2} Input Token}} & 
    \rotatebox{90}{\scalebox{0.8}[1]{\roundedbox{3.3.1} Probe Latent}} & 
    \rotatebox{90}{\scalebox{0.8}[1]{\roundedbox{3.3.2} Perturb Comp}} & 
    \rotatebox{90}{\scalebox{0.8}[1]{\roundedbox{3.3.3} Decipher Latent}} & 
    \rotatebox{90}{\scalebox{0.8}[1]{\roundedbox{3.4} Self-Reason}} & 
    \rotatebox{90}{\scalebox{0.8}[1]{\roundedbox{4.1} Attn. to Rel. Token}} & %
    \rotatebox{90}{\scalebox{0.8}[1]{\roundedbox{4.2.1} Steer Latent Vec}} & 
    \rotatebox{90}{\scalebox{0.8}[1]{\roundedbox{4.2.2} Modulate Neuron}} & %
    \rotatebox{90}{\scalebox{0.8}[1]{\roundedbox{4.2.3} Edit Model}} & 
    \rotatebox{90}{\scalebox{0.8}[1]{\roundedbox{4.3} Verify \& Output}} & 
    \rotatebox{90}{\scalebox{0.8}[1]{\roundedbox{4.4} Output w. Reason}} & 
    \rotatebox{90}{\scalebox{0.8}[1]{Ease Impl.}} & 
    \rotatebox{90}{\scalebox{0.8}[1]{\roundedbox{5.1} TDA Vis}} & 
    \rotatebox{90}{\scalebox{0.8}[1]{\roundedbox{5.2} Token Vis}} & 
    \rotatebox{90}{\scalebox{0.8}[1]{\roundedbox{5.3} Latent Vec Vis}} & 
    \rotatebox{90}{\scalebox{0.8}[1]{\roundedbox{5.4} Neuron Vis}} & 
    \multicolumn{2}{c@{\hskip 5pt}}{\rotatebox{90}{\scalebox{0.8}[1]{Publication}}} \\  
    \midrule 
    \citet{ferrando2025do}
    & \g &    &    &    
    &    &    &    &    & \g &    
    &    &    &    &    &    &    
    &    &    &    &    &    
    & \multicolumn{2}{l}{ICLR} \\
    \citet{theodorus2025finding}
    & \g &    &    &    
    &    &    &    &    & \g &    
    &    &    &    &    &    &    
    &    &    &    &    &    
    & \multicolumn{2}{l}{ICLR} \\
    \citet{muhamed2025decoding}
    &    & \g &    &    
    &    &    &    &    & \g &    
    &    &    &    &    &    &    
    &    &    &    &    &    
    & \multicolumn{2}{l}{NAACL} \\
    \citet{haerle2024scar}
    &    & \g &    &    
    &    &    &    &    & \g &    
    &    &    &    &    &    &    
    &    &    &    &    &    
    & \multicolumn{2}{l}{ArXiv} \\
    \citet{gallifant2025sparse}
    &    & \g &    &    
    &    &    &    &    & \g &    
    &    &    &    &    &    &    
    &    &    &    &    &    
    & \multicolumn{2}{l}{ArXiv} \\
    \citet{marks2025sparse}
    &    &    & \g &    
    &    &    &    &    & \g &    
    &    &    &    &    &    &    
    &    &    &    &    &    
    & \multicolumn{2}{l}{ICLR} \\
    \citet{jiang2024large}
    & \g &    &    &    
    &    &    &    &    & \g &    
    &    &    &    &    &    &    
    &    &    &    &    &    
    & \multicolumn{2}{l}{NAACL} \\
    \citet{hernandez2024linearity}
    &    &    & \g &    
    &    &    &    &    & \g &    
    &    &    &    &    &    &    
    &    &    &    & \g &    
    & \multicolumn{2}{l}{ICLR} \\
    \citet{ameisen2025circuit}
    &    &    &    &    
    &    &    &    &    & \g &    
    &    &    &    &    &    &    
    &    &    &    &    & \g 
    & \multicolumn{2}{l}{Anthropic} \\
    \citet{lindsey2025on}
    & \g & \g & \g &    
    &    &    &    &    & \g &    
    &    &    &    &    &    &    
    &    &    &    &    & \g 
    & \multicolumn{2}{l}{Anthropic} \\
    \citet{zhou2025llm}
    &    &    & \g &    
    &    &    &    &    & \g &    
    &    &    & \g &    &    &    
    &    &    &    &    &    
    & \multicolumn{2}{l}{ICLR} \\
    \citet{frikha2025privacyscalpel}
    &    &    &    & \g 
    &    &    &    &    & \g &    
    &    &    & \g &    &    &    
    &    &    &    &    &    
    & \multicolumn{2}{l}{ArXiv} \\
    \citet{hegde2024effectiveness}
    &    &    & \g &    
    &    &    &    &    & \g &    
    &    &    & \g &    &    &    
    &    &    &    &    &    
    & \multicolumn{2}{l}{SciForDL} \\
    \citet{he2025towards}
    & \g & \g & \g &    
    &    &    &    &    & \g &    
    &    &    & \g &    &    &    
    &    &    &    &    &    
    & \multicolumn{2}{l}{ArXiv} \\
    \citet{abdaljalil2025safe}
    & \g &    &    &    
    &    &    &    &    & \g &    
    &    &    & \g &    &    &    
    &    &    &    &    &    
    & \multicolumn{2}{l}{ArXiv} \\
    \citet{bayat2025steering}
    & \g &    &    &    
    &    &    &    &    & \g &    
    &    &    & \g &    &    &    
    &    &    &    &    &    
    & \multicolumn{2}{l}{ArXiv} \\
    \citet{wu2025interpreting}
    &    & \g &    &    
    &    &    &    &    & \g &    
    &    &    & \g &    &    &    
    &    &    &    &    &    
    & \multicolumn{2}{l}{ArXiv} \\
    \citet{o2024steering}
    &    & \g &    &    
    &    &    &    &    & \g &    
    &    &    & \g &    &    &    
    &    &    &    &    &    
    & \multicolumn{2}{l}{ArXiv} \\
    \citet{geva2022lm}
    & \g &    & \g &    
    &    &    &    &    & \g &    
    &    &    & \g &    &    &    
    &    &    &    & \g &    
    & \multicolumn{2}{l}{EMNLP} \\
    \citet{khoriaty2025don}
    &    & \g &    &    
    &    &    &    &    & \g &    
    &    &    & \g &    &    &    
    &    &    &    &    &    
    & \multicolumn{2}{l}{ArXiv} \\
    \citet{geva2022transformer}
    &    & \g &    &    
    &    &    &    &    & \g &    
    &    &    & \g &    &    &    
    &    &    &    &    &    
    & \multicolumn{2}{l}{EMNLP} \\
    \citet{yu2024mechanistic}
    & \g &    &    &    
    &    &    &    &    & \g &    
    &    &    &    & \g &    &    
    &    &    &    &    &    
    & \multicolumn{2}{l}{EMNLP} \\
    \citet{lee2025hudex}
    & \g &    &    &    
    &    &    &    &    &    & \g 
    &    &    &    &    &    &    
    &    &    &    &    &    
    & \multicolumn{2}{l}{ArXiv} \\
    \citet{akbar2024hallumeasure}
    & \g &    &    &    
    &    &    &    &    &    & \g 
    &    &    &    &    &    &    
    &    &    &    &    &    
    & \multicolumn{2}{l}{EMNLP} \\
    \citet{li2024safetyanalyst}
    &    & \g & \g & \g 
    &    &    &    &    &    & \g 
    &    &    &    &    &    &    
    &    &    &    &    &    
    & \multicolumn{2}{l}{ArXiv} \\
    \citet{betley2025tell}
    &    & \g &    & \g 
    &    &    &    &    &    & \g 
    &    &    &    &    &    &    
    &    &    &    &    &    
    & \multicolumn{2}{l}{ICLR} \\
    \citet{weng2023large}
    & \g &    &    &    
    &    &    &    &    &    & \g 
    &    &    &    &    & \g &    
    &    &    &    &    &    
    & \multicolumn{2}{l}{EMNLP} \\
    \citet{cheng2025think}
    & \g &    &    &    
    &    &    &    &    &    & \g 
    &    &    &    &    & \g &    
    &    &    &    &    &    
    & \multicolumn{2}{l}{ArXiv} \\
    \citet{dhuliawala2024chain}
    & \g &    &    &    
    &    &    &    &    &    & \g 
    &    &    &    &    & \g &    
    &    &    &    &    &    
    & \multicolumn{2}{l}{ACL} \\
    \citet{liu2025guardreasoner}
    &    & \g &    &    
    &    &    &    &    &    & \g 
    &    &    &    &    &    & \g 
    &    &    &    &    &    
    & \multicolumn{2}{l}{ArXiv} \\
    \citet{jiang2025comt}
    & \g &    &    &    
    &    &    &    &    &    & \g 
    &    &    &    &    &    & \g 
    &    &    &    &    &    
    & \multicolumn{2}{l}{ICASSP} \\
    \citet{ji2024chain}
    & \g &    &    &    
    &    &    &    &    &    & \g 
    &    &    &    &    &    & \g 
    &    &    &    &    &    
    & \multicolumn{2}{l}{AAAI} \\
    \citet{zhang2025safety}
    &    & \g &    &    
    &    &    &    &    &    & \g 
    &    &    &    &    &    & \g 
    &    &    &    &    &    
    & \multicolumn{2}{l}{ArXiv} \\
    \citet{kaneko2024evaluating}
    &    &    & \g &    
    &    &    &    &    &    & \g 
    &    &    &    &    &    & \g 
    &    &    &    &    &    
    & \multicolumn{2}{l}{ArXiv} \\
    \citet{prahallad2024significance}
    &    &    & \g &    
    &    &    &    &    &    & \g 
    &    &    &    &    &    & \g 
    &    &    &    &    &    
    & \multicolumn{2}{l}{ArXiv} \\
    \citet{li2024chain}
    &    & \g &    &    
    &    &    &    &    &    & \g 
    &    &    &    &    &    & \g 
    &    &    &    &    &    
    & \multicolumn{2}{l}{ArXiv} \\
    \citet{cao2024defending}
    &    & \g &    &    
    &    &    &    &    &    & \g 
    &    &    &    &    &    & \g 
    &    &    &    &    &    
    & \multicolumn{2}{l}{NaNA} \\
    \citet{rad2025refining}
    &    & \g &    &    
    &    &    &    &    &    & \g 
    &    &    &    &    &    & \g 
    &    &    &    &    &    
    & \multicolumn{2}{l}{ArXiv} \\
    \citet{moore2024reasoning}
    &    &    & \g &    
    &    &    &    &    &    & \g 
    &    &    &    &    &    & \g 
    &    &    &    &    &    
    & \multicolumn{2}{l}{ArXiv} \\
    \citet{sicilia2024eliciting}
    &    &    & \g &    
    &    &    &    &    &    & \g 
    &    &    &    &    &    & \g 
    &    &    &    &    &    
    & \multicolumn{2}{l}{NLP4PI} \\
    \citet{mou2025saro}
    &    & \g &    &    
    &    &    &    &    &    & \g 
    &    &    &    &    &    & \g 
    &    &    &    &    &    
    & \multicolumn{2}{l}{ArXiv} \\
    \citet{liu2025guardreasoner}
    &    & \g &    &    
    &    &    &    &    &    & \g 
    &    &    &    &    &    & \g 
    &    &    &    &    &    
    & \multicolumn{2}{l}{ArXiv} \\
    \citet{kwon2023finspector}
    &    &    & \g &    
    &    &    &    &    &    &    
    &    &    &    &    &    &    
    &    &    &    & \g &    
    & \multicolumn{2}{l}{IUI} \\
    \bottomrule
\end{tabular}
\rmfamily
\normalsize
\centering
\label{tab:overview-all-2}
\vspace{-5pt}
\end{table*}
\end{document}